\documentclass[twocolumn]{aastex7}

\newcommand{\Ha}{\textrm{H}\ensuremath{\alpha}\xspace}
\newcommand{\Hb}{\textrm{H}\ensuremath{\beta}\xspace}
\newcommand{\Hg}{\textrm{H}\ensuremath{\gamma}\xspace}

\newcommand{\OII}{[\textrm{O}~\textsc{ii}]\xspace}
\newcommand{\OIII}{[\textrm{O}~\textsc{iii}]\xspace}
\newcommand{\NII}{[\textrm{N}~\textsc{ii}]\xspace}

\newcommand{\logm}{\log(M_*/M_\odot)}
\usepackage{graphicx}
\usepackage{xspace}
\usepackage{subcaption}
\usepackage{amsmath}
\usepackage{multirow}
\usepackage{pgffor}
\newcommand{\oh}{\ensuremath{12+\log({\rm O/H})}\xspace}
\newcommand{\el}[1]{\ensuremath{\textrm{EL}_{#1}}}
\newcommand{\Msun}{\ensuremath{M_\odot}\xspace}
\defcitealias{Erb_08}{Erb08}

\graphicspath{{./}{}{./img/}{./individual_spec/}{./NGDEEP_fig1/}}
\usepackage[flushleft]{threeparttable}

\begin{document}

\title{A 13-Billion-Year View of Galaxy Growth: Metallicity Gradient Evolution from the Local Universe to $z=9$ with JWST and Archival Surveys}

\author[0000-0001-5951-459X]{Zihao Li}
\affiliation{Department of Astronomy, Tsinghua University, Beijing 100084, China}
% \href{mailto:zcai@tsinghua.edu.cn}{{\rm zcai@tsinghua.edu.cn}}
\affiliation{Cosmic Dawn Center (DAWN), Denmark}
% \href{mailto:zihao.li@nbi.ku.dk}{{\rm zihao.li@nbi.ku.dk}}
\affiliation{Niels Bohr Institute, University of Copenhagen, Jagtvej 128, DK-2200, Copenhagen N, Denmark}
\email[show]{zihao.li@nbi.ku.dk}
\author[0000-0001-8467-6478]{Zheng Cai}
\affiliation{Department of Astronomy, Tsinghua University, Beijing 100084, China}
% \href{mailto:zcai@tsinghua.edu.cn}{{\rm zcai@tsinghua.edu.cn}}
\email[show]{zcai@tsinghua.edu.cn}
\author[0000-0002-9373-3865]{Xin Wang}
% \href{mailto:xwang@ucas.ac.cn}{{\rm xwang@ucas.ac.cn}}
\affiliation{School of Astronomy and Space Science, University of Chinese Academy of Sciences, Beijing 100049, China}
\affiliation{National Astronomical Observatories, Chinese Academy of Sciences, Beijing 100101, China}
\affiliation{Institute for Frontiers in Astronomy and Astrophysics, Beijing Normal University, Beijing 102206, China}
\email[show]{xwang@ucas.ac.cn}
\author[0000-0001-7890-4964]{Zhaozhou Li}
\affiliation{Center for Astrophysics and Planetary Science, Racah Institute of Physics, The Hebrew University, Jerusalem, 91904, Israel}
\email{}
\author[0000-0003-4174-0374]{Avishai Dekel}
\affiliation{Center for Astrophysics and Planetary Science, Racah Institute of Physics, The Hebrew University, Jerusalem, 91904, Israel}
\affiliation{Santa Cruz Institute for Particle Physics, University of California, Santa Cruz, CA 95064, USA}
\email{}
\author[0000-0002-7767-8472]{Kartick C. Sarkar}
\affiliation{Center for Astrophysics and Planetary Science, Racah Institute of Physics, The Hebrew University, Jerusalem, 91904, Israel}
\affiliation{Raman Research Institute, Sadashivanagar, C. V. Raman Avenue, 560 080, Bangalore, India}
\email{}
\author[0000-0002-2931-7824]{Eduardo Bañados}
\affiliation{Max-Plank-Institut für Astronomie, Königstuhl 17, D-69117 Heidelberg, Germany}
\email{}
\author[0000-0002-1620-0897]{Fuyan Bian}
\affiliation{European Southern Observatory, Alonso de Cordova 3107, Casilla 19001, Vitacura, Santiago 19, Chile}
\email{}
\author[0000-0002-7080-2864]{Aklant K. Bhowmick}
\affiliation{Department of Physics, University of Florida, Gainesville, Florida 32601, USA}
\email{}
\author[0000-0002-2183-1087]{Laura Blecha}
\affiliation{Department of Physics, University of Florida, Gainesville, Florida 32601, USA}
\email{}
\author[0000-0001-8582-7012]{Sarah E. I. Bosman}
\affiliation{Max-Plank-Institut für Astronomie, Königstuhl 17, D-69117 Heidelberg, Germany}
\affiliation{Institute for Theoretical Physics, Heidelberg University, Philosophenweg 12, D–69120, Heidelberg, Germany}
\email{}
\author[0000-0002-6184-9097]{Jaclyn B. Champagne}
\affiliation{Steward Observatory, University of Arizona, 933 N Cherry Avenue, Tucson, AZ 85721, USA}
\email{}
\author[0000-0002-6336-3007]{Xiaohui Fan}
\affiliation{Steward Observatory, University of Arizona, 933 N Cherry Avenue, Tucson, AZ 85721, USA}
\email{}
\author[0000-0001-5160-6713]{Emmet Golden-Marx}
\affiliation{INAF-Astronomical Observatory of Padova vicolo dell’Osservatorio 5 35122 Padova, Italy}
\email{}
\author[0000-0003-1470-5901]{Hyunsung D. Jun}
\affiliation{Department of Physics, Northwestern College, 101 7th Street SW, Orange City, Iowa 51041, USA}
\email{}
\author[0000-0001-6251-649X]{Mingyu Li}
\affiliation{Department of Astronomy, Tsinghua University, Beijing 100084, China}
\email{}
\author[0000-0001-6052-4234]{Xiaojing Lin}
\affiliation{Department of Astronomy, Tsinghua University, Beijing 100084, China}
\affiliation{Steward Observatory, University of Arizona, 933 N Cherry Avenue, Tucson, AZ 85721, USA}
\email{}
\author[0000-0003-3762-7344]{Weizhe Liu}
\affiliation{Steward Observatory, University of Arizona, 933 N Cherry Avenue, Tucson, AZ 85721, USA}
\email{}
\author[0000-0002-4622-6617]{Fengwu Sun}
\affiliation{Center for Astrophysics | Harvard \& Smithsonian, 60 Garden St., Cambridge, MA 02138, USA}
\email{}
\author[0000-0002-6849-5375]{Maxime Trebitsch}
\affiliation{LUX, Observatoire de Paris, Université PSL, Sorbonne Université, CNRS, 75014 Paris, France}
\email{}
\author[0000-0003-4793-7880]{Fabian Walter}
\affiliation{Max-Plank-Institut für Astronomie, Königstuhl 17, D-69117 Heidelberg, Germany}
\email{}
\author[0000-0002-7633-431X]{Feige Wang}
\affiliation{Department of Astronomy, University of Michigan, 500 S State St, Ann Arbor, MI 48109, USA}
\email{}
\author[0000-0003-0111-8249]{Yunjing Wu}
\affiliation{Department of Astronomy, Tsinghua University, Beijing 100084, China}
\email{}
\author[0000-0001-5287-4242]{Jinyi Yang}
\affiliation{Department of Astronomy, University of Michigan, 500 S State St, Ann Arbor, MI 48109, USA}
\email{}
\author[0000-0002-0123-9246]{Huanian Zhang}
\affiliation{Department of Astronomy, Huazhong University of Science and Technology, Wuhan 430074, China}
\email{}
\author[0000-0002-0427-9577]{Shiwu Zhang}
\affiliation{Zhejiang Lab, Hangzhou, Zhejiang 311121, China}
\email{}
\author[0000-0001-5105-2837]{Mingyang Zhuang}
\affiliation{Department of Astronomy, University of Illinois at Urbana-Champaign, Urbana, IL 61801, USA}
\email{}
\author[0000-0002-3983-6484]{Siwei Zou}
\affiliation{Chinese Academy of Sciences South America Center for Astronomy, National Astronomical Observatories, CAS, Beijing 100101, China}
\affiliation{Departamento de Astronom\'ia, Universidad de Chile, Casilla 36-D, Santiago, Chile}
\email{}
%% Use the \collaboration command to identify collaborations. This command
%% takes an optional argument that is either a number or the word "all"
%% which tells the compiler how many of the authors above the command to
%% show. For example "\collaboration[all]{(DELVE Collaboration)}" wil include
%% all the authors above this command.
%%
%% Mark off the abstract in the ``abstract'' environment. 
\begin{abstract}
The galaxy gas-phase metallicity gradients have been extensively studied over the past four decades, both in the local and high-redshift universe, as they trace the baryon cycle and growth of galaxies. With the unprecedented spatial resolution and sensitivity of JWST, it is now possible to measure metallicity and its radial gradients out to redshifts as high as $z = 9$. Here, we present a sample of 455 spectroscopically confirmed galaxies from redshifts $1.7 \lesssim z \lesssim 9$ that are spatially resolved on sub-kiloparsec (kpc) scales by deep JWST NIRCam or NIRISS Wide Field Slitless Spectroscopy (WFSS). Synthesizing these new JWST observations with legacy observations from the literature, we observe that at redshift $z > 5$, galaxy centers are more metal-rich, exhibiting negative metallicity gradients of $\sim-0.4$ dex kpc$^{-1}$. These gradients flatten over time, reaching near-zero around $z \approx 2$, coinciding with the peak of the cosmic star formation rate. Beyond this point, the gradients become negative again at lower redshifts approaching $z=0$.
This evolution likely reflects transitions in galaxy formation modes: an inside-out growth phase dominated by intense central star formation with inefficient feedback and limited gas mixing during ``cosmic dawn", enhanced gas mixing due to feedback-driven wind and gas accretion at ``cosmic noon", and a later phase of slow evolution and reduced feedback toward the present day. These physical processes, including gas accretion and feedback, not only regulate star and galaxy formation on a cosmic scale but also shape the evolutionary pathways of individual galaxies over cosmic time.

\end{abstract}

%% Keywords should appear after the \end{abstract} command. 
%% The AAS Journals now uses Unified Astronomy Thesaurus (UAT) concepts:
%% https://astrothesaurus.org
%% You will be asked to selected these concepts during the submission process
%% but this old "keyword" functionality is maintained in case authors want
%% to include these concepts in their preprints.
%%
%% You can use the \uat command to link your UAT concepts back its source.

\section{Introduction}
The process of star formation and the rate at which stars form across the universe \citep[referred to as star formation rate density (SFRD),][]{Madau_14} depend on a complex interplay of gas inflows, gas outflows, and other mechanisms operating at different epochs in cosmic history. As such, the dominant modes of galaxy formation are also expected to vary with time \citep{Schreiber_15}.
These processes further influence the metal content in galaxies' interstellar medium (ISM), referred to as metallicity, as well as the spatial distribution of metals within galaxies over time. They provide insight into the overall growth and evolution of galaxies throughout cosmic history.

However, key physical processes that dominate galaxy formation and evolution at each epoch are still subject to debate. Traditionally, galaxy evolution has been understood as a gradual process, shaped by both internal dynamics and external interactions over long timescales. These include processes such as the gradual accretion of gas, the formation of stellar structures like bars, gravitational interactions with neighboring galaxies (galaxy harassment), and galaxy mergers \citep{Kormendy_04, Keres_05, Hopkins_09, Lang_14}.
Additionally, more dynamic and rapid events with shorter time scales, such as feedback from stellar winds and explosive supernovae (SNe) that last only a few million years (Myr), play a critical role in regulating star formation and the growth of galaxies \citep{Erb_15,Badry_16}. Over the decades, various theoretical models and simulations have been developed to explain galaxy evolution \citep{Larson_74,Lilly_13,Magrini_16,Hopkins_23}. Comprehensive models are required to account for the intricate processes driving the formation and distribution of chemical elements in galaxies, which is key to understanding their overall evolution \citep{Tinsley_80, Contreras_25,Lyu_25,Graf_24b}. 

Observing the cosmological evolution of metallicity gradients offers insights into the dominant processes at different epochs, as metal distribution is sensitive to gas transport and turbulence driven by these activities \citep{Sharda_23}. In the Milky Way, metallicity gradients have been extensively studied as a benchmark for understanding the chemical evolution of disk galaxies, providing a local reference point for interpreting extragalactic observations \citep{Molla_19,Lian_23,Graf_24,Johnson_24,Ratcliffe_25}.
Over the past four decades, significant advancements have been made in observing galaxies' metal contents and their distribution via multiple techniques such as long slit or multi-object fiber spectrograph  \citep{Pagel_81,Edmunds_84,Evans_86,Vilchez_88,Shields_90,Vila-Costas_92,vanZee_98,Considere_00,Magrini_07,Rupke_10a,Grasha_22}, integral field unit (IFU) observations \citep{Cresci_10, Swinbank_12, Queyrel_12, Jones_13, Troncoso_14, Carton_15, Ho_15, Wuyts_16, Leethochawalit_16, Belfiore_17, Lian_18, Poetrodjojo_18, Sanchez_18, Forster_18, Carton_18, Thorp_19,Bresolin_19, Zhuang_19, Curti_20a, Franchetto_21, Gillman_21, Lagos_22,Venturi_24, Khoram_25,Ju_24, Lit_25}, slitless spectroscopy \citep{Jones_15,Wang_17,Wang_20,Simons_21,Li_22,Wang_22, Cheng_24}, and radio telescopes \citep{Vallini_24}. 
Those works found a diversity of metallicity gradients across stellar masses and redshifts. They have reported negative metallicity gradients in local galaxies \citep[e.g.,][]{Ho_15} as well as at higher redshifts \citep[e.g.,][]{Forster_18}. Positive (inverted) gradients have also been observed, both in the local Universe \citep[e.g.,][]{Lit_25} and high-redshifts up to $z \sim 3$ \citep[e.g.,][]{Troncoso_14,Wang_22}. 

Various theoretical models have been proposed to explain certain behaviors of metallicity gradients. Negative gradients can arise in scenarios such as inside-out growth \citep{Boissier_00} and accretion disk with coplanar gas inflow \citep{Wangenci_22,Lyu_25}. \citet{Sharda_21a,Sharda_21b} suggest that negative gradients originate under equilibrium conditions, modeled through a more complex framework that accounts for the competition of radial advection, metal production, and gas accretion. The inside-out scenario is supported by analyzing radial distribution of stellar populations in low-redshift star-forming galaxies \citep{Perez_13,Goddard_17,Frankel_19}, as well as by simulations \citep{Tissera_16,Renaud_25}.

On the other hand, flat or positive gradients can arise in the case of pristine gas accretion \citep{Cresci_10,Molina_17,Li_22}, metal loss due to feedback-driven outflows \citep{Wang_19,Porter_22,Sunxd_24}, and mergers \citep{Rupke_10a,Rupke_10b,Torrey_12,Tissera_22}. The analytical model by \citet{Sharda_21a} suggests galaxies with positive gradients are out of equilibrium with extreme outflow metal loss, pristine gas accretion, or the re-accretion of metal-enriched gas \citep{Fu_13}. Observations with the much-enriched circumgalactic medium (CGM) on large scales at cosmic noon \citep{Cai_17,Zhang_23} also reflect that metal-enrichment could happen on a larger scale than previously expected. 

However, a comprehensive understanding of how metallicity evolves over cosmic time remains elusive. This is largely because earlier observations often lacked the spatial resolution and sensitivity needed to map metal distributions within galaxies at scales smaller than a few hundred parsecs (sub-kiloparsec scales), which is crucial to resolve internal structures and small-scale physical processes, and at greater distances corresponding to redshifts $z\gtrsim3$.
With several years of JWST observations, a large spectroscopic sample of high-redshift galaxies has now been assembled, enabling a comprehensive investigation of metallicity gradients across cosmic time \citep{Wang_22,Birkin_23,Venturi_24,Arribas_24,Ju_24}.

In this paper, we present a sample of 455 galaxies spanning redshifts from $z=1.7-9$ from JWST Wide Field Slitless Spectroscopic (WFSS) surveys, and measure the spatially resolved metallicities. Leveraging literature observations from archival surveys, we analyze the redshift evolution of metallicity gradients from $z=0$ to $z=9$. 

This paper is organized as follows. In Section \ref{sec:obs}, we describe the observations and data reduction. The methods are described in Section \ref{sec:method}. We present the results in Section \ref{sec:results} and discuss the physical implications of the observations. We summarize the main findings in Section \ref{sec:conclusion}.
Throughout this article, we assume a standard flat $\Lambda$CDM cosmology with $\Omega_m=0.3,\ \Omega_\Lambda=0.7$, and $H_0=70\ {\rm km\ s^{-1} Mpc}^{-1}$. Emission lines are indicated as follows: $\OIII\lambda5008$ := \OIII, $\OII \lambda\lambda3727, 3730$ := \OII, $\NII \lambda6585$ := $\NII$, if presented without wavelength values.

\section{Description of data}\label{sec:obs}
\subsection{Observations and data reduction}

In this work, we analyzed JWST observations from ``A SPectroscopic survey of biased halos In the Reionization Era" \citep[ASPIRE,][]{Wangf_23, Yang_23}, % GO-2078, PI: Feige Wang), 
``First Reionization Epoch Spectroscopically Complete Observations" \citep[FRESCO,][]{Oesch_23},  
%; GO-1895; PI: Pascal Oesch)  
and ``The Next Generation Deep Extragalactic Exploratory Public Survey" \citep[NGDEEP,][]{Bagley_23}.
%; GO-2079; PI: Steven Finkelstein, Casey Papovich and Nor Pirzkal). 
By combining these surveys, we built a large statistical sample of galaxies that have spatially-resolved grism spectroscopy at $z\sim2-8$ that can be used for metallicity gradient studies with sub-kpc resolution. The angular resolution reaches $0\farcs04-0\farcs14$ depending on the wavelength, corresponding to the physical size of $\sim0.5$ kpc in the redshift range $1.7<z<3.5$ and $\sim0.7$ kpc in the redshift range $5<z<9$. 

The ASPIRE survey contains NIRCam F356W R-grism observations of 25 Quasar fields, with direct imaging in the F356W, F200W, and F115W filters. FRESCO includes NIRCam F444W R-grism WFSS observations, with direct imaging in the F444W, F210M, and F182M filters. The NGDEEP observations include NIRISS WFSS R-grism and C-grism in the F200W, F150W, and F115W, with direct imaging in the F200W, F150W, and F115W NIRISS filters. The combined data products cover a spectral wavelength range $\lambda_{\rm obs}\in[1.0-2.2]\cup[3.1,5.0]\ \mu m$, enabling redshift coverage $z\in[1.3,3.5]\cup[5.3,9]$ for \OIII emitters.

The ASPIRE imaging is processed in the same way as described in \citet{Wangf_23, Yang_23}.
Briefly, the reduction was performed using version 1.8.3 of the JWST Calibration Pipeline with the reference files from version 11.16.15 of the standard Calibration Reference Data System.
After Stages 2 and 3, the images are aligned to Gaia DR3 and drizzled to a common pixel scale of 0\farcs031/pixel. For the FRESCO and NGDEEP samples, we use the public image products, including NIRCam and NIRISS, retrieved from the Dawn JWST Archive (DJA)\footnote{\url{https://dawn-cph.github.io/dja/}}. The DJA images are processed with Grism Redshift \& Line {Analysis tool} \citep[\textsc{Grizli,}][]{Brammer_23}.

We use JWST Calibration pipeline \texttt{CALWEBB} Stage 1 to calibrate individual NIRCam WFSS exposures, with reference files \texttt{jwst\_1090.pmap}. The $1/f$ noise is then subtracted using the routine described in \citet{Wangf_23}. The world coordinate system (WCS) information is assigned to each exposure with \texttt{assign\_wcs} step. The flat field is done with \texttt{CALWEBB} stage-2. We build the median backgrounds for the F356W and F444W filters based on all ASPIRE and FRESCO WFSS exposures, which are then scaled and subtracted from each exposure. We then measure the astrometric offsets between each of the short wavelength (SW) images and the fully calibrated direct mosaic to align each grism exposure with the direct image. 

The preprocessed NIRCam WFSS exposures are then processed with \textsc{Grizli}. We use the spectral tracing, grism dispersion, and sensitivity models described in \citet{Sun_2023}. The detection catalog for spectral extraction is built from the F356W (F444W) direct image for ASPIRE (FRESCO), and the continuum cross-contamination is subtracted by \textsc{Grizli} forward modeling using the reference image as the reference image for each grism exposure. The emission line maps are drizzled from the 2D grism to the image plane, matching the wcs of direct imaging, with \texttt{pixfrac = 1} and pixel size = 0\farcs06, to sample the NIRCam PSFs (FWHM=0\farcs115$\sim$0\farcs145) properly.

The NIRISS WFSS reduction is done similarly, with reference files \texttt{jwst\_1090.pmap} by \textsc{Grizli} full end-to-end processing, including \texttt{CALWEBB} Stage 1 calibration from raw exposures, flat fielding, $1/f$ noise subtraction, sky subtraction, and astrometry alignment. The contamination modeling and extraction of the spectra are performed with the latest \textsc{Grizli} NIRISS configuration file \citep{matharu_2022_7447545}. The emission line maps are drizzled from the 2D grism to the image plane with \texttt{pixfrac=1} and a pixel size of 0\farcs03, for properly sampling NIRISS PSFs (FWHM=0\farcs040$\sim$0\farcs066).

\subsection{Photometric catalog}

The sources in ASPIRE are identified using F356W as the detection image. We use \texttt{SourceXtractor++} \citep{Bertin_96}, a successor version of the original \texttt{SExtractor} \citep{sepp}, to construct the photometric catalog. The catalogs are extracted in F115W, F200W, and F356W bands after matching the PSFs to those of F356W (0\farcs1).
Fluxes measured in Kron apertures with ($k$=1.2, $R$=1.7) are corrected to total magnitudes using the ratio for Kron fluxes measured in larger apertures ($k=1.5, R=2.5$)
The uncertainties for each source's photometry are measured by placing 1000 random apertures the size of the Kron aperture across the image and measuring the root mean square (RMS). 

For the FRESCO and NGDEEP samples, we use the public catalogs retrieved from the Dawn JWST Archive (DJA).
The DJA catalogs are extracted with \texttt{SExtractor} \citep{Bertin_96}. The photometry extracted in circular apertures with a diameter of 0\farcs5 is corrected to the total fluxes within an elliptical Kron aperture. The details of the DJA image reduction and photometry are detailed in \citet{Valentino_23}.

\subsection{Sample selection}

We apply the emitter finder algorithm detailed in \citet{Wangf_23} to select galaxies with possible \OIII emission lines. Briefly, the algorithm first searches for peaks in 1D spectra and, assuming every peak as a candidate $\OIII\lambda5008$, measures the SNR at the expected position of $\OIII\lambda{4960}$. A good candidate has $\rm SNR_{\OIII\lambda4960}>2$ and a color excess in the corresponding image band. We note that there are $z>5$ \OIII emitters in ASPIRE and FRESCO that only have $\OIII\lambda5008$ detection because the $\OIII\lambda4960$ is too faint to be detected. 
We have ignored the sources where $\OIII\lambda4960$ is not detected 
because they will decrease the SNR in our stack. We remove candidates with cross-contamination in the 2D spectra by visual inspection. 

Since we only have \OIII and \Hb detection for our sample at $z>5$, we use empirical metallicity calibrations \citep[R3,][]{Sanders_23} to convert \OIII/\Hb ratio to metallicity (see more details in Section \ref{sec:z_measure}). The R3 calibration is double-valued at given line ratios, with a lower branch at $\mathrm{12 + \log(O/H)}\lesssim7.9$ and a higher branch at $\mathrm{12 + \log(O/H)}\gtrsim7.9$.
The metallicity estimations are unreliable if high-metallicity galaxies are mixed with the low-metallicity sample. One solution is to use the mass-metallicity relation to estimate the metallicity of galaxies and ensure the applicability of choosing the lower-branch solution.
\citet{Sarkar_24} provided an empirical mass-metallicity-redshift relation formulated as: $\mathrm{12 + \log(O/H)} = 6.29 + 0.237 \times \log(M_*/M_\odot) - 0.06 \times (1 + z)$. 
From this relation, galaxies at $5<z<9$ with stellar masses $\log(M_*/M_\odot)\lesssim9$ are expected to have metallicities $\mathrm{12 + \log(O/H)}\lesssim7.9$. 

From SED modeling (see details in Section \ref{sec:sed}), we find that most of our sample galaxies are below $\log(M_*/M_\odot)=9$, with only a small fraction above this threshold -- $12.6\%$ and $4.1\%$ for ASPIRE and FRESCO samples, respectively. To eliminate the bias of metallicity estimation by massive galaxies, we have removed galaxies with $\log(M_*/M_\odot)>9$ in our $z>5$ sample, and the low-mass sample includes 284 and 47 galaxies in ASPIRE and FRESCO, respectively. Since there is still a considerable number $(N=41)$ of massive $\log(M_*/M_\odot)>9$ galaxies in ASPIRE, we construct an additional massive sample with those galaxies, and we can measure their metallicity assuming the R3 upper branch solution (see Appendix \ref{sec:massbin}). Since we only have two galaxies with $\log(M_*/M_\odot)>9$ in FRESCO, which is insufficient to do a statistical study, we do not keep a $\log(M_*/M_\odot)>9$ bin in FRESCO.

In the NGDEEP sample, we additionally require the detection of \OIII, \OII, and \Hb, with the detection of either \Ha or \Hg, to ensure the usage of both the R3 and \OII/\Hb \citep[R2][]{Sanders_23} indices and usage of a reliable dust correction using the Balmer decrement. 
With the R2 calibration, we can resolve the double-branch ambiguity caused by the R3 calibration, since the \OII/\Hb ratio increases monotonically with metallicity at $12+\log(\rm O/H)\lesssim8.7$ \citep{Maiolino_08}.
We apply an SNR cut of three for both \OIII and \OII, which has been shown to yield reliable metallicity gradient measurements \citep{Wang_20}.

In our final galaxy sample, we incorporate 88 galaxies traced by \OIII with high SNR at $1.7<z<3.5$ in NGDEEP, 325 galaxies at $5.3<z<7$ in ASPIRE, and 42 galaxies at $7<z<9$ in FRESCO GOODS-S and GOODS-N. From line ratio diagnostics, these galaxies are negligibly impacted by AGN contamination (see Appendix \ref{sec:agn}).

\section{Methods}\label{sec:method}

\subsection{Stacking emission maps}

\begin{figure*}[ht!]
	\centering
	\includegraphics[width=\textwidth]{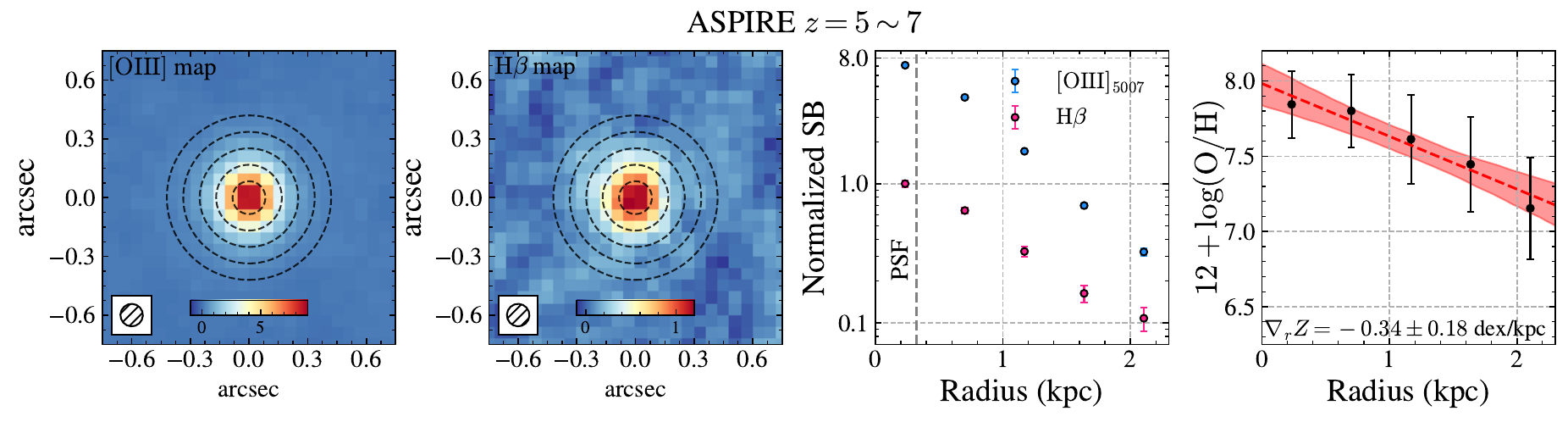}
	\includegraphics[width=\textwidth]{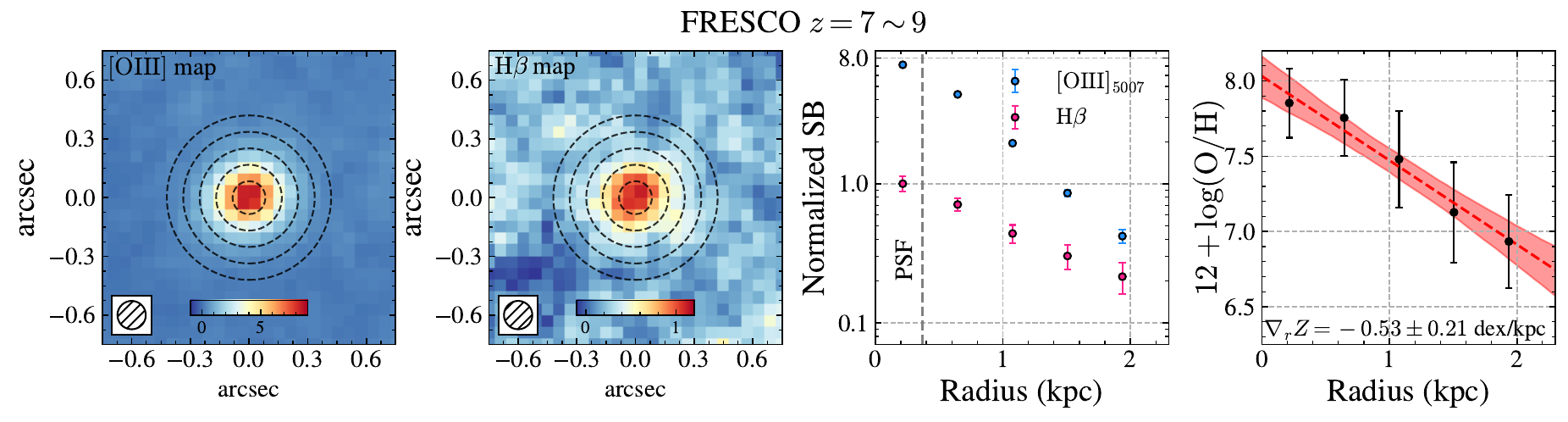}
	\caption{Stacking results of $5<z<7$ galaxies in ASPIRE (top) and $7<z<9$ galaxies in FRESCO (bottom). The first two columns show the median stack of emission maps (\OIII, \Hb\ ). As indicated by the color bar in the lower center of the line maps, the units correspond to surface brightness normalized by the peak \Hb\ flux. The black dashed annuli mark the region where we measure the line profiles. The FWHM of PSFs is shown on the bottom left in each line map. The third column shows the normalized SB profiles of \OIII and \Hb measured within different annuli. The vertical dashed line marks the angular resolution (half of the PSF FWHM). The fourth column shows the fitted metallicity gradient.  The black error bars include both statistical error and calibration error.
  The red shadow represents the $1\sigma$ confidence interval of the linear regression. Both samples in each redshift bin show steep gradients towards the center.
    \label{fig:zgrad-stacking}}
\end{figure*}

We create median maps of \OIII and \Hb by stacking individual galaxies in two redshift bins. As the galaxies in our sample have been observed under uniform conditions and have stable PSFs, stacking gives us the advantage of not requiring individual emission lines to be detected with high SNR. The stacked maps allow us to explore the regions with larger radii and lower surface brightness.
We first weight each map by \OIII flux measured in the 1D spectrum, to avoid excessive weighting toward bright sources \citep{Nelson_16, Wang_22a}.
To avoid introducing additional noise and systematics, we stack the galaxies without distorting the maps through rescaling, rotation, or de-projection.
% Before stacking the galaxies, we do not distort the maps either by rescaling, rotating, or deprojecting, not to introduce additional noise and systematics. 
We use \texttt{DAOStarFinder} \citep{larry_bradley_2024_13989456} to fit the centroid of each galaxy’s emission map using a Gaussian kernel, and then stack the centrally aligned emission maps.
% As noted by \citet{Forster_18}, this choice makes the morphological center position less sensitive to bright asymmetric or small-scale features such as clumps and is more robust when line emission is not centrally concentrated. 
Several relevant works also measure centers based on emission lines \citep{Forster_09, Wuyts_16, Forster_18}, although different methods may be applied. For example, \citet{Forster_18} defines the galaxy center as the morphological center of the outer isophote of the emission line map.
An alternative approach is to fit the center based on the continuum, but we did not find a significant difference in the stacked line profiles.
% It has been found that either rotating images to align major-axis of galaxies or deprojecting to source plane would provide qualitatively consistent results in similar studies\citep{Nelson_16}. 
% We do not rescale the map to physical units 
We create median stacks by measuring the median value at each pixel. 
The uncertainties of the radial profiles are computed by bootstrapping the stacks.
The physical sizes of the stacks are computed using the median redshift $z=6.28$ and $z=7.24$ in ASPIRE and FRESCO stacks, respectively. 
In Fig. \ref{fig:zgrad-stacking}, we show the stacked \OIII and \Hb maps in the first two panels, and the normalized radial profiles in the third panel.

\subsection{SED fitting}\label{sec:sed}

The SED modeling of our sample is performed with the Bayesian code \texttt{BEAGLE} \citep{Chevallard_16} with the broadband photometry and flux of \OIII\ lines as inputs. We adopt a delayed-$\tau$ star formation history (SFH), the SMC dust attenuation law, and a Chabrier initial mass function (IMF) \citep{Chabrier_03} with an upper limit of 100${\rm M}_\odot$. 
% which are broadly suggested in high-redshift star-forming galaxies (reference XX). 
We assume a flat prior in log-space for the characteristic star formation timescale, $\tau$, spanning from $10^7$ to $10^{10.5}$ years, and for stellar masses from $10^4M_\odot$ to $10^{12}M_\odot$. We set the effective optical depth in the V band in the range of $\hat\tau_{\rm V}\in[0,0.5]$ with a log-uniform prior. In FRESCO, the spectral fitting is performed using three JWST NIRCam bands (F182M, F210M, and F444W) and eight HST bands (F435W, F606W, F775W, F810W, F105W, F125W, F140W, F160W).  For the NGDEEP galaxy sample, we used eight JWST NIRCam bands (F090W, F182M, F210M, F277W, F335W, F356W, F410M, F444W), three JWST NIRISS bands (F115W, F150W and F200W), and eight  HST bands (F435W, F606W, F775W, F810W, F105W, F125W, F140W, F160W). For ASPIRE, we used three JWST NIRCam bands (F115W, F200W, F356W). 

We note that for the ASPIRE and FRESCO samples, we lack photometric bands (e.g., JWST MIRI) to fully constrain the SED shape redward of the 4000 \AA\ break, which is sensitive to age and dust properties of stellar populations. 
This limitation increases uncertainties in the derived physical properties, consistent with findings by \citet{Liq_24} who report that stellar mass and SFR uncertainties can increase by $\sim 0.1$ dex without MIRI coverage.
Although the SED code gives small errors on stellar masses, they only represent model-based statistical uncertainty. There are also notable systematic uncertainties in the assumed models, including star formation history, initial mass function, and dust attenuation laws. Stellar masses of young galaxies can differ significantly with different SFHs. Since young stars dominate the observed rest-UV and optical SED, they outshine potential older stellar populations that may be present. Consequently, the inferred early star formation history, along with the estimated age and stellar mass of a given system, can be sensitive to the choice of SFH model.
The parametric SFHs, like the constant and delayed-$\tau$, tend to underestimate the stellar masses compared with the non-parametric SFH, as they are not flexible enough to capture early star formation \citep{Whitler_23, Reyes_24}. Since many recent works on mass-metallicity relations have adopted parametric SFH to estimate stellar masses \citep{Matthee_23, Curti_24, Sarkar_24}, and our sample selection directly relies on the MZR provided by \citet{Sarkar_24}, the adherence of the parametric SFH model can be more consistent with their formalism. 

On the other hand, the choice of dust attenuation models also influences the fitted results. The empirical dust attenuation curves are derived based on nearby galaxies. The Calzetti law \citep{Calzetti_94} was derived from a sample of local
starburst galaxies, while the SMC law was measured with the Small Magellanic Cloud, showing a sharp rise at short wavelengths \citep{Weingartner_01}. However, as the dust properties may evolve through cosmic times, the local templates may not be applicable for all redshifts. \citet{Markov_23} found that the estimated stellar masses in a galaxy sample at $z=7-8$ can differ up to $\sim0.3$ dex depending on whether the Calzetti or SMC attenuation law is assumed. \citet{Markov_25} showed that the median dust attenuation slopes of galaxies at $z\sim2-3.5$ approach the SMC slope, but gradually flatten with increasing redshift beyond $z>3.5$.
Our choice of the SMC dust law is appropriate for our $z=1.7-3.5$ sample, but it may not provide the best fit for our $z>5$ samples. A more flexible model may better characterize the dust properties at different redshifts \citep[e.g., see][]{Markov_23,Vega_25}, but it is beyond the scope of this paper. Therefore, we caution potential uncertainties raised by different dust templates.

According to our SED modeling, the stellar masses in our final NGDEEP sample range from $10^{7}\Msun$ to $10^{10}\Msun$ with a median mass of $10^{8.43}\Msun$, and the stellar masses in ASPIRE and FRESCO range from $10^{7}\Msun$ to $10^{10}\Msun$ with a median mass $10^{8.16}\Msun$ (Table \ref{tab:info}).

\subsection{Metallicity measurements}\label{sec:z_measure}
We jointly constrain metallicity, nebular dust extinction ($A_v$), and de-reddened \Hb\ flux ($f_{H\beta}$), by the Bayesian inference method for forward modeling \citep{Li_22, Wang_17}. The likelihood function is defined as 
\begin{equation}
    \mathcal{L}\propto\exp\left(-\frac{1}{2}\cdot\sum_i \frac{\left(f_{\el{i}} - R_i \cdot f_{H\beta}\right)^2}{(\sigma_{\el{i}})^2 + (f_{H\beta})^2\cdot(\sigma_{R_i})^2}\right),\label{eq:mcmc}
\end{equation}
where $f_{\el{i}}$ and $\sigma_{\el{i}}$ represent the de-reddened emission-line (e.g., \OIII, \OII, \Hg, \Hb, \Ha) flux.
The de-reddened line fluxes are corrected from observed line ratios assuming the SMC dust law with the normalization $A_{\rm v}$ as a free parameter in the fitting.
$R_i$ corresponds to the expected line-flux ratios between EL$_i$ and H$\beta$ (i.e., $R_i$ can be the Balmer decrement of $\Hg/\Hb=0.6,\Ha/\Hb=2.86$, assuming case B recombination, and the metallicity diagnostics of $\OIII/\Hb$ and $\OII/\Hb$), and $\sigma_{R_i}$ is their intrinsic scatter. We perform MCMC sampling using the package $\textsc{Emcee}$ \citep{2013PASP..125..306F}. Potential stellar absorption may impact the dust corrections. The typical equivalent width of \Hb is found to be $\mathrm{EW(\Hb)}\sim1-6~\text{\AA}$ in local \Hb absorption line galaxies\citep{McDermid_15}. \citet{Carnall_23} reported $\mathrm{EW(H_\delta)}=7.9~\text{\AA}$ for a quiescent galaxy at $z=4.658$. However, for star-forming galaxies dominated by young O and B stars, the equivalent widths of Balmer absorption lines are much weaker. The low-mass star-forming galaxies are found to exhibit high \OIII and \Hb equivalent width both at $z\sim2$ and higher redshift $z\sim6$ \citep{van_11, Matthee_23}, with \Hb emission reaching $\mathrm{EW(\Hb})\sim100~\text{\AA}$. The Balmer absorption line-strength contributes only to $\sim1-10\%$ of emission line-strength and can be negligible. Several recent works also correct for dust by the observed Balmer emission line ratios without taking stellar absorption into account \citep{Curti_23a, Matharu_23}.

For the $z>5$ galaxy sample, we adopt the metallicity diagnostics presented by \citep{Sanders_23}. 
\citet{Sanders_23} built the empirical calibration using a sample of $z=2-9$ galaxies with $\OIII\lambda{4363}$ detected using the direct $T_e$
method, which better characterizes the properties of high-redshift galaxies. We also refer to \citet{Scholte_25,Chakraborty_25} who provide similar $T_e$ calibrations.
One caveat for the $\OIII/\Hb$ (R3) diagnostics is that it yields a double-branched solution if given a line ratio. For our NGDEEP sample, we have detections of both $\OII\lambda{3727}$ and $\OIII\lambda{5007}$.  With the additional $\OII\lambda{3727}$ 
we can distinguish the two solutions using  $\OII_{3727}/\Hb$ (R2). 
However, in the ASPIRE and FRESCO samples, $\OII_{3727}$ is out of the spectral coverage. Thus we assume the lower branch solution with $12+\log(\rm O/H)\lesssim7.9$, given the fact that most galaxies at similar redshifts of $z\ge 6-7$ have had their metal-poor nature verified using the $T_e$ method \citep{Sanders_23,Curti_23a,Nakajima_23,Trump_23,Jones_23,Chakraborty_25}.

We rely on the strong-line diagnostics from \citet{Bian_18} for sample galaxies at $1.7<z<3.5$, which was built from local analogs of $z\sim2$ galaxies. As \citet{Sanders_23} is built from a sample of high-redshift galaxies with direct metallicity measurements, it is more applicable for our sample of relatively metal-poor $z>5$ galaxies.  However, this model does not well constrain potential higher metallicity galaxies as their calibration is fitted with insufficient numbers of high metallicity galaxies. 
Because of this issue with \citet{Sanders_23}, we choose to use \citet{Bian_18} as it is robust for higher metallicity galaxies and is widely used for galaxies at $z\approx 2$ \citep{Wang_20}. 

In Appendix \ref{sec:syst}, we test the systematic uncertainties arising from different strong-line calibrations and find that alternative methods \citep{Maiolino_08,Curti_23a,Nakajima_22} yield consistent metallicity gradient measurements with our fiducial measurements using \citet{Sanders_23} and \citet{Bian_18}.
\subsection{Gradient measurements}
To securely measure metallicity gradients in the NGDEEP sample, we apply Voronoi tessellation \citep{Cappellari_03} to each galaxy's \OIII map to bin emission maps into subregions, each having $\mathrm{S/N}>5$ in \OIII. The Voronoi tessellation is superior to averaging the signal in radial annuli because the former keeps the information of azimuthal variations in metallicity. After the binning, the number of resolution elements is $\gtrsim50$ in our NGDEEP sample. Increasing the SNR threshold leads to coarser binning, which averages out spatial variations, whereas lowering the threshold yields more bins with low SNRs, thereby increasing the uncertainty in our measurements. An SNR cut of 5 has been tested to be an appropriate threshold for recovering robust gradient measurements, which balances both SNR and resolution elements. \citep{Wang_20}.
However, for our stacked sample, the azimuthal variations have already been averaged during stacking, so we simply average the flux in different annuli for $z>5$ samples. We evenly divide the maps into five concentric annuli with radii from $r=0$ to $r=0\farcs42$. Due to the different redshifts of the ASPIRE and FRESCO samples, the physical sizes of the corresponding annuli are slightly different (see Fig. \ref{fig:zgrad-stacking}).

We use a simple linear least-squares method to fit the metallicity gradients with the following formula:
\begin{equation}
    \oh=\theta_0+\theta_1 r
\end{equation}
where $\theta_0$ is the intercept, $\theta_1$ is the slope of the fitted line, and $r$ is the galactocentric radius in kiloparsecs. 

For the NGDEEP galaxies, we de-project the radius to the source plane. We assume the galaxies are circular disks, and the observed elasticity is due to projection effects. The inclination angle, $i$, is estimated by the axis ratio from S\'ersic fitting: $\cos(i)=b/a$. The distance of each position to the center is then corrected by inclination. We note the assumption of disks may not apply for every galaxy at high-redshift, especially given the recent discovery of less disky galaxies towards high-redshift \citep{Simons_17, Pandya_24}. While some works have corrected for projection \citep{Li_22, Wang_20, Forster_18}, others directly use the projected distance to measure gradients \citep{Simons_21, Curti_20a}. 
We find a $\rm \sim0.1\ dex\ kpc^{-1}$ difference between the projected gradients and de-projected gradients, comparable to the 1-sigma statistical uncertainties. We use the de-projected results of the NGDEEP sample throughout this article to remain consistent with our previous work \citep{Li_22, Wang_20}.
The stacking method for the $z>5$ sample averages the projection effects, so we do not correct for projection for the $z>5$ sample. The systematic uncertainties of the individual and the stacked measurements are estimated by mock observations in Appendix \ref{sec:uncertain}. We find good agreement between the stacked measurements and the median of the individual samples.

In the fourth panel of Fig. \ref{fig:zgrad-stacking}, we show the measurements on median stacks, and the individual measurements of the NGDEEP sample are shown in Figure \ref{fig:ngdeep-ind}. The metallicity gradients measured from the median stacks are reported in Table \ref{tab:zgrad_info}, and the individual measurements of the NGDEEP sample are reported in Table \ref{tab:info}.  

We note that variation in R3 ratios could also be attributed to changes in the ionization parameter (U) \citep{Garg_23}, which potentially leads to a bias in metallicity measurements. In Appendix \ref{sec:ion}, we show that the variations in R3 ratios are more significant than can be accounted for by changes in the ionization parameter, and potential variation in U only leads to an uncertainty of $\sim0.1~\rm dex~kpc^{-1}$. The steep gradients observed in Figure \ref{fig:zgrad-stacking} are therefore robust.

\begin{deluxetable}{ccc}
\tabletypesize{\small}
\tablecaption{Metallicity gradients from median stacks in two $z>5$ redshift bins.\label{tab:zgrad_info}}
\tablehead{
\colhead{Survey} & \colhead{Redshift} & \colhead{$\nabla_r\log(\mathrm{O/H})\ [\mathrm{dex~kpc^{-1}}]$}
}
\startdata
ASPIRE & $5.3<z<7.0$ & $-0.34 \pm 0.18$ \\
FRESCO & $7.0<z<9.0$ & $-0.53 \pm 0.21$ \\
\enddata
\end{deluxetable}

% There is a concern about the systematic bias when the stacks are coadded without de-projection. We further tested stacking the NGDEEP sample. We evenly divided the NGDEEP sample into two redshift bins of $1.7<z<2.3$ and $2.3<z<3.5$. We found that the metallicity gradients from stacked emission maps are consistent with the median metallicity gradients from individual measurements. The stacked results are listed in Table \ref{tab:zgrad_info}. 

\section{Results and Discussion}\label{sec:results}

\subsection{Redshift evolution of metallicity gradients}
\begin{figure*}[t!]
 \centering
 \hspace{1cm}\includegraphics[width=\textwidth,trim=0cm 0cm 0cm 0cm,clip]{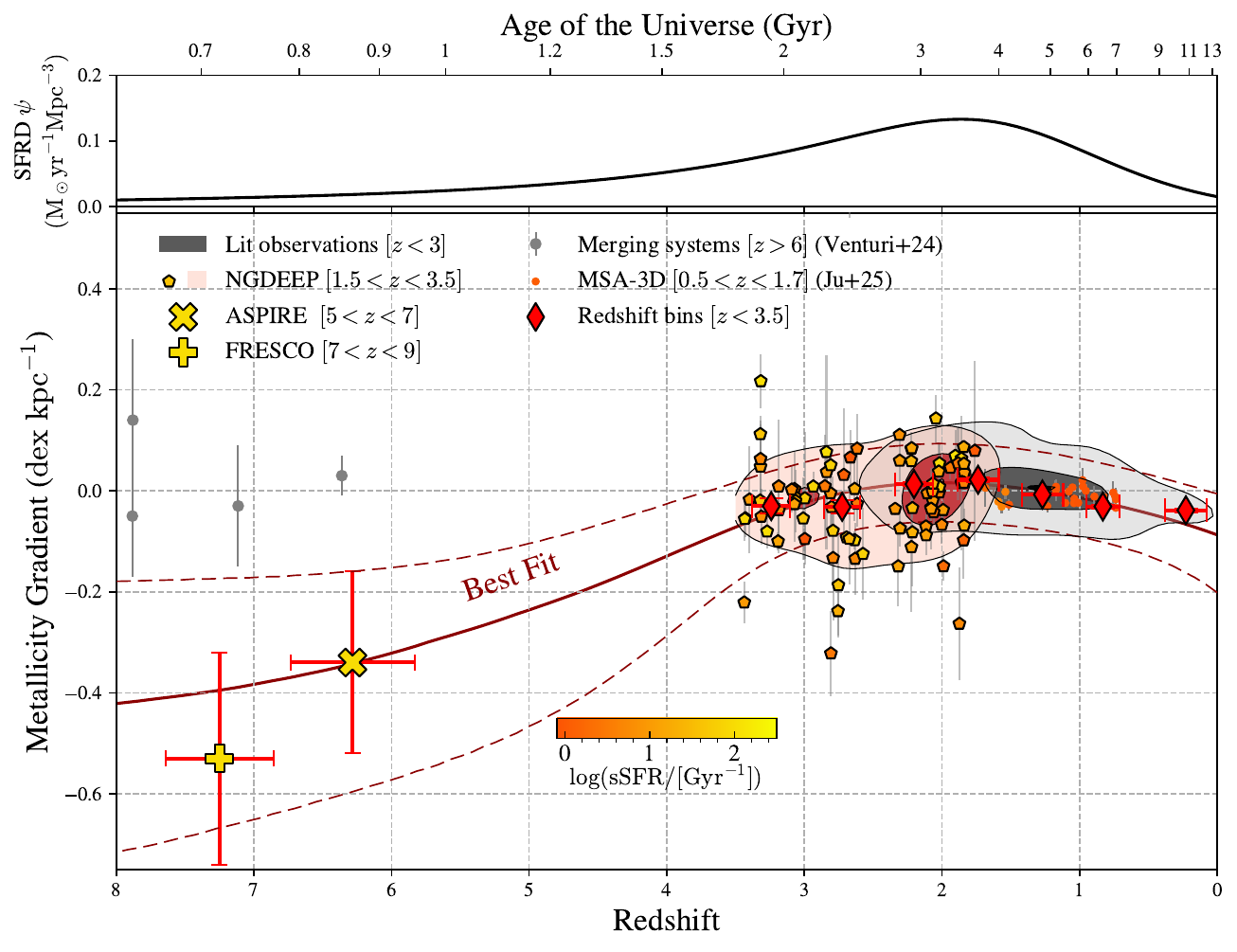} \\
 
 \caption{The redshift evolution of metallicity gradients. The thin red diamonds show the weighted mean in redshift bins at $z=[0,3.5]$ from both this work and the literature. Red ``$\times$" and ``+" represent the median stacks of the ASPIRE sample in $z\approx 6$ and the FRESCO sample in $z\approx 7$. 
 The NGDEEP sample at $z\approx 1-3$ is denoted by pentagons, color-coded by specific star formation rate (sSFR). Gray contours encompass individual measurements at $z<3$ in the literature using HST \citep{Wang_17,Wang_20,Simons_21}, and ground-based AO-assisted surveys \citep{Swinbank_12,Jones_13,Forster_18}. We also include seeing-limited observations \citep{vanZee_98,Queyrel_12,Curti_20a,Carton_18,Wuyts_16}.
 The red contour only encompasses our observations at $1.7<z<3.5$, overlapping with literature observations at $z\sim2$. Gray circles with error bars represent interacting galaxies at $z>6$ recently observed with JWST NIRSpec \citep{Venturi_24}. Recent JWST MSA-3D observations \citep{Ju_24} are noted in circles, color-coded by sSFR.
 The dark-red line shows our best fit of all stacks, with dashed lines in the same color showing the $1\sigma$ confidence interval of the fit. 
 In the upper panel, we plot the cosmic star formation rate density (SFRD \citep{Madau_14}), where the evolution of metallicity gradients also peaks at $z\sim2$.
 \label{fig:zgrad-z}}
 
\end{figure*}

To have a more complete study of the evolution of the metallicity gradient, we have added existing measurements from a variety of sources, including a wealth of JWST observation \citep{Ju_24}, HST observations \citep{Wang_17, Wang_20, Wang_22, Simons_21} and ground-based adaptive optics (AO)-assisted observations \citep{Forster_18,Jones_13,Swinbank_12}. These measurements are made mainly at $z\lesssim2$.
Additionally, we included samples obtained under seeing-limited conditions \citep{vanZee_98, Queyrel_12, Curti_20a, Carton_18, Wuyts_16} for a more comprehensive analysis. Although different metallicity calibrators \citep[e.g., R23, N2, O3N2,][]{Maiolino_08, Curti_20b} are used in different works, the systematic difference should not change the general trend in gradients \citep{Rupke_10b, Wang_19}, despite some potential impacts. Specifically, \citet{Poetrodjojo_21} found that metallicity gradients measured using different line diagnostics show deviations; however, different measurements generally correlate well with each other, with Spearman’s rank coefficients greater than 0.6.
Our final collected literature samples cover the redshift range $0<z<2.5$.

We note a difference between our $z>6$ results and recent JWST NIRSpec observations of \citet{Venturi_24} (Fig.~\ref{fig:zgrad-z}).  
The results in \citet{Venturi_24} are based on the gradient measurement of four galaxies at $z>6$, selected with the largest offset between UV/optical and far-infrared emission lines \citet{Carniani_21}. Such offsets may indicate nontypical dust geometry and/or galactic structures so that the dust and star-forming regions are not uniformly distributed, possibly induced by merger events \citep[also see discussions in][]{Garcia_25}. Therefore, their sample differs from our main-sequence galaxy sample at $5<z<9$, which was unbiasedly selected based on NIRCam WFSS observations.
These interacting systems are expected to undergo violent gas mixing \citep{Tissera_22}, and, as such, the metal distribution is expected to be flatter than normal star-forming galaxies. These effects are likely to explain the differences in our sample.
Thus, we have not included these systems in our high-redshift sample for statistical analysis.

In Fig.~\ref{fig:zgrad-z}, we show the redshift evolution of the metallicity gradients from $z=8$ to $z=0$. We see that the cosmic evolution of the metallicity gradient shows ascending and descending phases before and after $z\approx2$. 
Galaxies show steep negative gradients at the earliest times ($z>5$). Such a steep gradient has been found in local extremely metal-poor galaxies \citep[EMPGs,][]{Kashiwagi_21} and is indicative of inside-out growth and inefficient gas mixing. The metallicity gradients flatten over time until $z\sim2$ and steepen again toward the present.
To describe this trend, we first applied a linear regression analysis for individual galaxies. 
We divide the individual galaxies into two redshift bins, $1.75<z<3.5$ and $0<z<1.75$.
The ascending phase with time can be described as $-0.046\pm0.0
12~\mathrm{dex~kpc^{-1}}/\delta z$ from $z=3.5$ to $z=1.75$, followed by a descending phase of $0.040\pm0.006~\mathrm{dex~kpc^{-1}}/\delta z$ from $z=1.75$ to $z=0$. We applied the likelihood-ratio test to see if this evolution is significant compared with a non-evolving metallicity gradient. We find a $p$-value $=1.3\times10^{-22}$, indicating a strong preference for the model with redshift evolution.

To better characterize the rising and descending phases for the metallicity gradients, we applied a double power law model:
\begin{equation}
    \nabla_r \log(\mathrm{O/H})(z)=\gamma_0 \frac{(1 + z)^{\gamma_1}}{1 + \left[\left(1 + z\right)/\gamma_2 \right]^{\gamma_3}}-\gamma_4\label{eq:zgrad-eq}
\end{equation}
Fitting the above function to our observed data (Appendix \ref{sec:zfit}), we obtain the following constraints: $\gamma_0=0.40^{+0.30}_{-0.19}$, $\gamma_1=0.27^{+0.16}_{-0.09}$, $\gamma_2=5.17^{+1.15}_{-0.59}$, $\gamma_3=5.23^{+2.42}_{-2.01}$, $\gamma_4=-0.50^{+0.20}_{-0.31}$.
This fitting provides a first glimpse of the redshift evolution of the metallicity gradient with an ascending phase scaling as $\nabla_r \log(\mathrm{O/H})(z)\propto(1+z)^{-5.23}$ at $2\lesssim z\lesssim8$, and transforms to a slowly descending phase 
to the present day, scaling as $\nabla_r \log(\mathrm{O/H})(z)\propto(1+z)^{0.27}$. This analytical form suggests the transition to happen around $z\approx2.1$. Intriguingly, it is close to the peak of SFRD \citep{Madau_14}. Similar transitions at the same epoch include the break of the fundamental metallicity relation \citep{Curti_24} at $z>2$ and the break of mass-metallicity relations of damped Lyman alpha systems (DLA) at $z>2.6$ \citep{Moller_13}. They point to certain mode transitions about galaxy formation and chemical enrichment at cosmic noon.

Since galaxies are intrinsically smaller at higher redshift \citep{Morishita_24}, a fixed metallicity change implies steeper gradients in smaller systems. Given the correlation between galaxy size and stellar mass, expressing gradients relative to the effective radius $\rm R_e$ helps reduce dependence on mass and size \citep{Ho_15,Bresolin_19}. In addition, the normalization of the effective radius helps reduce the impact from the beam-smearing effects of seeing limited data \citep{Wuyts_16}.
In Appendix \ref{sec:Re}, we discuss the metallicity gradients measured with the effective radius, and the trend of redshift evolution is consistent.

\subsection{Comparison with hydrodynamical simulations}
\begin{figure*}[t]
 \centering
 \includegraphics[width=\textwidth]{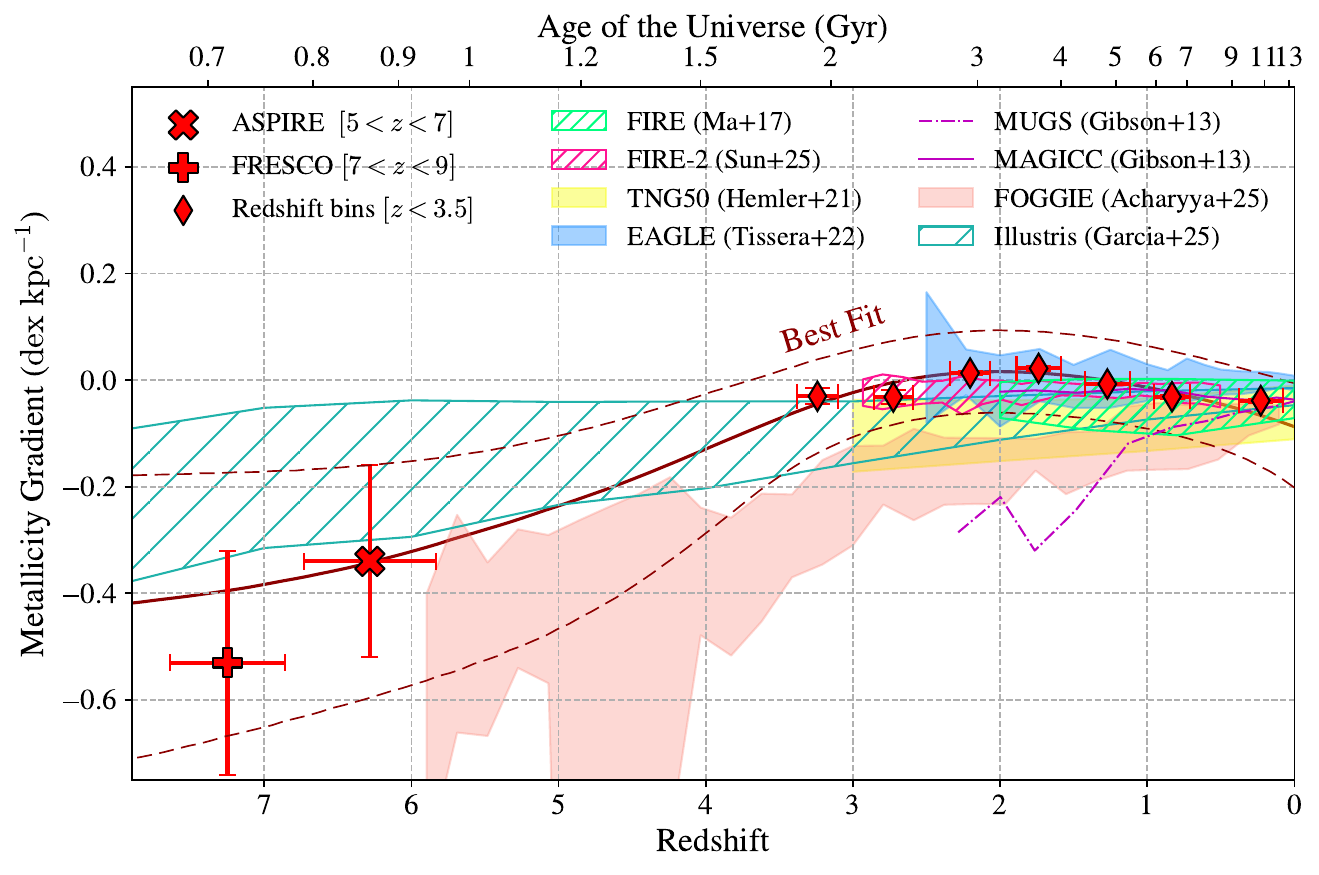} \\
 \caption{The comparison between the observed best-fit redshift evolution of metallicity gradients and predictions from different suites of cosmological simulations. The simulations including FIRE \citep{Ma_17}, FIRE-2 \citep{Sunxd_24}, TNG50 \citep{Hemler_21}, EAGLE \citep{Tissera_22}, MUGS/MAGICC \citep{Gibson_13}, FOGGIE \citep{Acharyya_25}, and Illustris \citep{Garcia_25} are marked in different colors.
\label{fig:zgrad-z-sim}}
 
\end{figure*}
In Fig. \ref{fig:zgrad-z-sim}, 
we compare our observations with simulations from MUGS \citep{Gibson_13}, MAGICC \citep{Gibson_13}, FIRE \citep{Ma_17}, FIRE-2 \citep{Sunxd_24}, TNG50 \citep{Hemler_21}, EAGLE \citep{Tissera_22}, FOGGIE \citep{Acharyya_25}, and Illustris \citep{Garcia_25}.
Those simulations adopt different settings for physical processes, including the treatment of feedback (See Appendix \ref{sec:deg_sim}).

The FOGGIE simulation shows galaxies evolving from steep metallicity gradients at $z= $ to shallower gradients over time. They attribute the flattening to an accretion-dominated phase towards $z\sim1$. While this evolutionary trend from $z > 5$ to $z \sim 2$ is qualitatively similar to our observations, the FOGGIE simulations predict gradients that are systematically steeper overall.
\citet{Acharyya_25} noted that the supernova feedback prescription in the FOGGIE simulations is insufficient to expel enough metal-rich gas, which may lead to the formation of steeper gradients compared with observations, especially at $z\sim2$.
The Illustris simulation shows a similar trend to FOGGIE, in which the galaxies evolve initially with steep gradients and flatten over time. \citet{Garcia_25} obtained the redshift evolution from Illustris simulation as $-0.016\pm0.004~\mathrm{dex~kpc^{-1}}\delta z$. In addition, from EAGLE \citep{McAlpine_16}, IllustrisTNG \citep{Nelson_19} and SIMBA \citep{Dave_19} simulations, \citet{Garcia_25} found similar redshift evolutions in the range of $-0.015$ to $-0.028~\mathrm{dex~kpc^{-1}}~\delta z$. While these values are broadly consistent with our observed ascending phase, the simulations do not reproduce the peaking gradients around $z \sim2$. \citet{Garcia_25} suggest that simulations with smooth stellar feedback may not efficiently mix metals, which could contribute to the discrepancy between simulations and observations around $z \sim 2$.

At $z \lesssim 3$, our observations align more closely with the EAGLE, FIRE, FIRE-2, and MAGICC simulations, whereas TNG50 predicts slightly more negative gradients and MUGS yields overly steep gradients at $z \gtrsim 1$. 
Differences in the evolution of metallicity gradients in these simulations can be interpreted as a result of the use of different feedback models. The feedback drives outflows that transport metal-enriched gas to larger radii or out of the galaxy, where it mixes with metal-poor gas in the ISM and CGM. This process redistributes the metals and results in a flat metallicity gradient.
The more bursty feedback explicitly modeled in FIRE simulation produces flat gradients at $z\gtrsim1$ \citep{Ma_17}, while the smooth feedback in TNG50 allows for more negative gradients \citep{Hemler_21}. MAGICC is a variation of MUGS simulation, running with the same code, except for the enhanced thermal feedback energy from SNe and the inclusion of radiation from massive stars in the MAGICC simulation \citep{Gibson_13}. The enhanced feedback schemes in MAGICC produce flatter gradients at $z\sim2$. However, the redshift evolution in the FIRE simulations is less evident, as they exhibit consistently flat gradients across all redshifts, likely due to the episodic nature of the strong feedback model used in FIRE \citep{Ma_17,Sunxd_24}. EAGLE simulations also suggest that galaxy mergers or interactions that trigger gas inflows and starbursts could flatten or invert metallicity gradients \citep{Tissera_22}. Therefore, the observed metallicity gradients at $z\sim2$ generally prefer models with strong feedback as a consequence of bursty star formation, though the feedback strength should not result in persistently flat gradients. Thus, the evolution of the metallicity gradient indicates an evolving feedback strength through cosmic time. There is evidence that the energy and momentum injection from massive stars peaks at $z\approx2$ \citep{Heckman_23}. At this redshift, outflows are found to efficiently redistribute large proportions of metals from galaxies \citep{Sanders_23a}, leading to flat/positive gradients \citep{Wang_19}. After this epoch, feedback strength should diminish as a consequence of decreased star formation rate \citep{Yu_23}, allowing negative gradients to develop.
% Similarly, the merger rate also peaks at $z\sim2$\citep{O'Leary_21}, further leading to efficient gas mixing\citep{Tissera_22}. 

In addition to outflow launched by feedback, inflows due to cosmic gas accretion from the intergalactic medium (IGM) may also mix with the ISM gas and redistribute metal contents.
Cosmic accretion history in simulations suggests that cold-mode accretion dominates the total accretion rate at $z\sim2$, and hot-mode accretion becomes increasingly important at lower redshifts \citep{Keres_2009, Voort_11}.  
At these lower redshifts, hot-mode accretion forms a shock shell around the massive halos $M_h\gtrsim10^{12}M_\odot$ \citep{Birnboim_03, Dekel_09}. As the cooling time is longer than the infall time, the shock-heated gas cannot cool and collapse inward efficiently, thereby limiting its ability to fuel star formation readily.
However, the cold-mode accretion is predicted to penetrate deep inside the halo without being shock-heated because the cooling time is shorter than the shock compression time.
Such cold flows reduce the metallicity in the galactic center via filaments penetrating halos, which causes the flattening/inverting of the metallicity gradients at cosmic noon \citep{Cresci_10, Li_22}.

The joint effect of feedback and gas accretion may contribute to the flat/positive gradients observed during cosmic noon. Another contributing factor is mergers and gravitational interactions, which can also drive gas inflows and enhance both star formation and feedback \citep{Perez_24}. In particular, stochastic gas accretion associated with these mergers may play an important role in low-mass galaxies, in the context of hierarchical structure formation.
\citet{Torrey_12} found that gravitational tidal forces during galaxy mergers or close interactions can generate significant inflows, transporting metal-poor gas inwards and redistributing metals within the disk. \citet{Tissera_22} showed that such merger-driven inflows are common in high redshifts, and can account for a notable fraction $(\sim20\%)$ of the positive metallicity gradients at $z \lesssim 2$ observed in the EAGLE simulations.

In contrast, measurements at $z\sim6-8$ reveal steeper gradients compared to galaxies at $z\sim2$, possibly indicating reduced gas mixing at cosmic dawn allowed for the existence of steeper gradients in early galaxies. 
Recent works have revealed numerous instances of star-forming galaxies that deviate from the main sequence of the fundamental metallicity relation (FMR) established at $z\lesssim2$ \citep{Lara-lopez_10}. These deviations, characterized by metal deficiency at $z\gtrsim7$ \citep{Heintz_23, Nakajima_23}, imply a continuous pristine gas inflow that effectively dilutes the metal abundances in these galaxies. We indeed observe a metallicity deficiency in the ASPIRE sample, which is consistent with the analytical model featuring intense gas inflows, as discussed in our companion work \citep{Li_25}.
With continuous gas inflow diluting galaxies, steep metallicity gradients may exist when star formation efficiency is high and the local metal production is high at the center. Theoretical models \citep{Wangenci_22, Sharda_21a} have suggested the importance of in situ metal production in building negative metallicity gradients. The observed negative metallicity gradients indicate that in situ star formation and metal production, driven by intense central starbursts, may dominate the metal distribution, while metal mixing by gas inflows and feedback-driven outflows likely plays a minor role.

At low-redshift ($z \lesssim 1$), metallicity gradients tend to become negative again. The underlying mechanisms likely differ from those at $z > 5$, as star formation is significantly suppressed in hot halos at $z = 0$, due to inefficient cooling \citep{van_12,Correa_18} and reduced gas accretion rate \citep{van_11,Faucher_11}. From FIRE-2 simulation, \citet{Graf_24b} suggest that as galaxies evolved, declining gas accretion \citep{Trapp_22}, reduced star formation \citep{Yu_23} and feedback \citep{Orr_22}, and the settling of disks with reduced velocity dispersion and gas turbulence \citep{Bird_21} led to a reduction in metal mixing, allowing negative gradients to develop and persist. 
Comparing observations in the local Universe and at $z\sim2$ highlights a transition from strong turbulence, gas inflows, and feedback-driven mixing at high-redshift to more settled, secular evolution by $z \sim 0$.

\subsection{Mass-Metallicity Gradient Relation (MZGR)}\label{sec:mzgr}

\begin{figure*}[!t]
 \centering
\includegraphics[width=\textwidth,trim=0cm 0cm 0cm 0cm,clip]{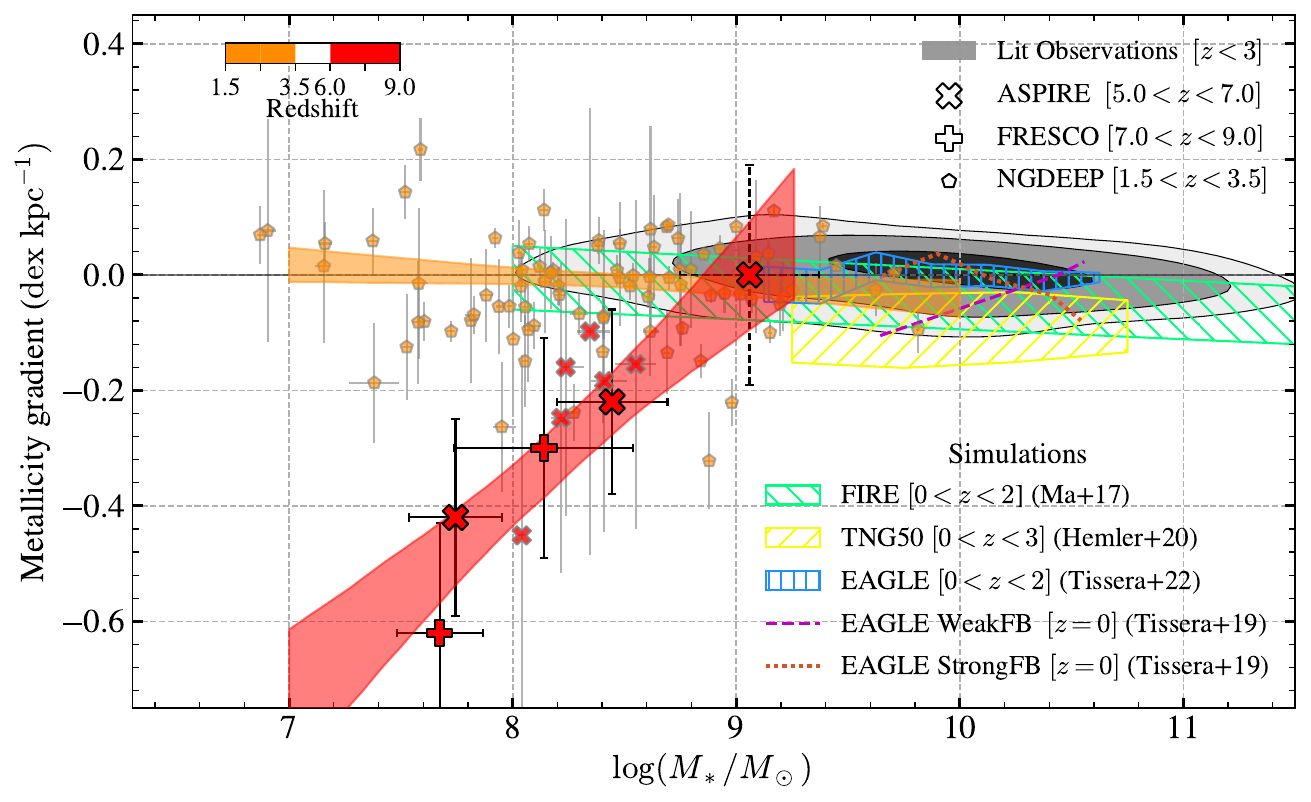} \\
 \caption{The mass dependence of metallicity gradients. Galaxies in NGDEEP, ASPIRE, and FRESCO are labeled with a pentagon, ``$\times$" and ``$+$", respectively, color-coded by redshift. The highest mass bin in ASPIRE is marked with a dashed error bar; for this bin, the metallicity gradients were derived using the higher-branch solution. The red and orange shaded areas represent the $1\sigma$ confidence intervals for the linear fit to the stacked data at $z > 5$ and the individual NGDEEP measurements within $1.7 < z < 3.5$, respectively.
 The literature observations \citep{Swinbank_12, Jones_13,Wuyts_16,Wang_17, Carton_18,Forster_18,Wang_20, Curti_20a,Simons_21,Ju_24} are included in gray-filled regions with 1, 2 and 3$\sigma$ levels of the density contours of the distribution. Green, yellow, and blue hatched areas represent the $1\sigma$ interval from FIRE \citep{Ma_17}, TNG50 \citep{Hemler_21}, and EAGLE \citep{Tissera_22} simulations, respectively. The dashed magenta and dotted orange lines show the median metallicity gradients of EAGLE galaxies with strong feedback (StrongFB) and weak feedback (WeakFB) prescriptions \citep{Tissera_19}.
 \label{fig:zgrad-mass}}
\end{figure*}

We explore how metallicity gradients change with galaxy mass, in terms of the mass-metallicity gradient relation \citep[MZGR, also see][]{Spolaor_09,Sharda_21c}. We divided our FRESCO, ASPIRE samples into different mass bins and applied the same stacking method  (Appendix \ref{sec:massbin}, Fig. \ref{fig:massbin}, \ref{fig:mass-ge9}).

In Fig. \ref{fig:zgrad-mass}, we show the metallicity gradients as a function of stellar mass (Mass-Metallicity Gradient Relation, MZGR), and compare the observations with the FIRE \citep{Ma_17}, TNG50 \citep{Hemler_21}, and EAGLE \citep{Tissera_22} simulations. The mass limits for JWST data in our work are deep enough to compare the change of metallicity gradients over redshift.

We observe that they follow distinct log-linear relations at $z\sim2$ and $z>5$. For galaxies at $z>5$, we find a notable positive correlation in MZGR quantified as $ \nabla_r\log(\mathrm{O/H})_{5<z<9}= 0.30\pm0.19 \times \log(M_*/M_\odot) -2.78\pm1.54$. This indicates that lower mass galaxies exhibit steeper negative gradients.
However, literature data (gray contours) suggest a weak negative mass dependence of the metallicity gradient within the redshift range $0<z<3.5$.
Our observations from the NGDEEP sample at $1.7<z<3.5$ also reveal a marginal anti-correlation, quantified as $\nabla_r\log(\mathrm{O/H})_{1.7<z<3.5}= -0.020\pm0.018 \times \log(M_*/M_\odot) +0.157\pm0.156$.
This dependence on mass indicates the presence of different modes during those two periods.

A positive dependence on stellar mass is anticipated in secular galaxy evolution models, such that gradients flatten over time due to inside-out growth with the accumulation of stellar mass and metals \citep{Pilkington_12}. The positive mass dependence was previously observed in local galaxies \citep{Ho_15, Bresolin_19}, which is also predicted by EAGLE simulations at $z=0$ with weak feedback \citep[magenta dashed line in Fig. \ref{fig:zgrad-mass},][]{Tissera_19}.
\citet{Baker_24} also demonstrates the early signs of inside-out growth, with a galaxy at $z=7.4$ that features a compact core and an extended star-forming disk. 
% along with an increasing specific star formation rate (sSFR) as the radius increases. 
Our observed positive mass dependence further supports inside-out growth at the cosmic dawn. Low-mass galaxies form earlier with concentrated cores, gradually building up outer disks over time. As they evolve, they accumulate stellar mass and metals at larger radii. The flattening of metallicity gradients in higher-mass galaxies highlights the dynamical nature of galaxy growth during the cosmic dawn.
\begin{figure*}[!t]
 \centering
 \includegraphics[width=\textwidth]{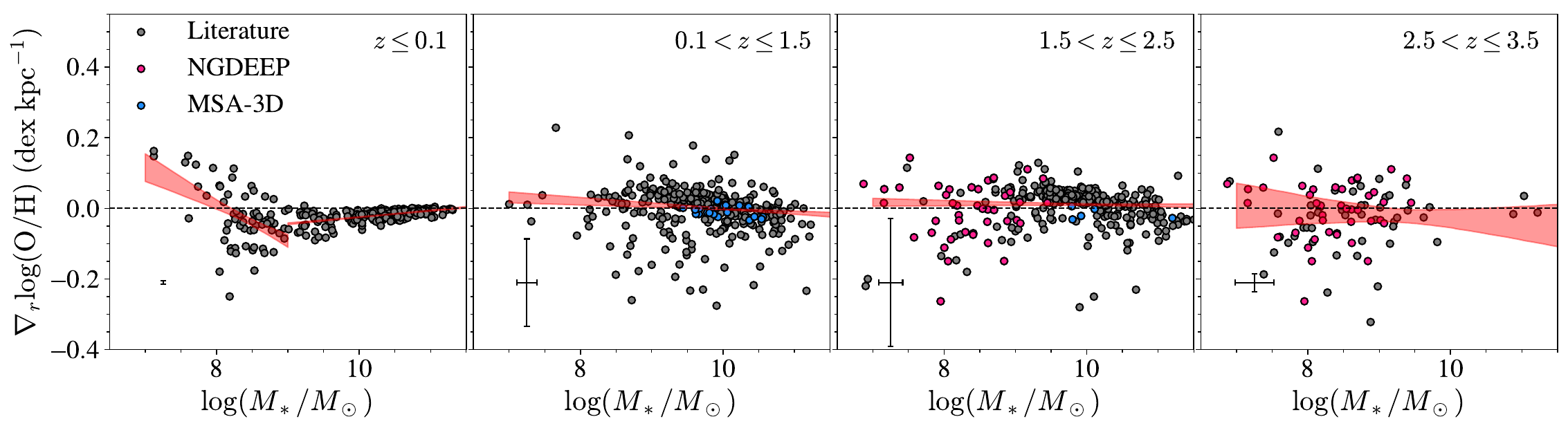}
 \caption{The mass-metallicity-gradient-relation for individual galaxies in four redshift bins.
 The red shadows represent the $1\sigma$ confidence interval of the linear regression. For $z\leq0.1$ measurements, regressions are performed within two stellar mass bins: $\log(M_*/M_\odot)\leq9$ and $\log(M_*/M_\odot)>9$, while high-redshift measurements are performed across the entire mass range.
 The error bar on the bottom left of each figure represents median uncertainty in stellar masses and gradients. The pre-JWST observations are shown in gray circles \citep{Swinbank_12,Jones_13,Ho_15,Sanchez_16,Wuyts_16,Forster_18,Carton_18,Bresolin_19,Curti_20a,Simons_21,Lit_25}; NGDEEP and recent JWST MSA-3D \citep{Ju_24} observations are highlighted in red and blue, respectively.}
 \label{fig:zgrad-mass-fit}
\end{figure*}
On the other hand, a negative relation at $z\sim1-2$ has been observed by \citet{Wang_20, Gillman_21, Simons_21}, and they interpreted it as the effect of feedback-driven gas outflows, which is more pronounced in low-mass galaxies due to less restraint by lower gravitational potential \citep{Badry_16,Heckman_23}. 
FIRE simulations suggest that efficient gas outflow rates could efficiently mix metals, causing significant fluctuations in metallicity gradients \citep{Ma_17}. A similar negative relation is also found in EAGLE simulations with enhanced SNe feedback \citep[][see orange dotted line in Fig. \ref{fig:zgrad-mass}]{Tissera_19}. From TNG50 simulations, \citet{Garcia_23} indicates that when the enrichment timescale is shorter than the gas mixing timescale, galaxies can maintain steep gradients. Additionally, if feedback processes effectively generate high-velocity gas outflows and turbulence, this can reduce the gas mixing timescale, leading to shallower gradients. The flattening of metallicity gradients toward low-mass galaxies arises as the result of stronger feedback effects with lower gravitational potential. The two distinct MZGR at cosmic noon and cosmic dawn then indicate the weak feedback at $z>5$ and strong feedback at $z\sim2$.

\subsection{Redshift evolution of MZGR}

To further investigate whether the mass dependence of metallicity gradients evolves with redshift, we divide the sample of NGDEEP and literature observations into four redshift bins: $z\leq0.1$, $0.1<z\leq1.5$, $1.5<z\leq2.5$, and $2.5<z\leq3.5$. Specifically, we have included the measurements from the local universe at $z=0$, reported by \citet{Ho_15, Bresolin_19,Lit_25}.
We perform a linear regression on these measurements using the package \textsc{Linmix} \citep{Kelly_07}, with errors on both metallicity gradients and stellar masses taken into account. The regression is formulated as:
\begin{equation}
    \nabla_r\log(\mathrm{O/H})(M_*)=\alpha+\beta\log(M_*/M_\odot)+\textrm{N}(0,\sigma),\label{eq:mzgr}
\end{equation}
where $\alpha$ is the intercept of the linear function and $\beta$ is the slope of the linear function. $\rm N(0,\sigma^2)$ represents a normal distribution with $\sigma$ representing the intrinsic scatter about the regression in units of $\rm dex~kpc^{-1}$. \citet{Lit_25} recently reported an inverted MZGR for low mass galaxies at $z=0$ with $\logm<9$, contrary to high mass trends. This turnover pattern was also reported by \citet{Belfiore_17,Poetrodjojo_21,Khoram_25}, although the exact turnover point remains uncertain, in the range of $\logm\sim9-10$.
Here we assume the turnover point at $\logm=9$ and divide the $z=0$ sample into low-mass ($\logm<9$) and high-mass ($\logm\geq9$) bins, which are fitted separately. For higher-redshift observations, no compelling evidence of such a turnover pattern has been reported so far. Therefore, we perform a linear fit across the entire mass range for the other three high-redshift bins.

\begin{deluxetable}{ccccc}
\tablecaption{Fitted MZGR parameters in Eq. \ref{eq:mzgr}, and the Spearman coefficients ($r,p$).\label{tab:reg_info}}
\tablewidth{0pt}
\tablehead{
\colhead{Redshift} & \colhead{$\beta$ [$\rm dex~kpc^{-1}]$} & \colhead{$\alpha$ [$\rm dex~kpc^{-1}$]} & \colhead{$\sigma$ [$\rm dex~kpc^{-1}$]} & \colhead{$r,p$}
}
\startdata
$z=0^\textrm{a}$ & $-0.100\pm0.026$& $0.811\pm0.218$&$0.063\pm0.007$&$-0.511,0.000$\\
$z=0^\textrm{b}$ & $0.019\pm0.002$& $-0.211\pm0.019$&$0.013\pm0.001$&$0.571,0.000$\\
$0.1<z\leq1.5$ & $-0.009\pm0.004$& $0.084\pm0.037$&$0.036\pm0.002$&$-0.173,0.002$\\
$1.5<z\leq2.5$ & $-0.002\pm0.003$& $0.032\pm0.029$&$0.031\pm0.002$&$0.027,0.621$\\
$2.5<z\leq3.5$ & $-0.010\pm0.022$& $0.064\pm0.196$&$0.121\pm0.015$&$-0.060,0.680$\\
$5.0\leq z\leq9.0^{c}$ & $0.304\pm0.190$ & $-2.784\pm1.539$ & $0.131\pm0.125$ & $0.949, 0.014$ \\
\enddata
\tablecomments{$^{a}$ Fitted within low mass bin $\logm<9$. $^{b}$ Fitted within high mass bin $\logm\ge9$. $^{c}$ In this redshift range, the MZGR is fitted on median stacks (Fig. \ref{fig:zgrad-mass})}
\end{deluxetable}

In Fig. \ref{fig:zgrad-mass-fit}, we show the fitting results from the four redshift bins. 
The fitted parameters are shown in Table \ref{tab:reg_info}. In the local universe, the massive galaxies ($\logm>9$) show a positive MZGR, indicating secular evolution. While for the finding of \citet{Lit_25}, we see that low mass galaxies ($\logm<9$) exhibit a negative mass dependence. This suggests that in the local universe, more massive, evolved galaxies predominantly undergo secular evolution, whereas low-mass dwarf galaxies remain influenced by metal mixing and transport processes, such as outflows driven by ongoing star formation.
At redshift $z\sim2$, we find the slopes are consistent with the value previously reported by \citet{Wang_20} for a sub-sample in the redshift range $1.2<z<2.5$. However, unlike local ($z = 0$) galaxies, the $z\sim2$ population exhibits a negative MZGR within the same mass range of $\logm\gtrsim8$. This negative MZGR suggests enhanced gas mixing due to feedback-driven wind at cosmic noon, as we have discussed in Section \ref{sec:mzgr}.

We find a mild evolution of the MZGR slope $\beta$. The MZGR is positive at $z=0$ for galaxies at $\logm>9$, and it becomes negative at higher redshifts. The significance of the negative MZGR slope decreases with increasing redshift. At $z=0$, 
The evolution of the MZGR slope coincides with the evolution of the metallicity gradient (Fig. \ref{fig:zgrad-z}). This further indicates the evolution of feedback strength: with strong feedback at $z\sim2$, the feedback promotes more positive gradients. Meanwhile, it has a stronger impact on low-mass galaxies, reflected in more negative MZGR. At higher redshift bins $2.5<z\leq3.5$, it starts to exhibit less negative MZGR, closer to the prediction without efficient outflows, and finally, the MZGR is significantly positive at $5.0\leq z\leq9.0$.

\begin{figure*}[!t]
  \begin{subfigure}{0.48\textwidth}
  \includegraphics[width=\linewidth]{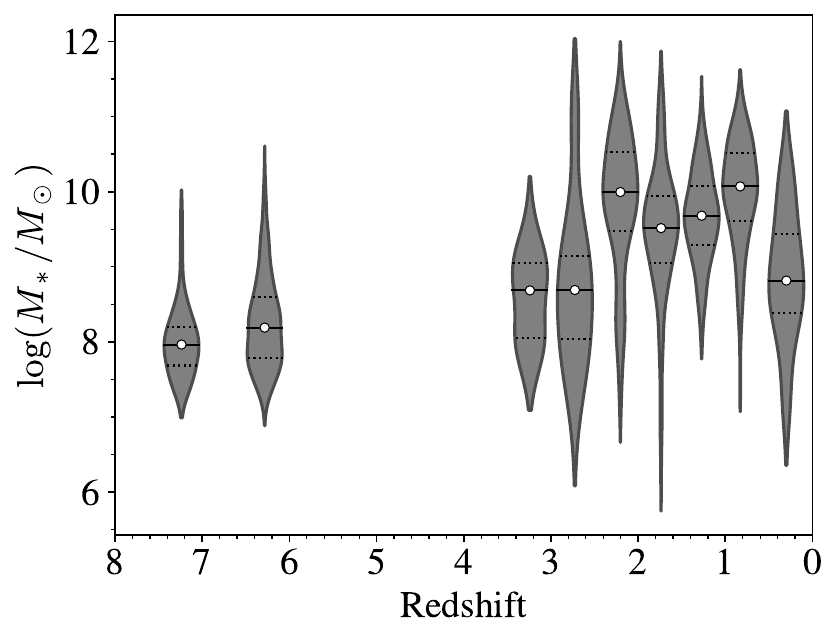}
    \caption{}\label{fig:mass-z}
  \end{subfigure}
  \hspace*{\fill}
  \begin{subfigure}{0.5\textwidth}
    \includegraphics[width=\linewidth]{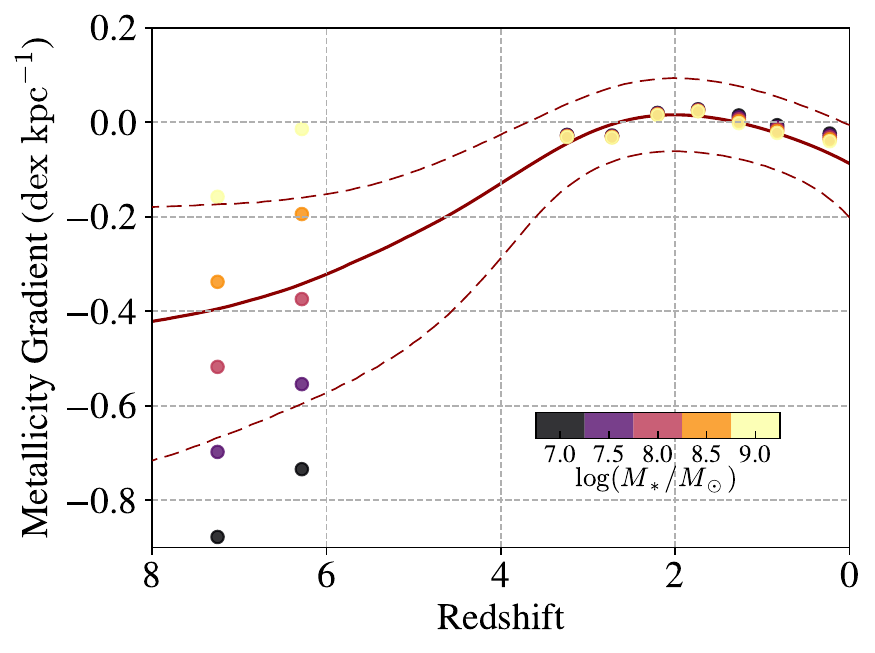}
    \caption{} \label{fig:mass-corr}
  \end{subfigure}
\caption{(a) The violin plot showing the mass distribution of our full sample at each redshift bin. The white circles represent the median stellar mass and redshift of each bin. The dotted lines represent $1\sigma$ intervals of stellar mass distributions. (b) The redshift evolution of metallicity gradients with/without normalization by stellar masses. The red circles are the observed metallicity gradients, with the dark-red lines representing the best fitting, the same as Fig. \ref{fig:zgrad-z}. The circles represent metallicity gradients normalized to a fixed stellar mass, with colors indicating stellar masses normalized relative to the observed MZGR at each redshift.}
\end{figure*}

In Fig. \ref{fig:mass-z}, we show the mass distribution as a function of redshift in our sample. 
Since the galaxy sample is collected from a variety of works with different selection criteria, they have different stellar masses in different redshift bins. For example, the galaxies in $0.5<z<2.5$ bins are more massive ($\approx1$ dex) than others, which corresponds to an $\rm -0.01~dex~kpc^{-1}$ reduction in metallicity gradients using the MZGR derived above. To quantify the impact of different stellar masses on our observed redshift evolution, we normalize the gradients at each redshift bin to the same stellar masses based on the observed MZGR. In Fig. \ref{fig:mass-corr}, we compare the observed and mass-normalized metallicity gradients. We find that the overall trend of redshift evolution shown in Fig. \ref{fig:zgrad-z} remains consistent.
To investigate the intrinsic scatter for galaxies with different stellar masses, we divide the sample galaxies into low-mass ($M_* < 10^9 \Msun$) and high-mass ($M_* > 10^9 \Msun$) bins. At redshift $1.5 < z \leq 3.5$, the intrinsic scatter for low-mass galaxies is found to be $\sigma_{\rm MZGR} = 0.098 \pm 0.02$, while for massive galaxies, it is smaller, with $\sigma_{\rm MZGR} = 0.023 \pm 0.002$. In contrast, at lower redshifts ($0.01 < z \leq 1.5$), the scatter is reduced, with $\sigma_{\rm MZGR} = 0.045 \pm 0.008$ for low-mass galaxies and $\sigma_{\rm MZGR} = 0.037 \pm 0.002$ for massive galaxies.
The increasing scattering of metallicity gradients towards the low-mass end was also found by \citet{Wang_20}, who suggests more efficient feedback for low-mass galaxies, as shallower gravitational potentials are more susceptible to gas disruption \citep{Zhuang_19}. 
In addition, the lowest mass galaxies have the highest specific star formation rate (sSFR) \citep{Wang_20}, which could be triggered by either cold mode accretion \citep{Woods_14}, or merger-induced gas inflow \citep{2008MNRAS.385L..38M,2013MNRAS.430.1158S}. 
The latter was found to be dominant in dwarf galaxies \citep{Wang_20}. The combined effects of metal production in merger-induced starbursts and gas dilution could lead to diverse chemical structures, reflected by large scatter in metallicity gradients at $z\sim2$. When we evenly divide the NGDEEP sample into high- and low-sSFR bins, we find that the high-sSFR subsample exhibits twice the intrinsic scatter compared to the low-sSFR subsample, reflecting the effects of those processes.
The reduced scatters at $z<1.5$ also suggest the impact of feedback on low-mass galaxies weakens at low redshifts.

\subsection{Toy models for metallicity gradients}
\begin{figure*}[t!]
\begin{subfigure}{0.44\textwidth}
  \includegraphics[width=\linewidth]{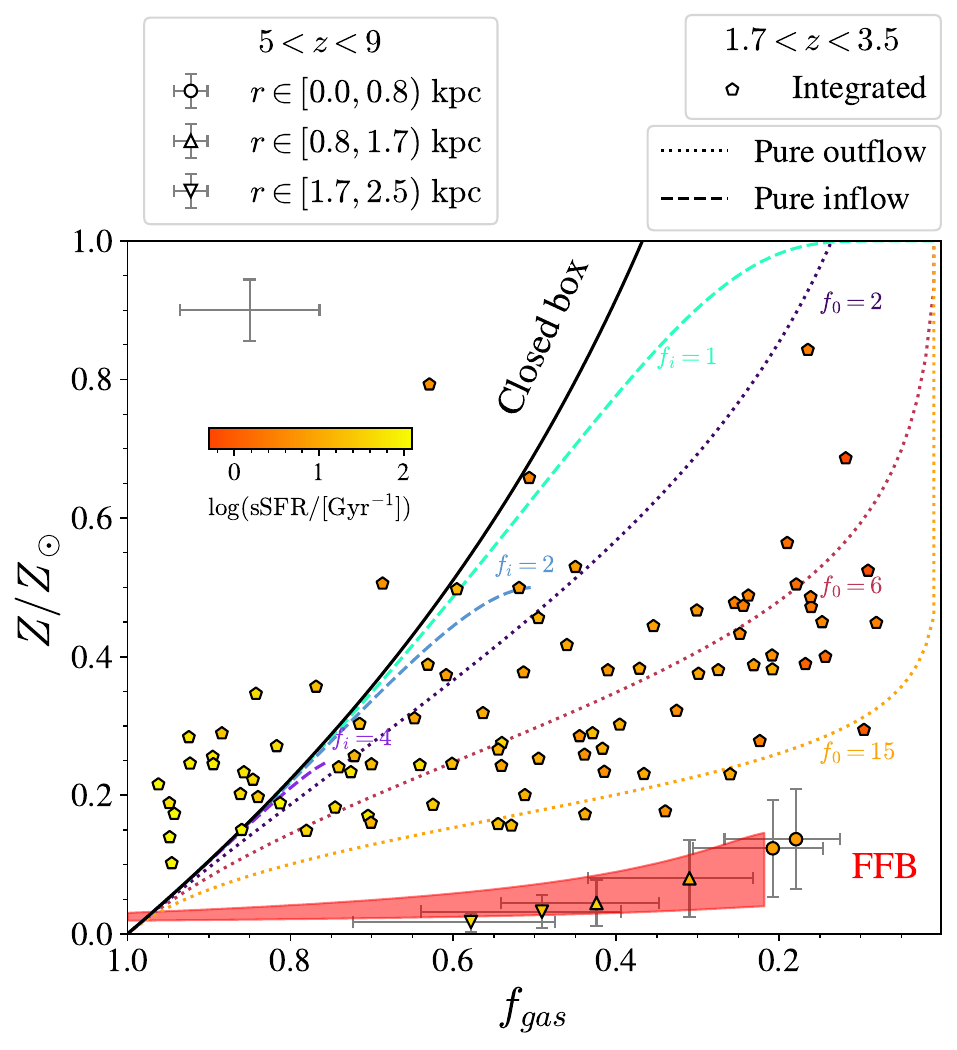}
    \caption{}\label{fig:fgas-z-a}
  \end{subfigure}
  \hspace*{\fill}
  \begin{subfigure}{0.49\textwidth}
    \includegraphics[width=\linewidth]{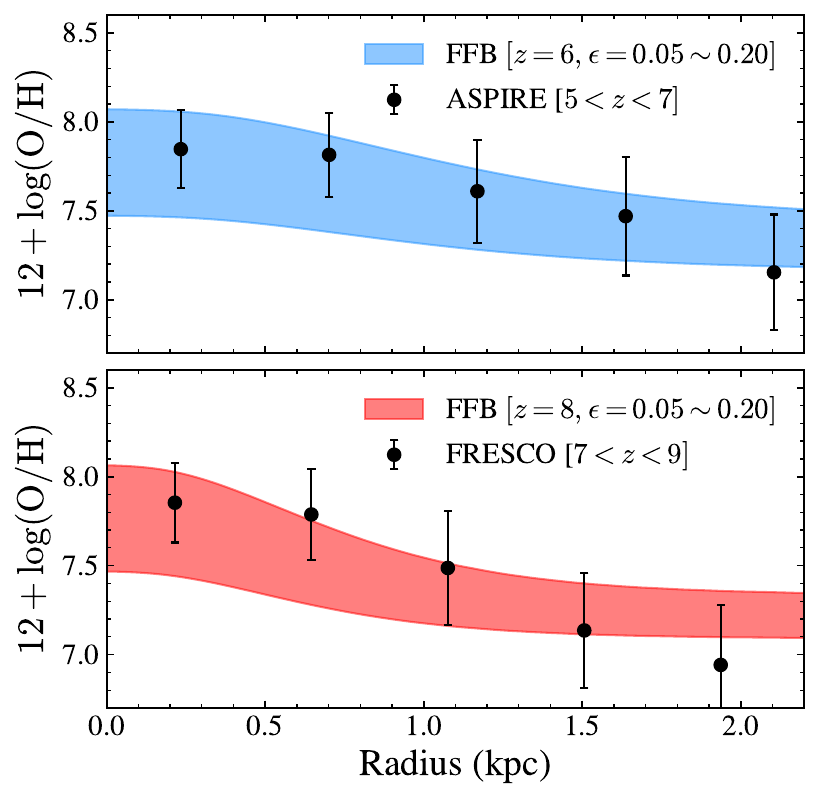}
    \caption{} \label{fig:fgas-z-b}
  \end{subfigure}
\caption{(a) The measured gas fraction and metallicity compared with the \citetalias{Erb_08} model \citep{Erb_08} and the FFB model \citep{Dekel_23,Lizhaozhou_23}.
We consider \citetalias{Erb_08} models with pure gas outflow ($f_i=0$) and with pure inflow ($f_o=0$). The curves for the \citetalias{Erb_08} models are color-coded by inflow/outflow rate relative to SFR, with $f_i$ denoting the inflow model (dashed curves) and $f_o$ denoting the outflow model (dotted curves). 
The solid black curve represents the closed-box model without outflow and inflow. The red shaded region shows the predictions of the FFB model \citep{Lizhaozhou_23} at redshift $z=8$ with a star formation efficiency of $\epsilon=0.05\text{--}0.20$ and a mass loading factor of 0.95, the maximum allowed value. 
The points in ASPIRE ($5<z<7$) and FRESCO ($7<z<9$) are estimated within different radial annuli (see Fig.~\ref{fig:fgas}), while the pentagon points represent integrated measurements for the NGDEEP sample at $z=1.7\text{--}3.5$. The symbols are color-coded by sSFR.
The gas fraction uncertainties of the ASPIRE and FRESCO galaxies include the uncertainty from unconstrained dust attenuation in estimating SFR maps.
(b) The measured metallicity gradients at redshift $z=5\text{--}7$ and $z=7\text{--}9$ compared with the FFB model.
The shaded regions represent the metallicity profiles predicted by the FFB model at redshifts 6 (blue) and 8 (red) with star formation efficiency $\epsilon=0.05\text{--}0.20$, which is in line with our observations.}
\label{fig:fgas-z}
\end{figure*}
To further understand how inflow and outflow take effect in our sample galaxies, we measure the gas fraction and metallicity in comparison with a simple analytical model \citep{Erb_08} (hereafter \citetalias{Erb_08} model) and the Feedback-Free starburst (FFB) model \citep{Dekel_23, Lizhaozhou_23,Dekel_25}. 
The \citetalias{Erb_08} model assumes that the surface density of star formation is proportional to the surface density of the gas \citep[Kennicutt-Schmidt law,][]{Schmidt_59, Kennicutt_89} and that inflows and outflows regulate the gas required by sustained star formation. This model has established basic guidelines for the relation between mass, metallicity, and gas fraction at $z\lesssim3$. Such a ``bathtub" model is widely used to compare with observations \citep{Cresci_10, Yabe_15, Zhuang_19, Wang_19}, although we acknowledge that it assumes no radial gas flows and constant star-formation efficiency (SFE). 
We consider two simple cases for the \citetalias{Erb_08} model, one with pure inflow (with inflow rate defined as inflow gas mass per star formation $f_i=\dot{M}_{\mathrm{in}}/\mathrm{SFR}$) and the other with pure outflow (with outflow rate defined as outflow gas mass per star formation $f_o=\dot{M}_{\mathrm{out}}/\mathrm{SFR}$). 
The FFB scenario is proposed to characterize galaxies at $z \sim 10$ where the density of gas in star-forming clumps is above a threshold of $\sim 3\times 10^3$ cm$^{-3}$ with gas metallicity $\lesssim 0.2Z_\odot$, where $Z_\odot$ is the solar metallicity. 
The FFB model predicts efficient star formation in thousands of globular-cluster-like clouds occurring on a free-fall timescale before the onset of effective stellar and supernova feedback. This allows multiple generations of starbursts, each with a high star-formation efficiency, separated by relatively quiescent periods.
After each burst, SN feedback from massive stars will inevitably occur, yielding metals and enriching the galaxy. In Appendix \ref{sec:halo}, we show that our $5<z<9$ galaxy sample satisfies the FFB criteria and is therefore suitable for applying this model.
% We applied a lower yield in FFB model, so that a fraction of metals produce by SNe is lost out of the galaxy to simulate the inefficient mixing scenario.

% \allowbreak % allow linebreak for inline equations

We note that FFB is a simple analytical model aiming to illustrate the idea of efficient star formation with inefficient metal mixing. We refer to more sophisticated models \citep[e.g.,][]{Fu_13,Molla_19,Henriques_20, Sharda_21c,Sharda_21a,Sharda_21b,Sharda_23}, which consider more physical processes such as metal diffusion, advection, and more detailed feedback processes. \citet{Bellardini_21} demonstrated that neglecting diffusion leads to pronounced gradient steepening, whereas larger diffusion coefficients promote gradient flattening. Our simplified model thus illustrates the conditions under which metal mixing is inefficient in the absence of mechanisms such as diffusion.

Here, we consider a case of the steady-state solution for gas metallicity involving both high inflow rates and SFR, along with the resultant outflows in the FFB scheme \citep{Dekel_23, Lizhaozhou_23}.  
In reality, the stars of FFB galaxies are expected to form in multiple generations of bursts rather than a continuous process; therefore, this model should be considered a time-smoothed approximation.
According to the conservation of gas mass among the ISM, gas inflow, SFR, and outflow ($M_\mathrm{ism}$, $M_\mathrm{in}$, $\mathrm{SFR}$, $M_\mathrm{out}$), and the conservation of the corresponding metal mass, we have
\begin{align}
\dot M_\mathrm{ism} + \dot M_\mathrm{out} &= 
    \dot{M}_{\mathrm{in}} - (1 - f_{\mathrm{sn}}) \mathrm{SFR}\\
Z_{\mathrm{ism}}  \dot{M}_{\mathrm{ism}} + Z_{\mathrm{out}} \dot{M}_{\mathrm{out}} &= Z_{\mathrm{in}}  \dot{M}_{\mathrm{in}} - (Z_{\mathrm{sf}} - Z_{\mathrm{sn}} f_{\mathrm{sn}}) \mathrm{SFR}
\end{align}
% \begin{align}
% \dot M_\mathrm{ism} + \dot M_\mathrm{out} &= 
%     \dot{M}_{\mathrm{in}} - (1 - f_{\mathrm{sn}}) \mathrm{SFR} \\
% Z_{\mathrm{ism}} \dot{M}_{\mathrm{ism}} +Z_{\mathrm{out}} \dot{M}_{\mathrm{out}} &= Z_{\mathrm{in}} \dot{M}_{\mathrm{in}} 
%      \\
% &\quad - (Z_{\mathrm{sf}} - Z_{\mathrm{sn}} f_{\mathrm{sn}}) \mathrm{SFR} \nonumber 
% \end{align}
where $f_\mathrm{sn}$, $Z_\mathrm{sn}$, and $Z_\mathrm{sn}$ are the mass fraction, metallicity of supernova ejecta, and metallicity of gas forming stars, respectively. 
The second equation assumes a steady state solution without metallicity evolution, i.e., $\dot{Z}_\mathrm{ism}=\dot{Z}_\mathrm{out}=\ldots=0$. 
Below, we estimate the metallicity and spherically averaged density profile, $\rho(r)$, of each component.

For a halo with an accretion rate of $\dot M_\mathrm{halo}$,
the gas accretion rate and SFR are $\dot M_\mathrm{in}=f_{b}\dot M_\mathrm{halo}$
and $\mathrm{SFR}=\epsilon \dot M_\mathrm{in}$, respectively.  In this equation,  $f_\mathrm{b}$ is the cosmic baryonic fraction
and $\epsilon$ is the star formation efficiency.
The outflow rate is regulated by the mass loading factor, $\eta\equiv \dot M_\mathrm{out}/\mathrm{SFR}$.
We introduce a parameter $f_\mathrm{out}=\eta/\eta_{\max}\leq 1$, 
where $\eta_{\max}$ is the maximum mass loading factor allowed for non-negative $\dot M_\mathrm{ism}$, $\eta_{\max}=\epsilon^{-1}-1+f_\mathrm{sn}$.
We then have $\dot M_\mathrm{out} = f_\mathrm{out}\eta_{\max}\mathrm{SFR}$ and 
$\dot M_\mathrm{ism} = (1-f_\mathrm{out}) \eta_{\max}\mathrm{SFR}$.
We obtain the stellar mass and gas mass in ISM by integrating over time: 
$M_\mathrm{star} =(1-f_\mathrm{sn})\epsilon f_b  M_\mathrm{halo}$
and $ M_\mathrm{ism} = (1-f_\mathrm{out}) \eta_{\max}\epsilon f_b M_\mathrm{halo}$.
Given the observed galaxy size $R_\mathrm{e}$ with a typical S\'ersic index of $n_\mathrm{s}=1$, we estimate the density profile of stellar mass and ISM gas, $\rho_\mathrm{star}(r)$ and $\rho_\mathrm{ism}(r)$.

% We then estimate the spherically averaged density profile of the inflow, $\rho_\mathrm{in}(r)$,
% according to $\dot M_\mathrm{in}$ and the inflow velocity\citep{Lizhaozhou_23},
% and the profile of the outflow, $\rho_\mathrm{out}(r)$, according to $\dot M_\mathrm{out}$ and the SFR \citep{Chevalier_Clegg_1985}.

The model considers star formation occurring in the shock region with enhanced gas density at $R_\mathrm{shell}\simeq2R_\mathrm{e}$, where the inflow encounters the outflow \citep{Dekel_23, Lizhaozhou_23}.
We estimate the spherically averaged density of the inflow at radius $r$ using mass flux $\dot M'_\mathrm{in}(r)$ and inflow velocity $v_{{\mathrm{in}}}(r)\sim v_\mathrm{vir}$ \citep{Aung2024}: $\rho_{{\mathrm{in}}} (r) = {\dot{M}'_{{\mathrm{in}}}(r)}/{4 \pi r^2 v_{{\mathrm{in}}} (r)}$.
For $r>R_\mathrm{shell}$, we assume a constant inflow flux: $\dot M'_\mathrm{in}(r)=\dot M_\mathrm{in}$. Within $R_\mathrm{shell}$, the flux decreases due to star formation and conversion to ISM, approximated by $\dot M'_\mathrm{in}(r)=(\dot M_\mathrm{in}-\mathrm{SFR})(r^3/R_\mathrm{shell}^3)$.
After their birth, stars redistribute within the galaxy and launch outflows.
We adopt the asymptotic solution for outflow density from \citet{Chevalier_Clegg_1985}, 
\begin{equation}
    \rho_{{\mathrm{out}}} (r) = 
    \begin{cases}
        0.3 \left(\dfrac{\dot{M}_{{\mathrm{out}}}}{\dot{E}_{{\mathrm{sn}}}}\right)^{1 / 2} \dfrac{\dot{M}_{{\mathrm{out}}}}{R_{{\mathrm{shell}}}^2} & \text{for } r<R_\mathrm{shell},\\
        0.05 \left(\dfrac{\dot{M}_{{\mathrm{out}}}}{\dot{E}_{{\mathrm{sn}}}}\right)^{1 / 2} \dfrac{\dot{M}_{{\mathrm{out}}}}{r^2} & \text{for } r>R_\mathrm{shell},
    \end{cases}
\end{equation}
where $\dot{E}_{{\mathrm{sn}}}$ is the SN energy returned by given SFR \citep{Lizhaozhou_23}.

Assuming star formation from inflow gas without mixing \citep{Lizhaozhou_23}, $Z_\mathrm{sf}=Z_\mathrm{in}$, and the outflow being launched from ISM, $Z_\mathrm{ism}=Z_\mathrm{out}$, we get
\begin{equation}
Z_{\mathrm{ism}} = \frac{Z_{\mathrm{in}}  (1 - \epsilon) + Z_{\mathrm{sn}} f_{\mathrm{sn}} \epsilon}{(1 - \epsilon) + f_{\mathrm{sn}} \epsilon}.
\label{eq:z_ism}
\end{equation}

We then obtain the spherically averaged gas-phase metallicity profile as
\begin{equation}
Z (r) = \frac{Z_{\mathrm{in}} \rho_{\mathrm{in}}(r) + Z_{\mathrm{ism}} 
[\rho_{\mathrm{out}}(r) + \rho_{\mathrm{ism}}(r)]}{\rho_{\mathrm{in}}(r) +
\rho_{\mathrm{out}}(r) + \rho_{\mathrm{ism}}(r)},    
\end{equation}
and compute the projected metallicity and gas fraction profiles ($\Sigma_\mathrm{gas}(r)/\Sigma_\mathrm{star+gas}(r)$, where $\Sigma$ represents the surface density) accordingly (Fig. \ref{fig:fgas-z}).
To mimic the observation, we compute the projected profiles for a galaxy sample with varying $R_\mathrm{e}$. Our sample can be approximated by a lognormal distribution with a scatter of $\sigma_{\ln R_\mathrm{e}}=0.5$ and a median value of $R_\mathrm{e}=0.53$ kpc for $z=5-7$ and $0.36$ kpc for $z=7-9$.

We adopt $\epsilon=0.05{\sim}0.2$, $Z_\mathrm{in}=0.01Z_\odot$, $Z_\mathrm{sn}=3Z_\odot$, $f_\mathrm{sn}=0.2$, and $f_\mathrm{out}=0.95$ (corresponding to $\eta=18{\sim}4$), which reproduce the observation well. 
Eq.~\ref{eq:z_ism} gives $Z_{\mathrm{ism}}=0.04Z_\odot$ for $\epsilon=0.05$ and $0.15Z_\odot$ for $\epsilon=0.2$.
The gas-phase metallicity is dominated by the enriched ISM in the central region and by the metal-poor inflows in the outskirts. 

Different from \citetalias{Erb_08}, the FFB model introduces radially dependent inflows and outflows.
Assuming steady winds driven by supernovae in the FFB scheme, we consider a toy model for a steady solution of gas metallicity involving high inflow rates, SFR, and resultant outflows as a time-average approximation. 
At high redshift, galaxies are fed by cold gas streams that penetrate deep into halos without significant mixing until joining the central galaxy \citep{Dekel_09}. 
These metal-poor cold inflows dominate the gas mass outside the galaxy, leading to low metallicity at large radii. These radially dependent gas flows are similar to the accretion disk model proposed by \citet{Wangenci_22}, and are consistent with an inside-out growth scenario.

We also estimate the gas fraction $f_{\rm gas}$ in our observation samples (see Appendix \ref{sec:gas}) and compare it with predictions from different models.
As shown in Fig. \ref{fig:fgas-z-a}, galaxies at cosmic noon can reasonably be predicted by the \citetalias{Erb_08} models with a combination of inflow and outflow. We note that a few outliers in the NGDEEP sample with high sSFR exhibit metallicities higher than those predicted by the closed-box model. Since the model does not consider the gas recycling, those points may be attributed to the re-accretion of metal-enriched gas, which increases the metallicity. \citet{Alcazar_17} showed that in lower-mass galaxies, a larger fraction of the gas supply originates from wind recycling, with these systems re-accreting more gas relative to their stellar mass. As a result, the metallicity at high gas fractions can be higher than predicted by the closed-box model. This indicates efficient gas recycling in low-mass galaxies driven by high sSFR. The re-accretion of metal-enriched gas onto the galactic disk further contributes to flat/positive metallicity gradients \citep{Schonrich_17}.
\begin{figure*}[!t]
    \includegraphics[width=\textwidth]{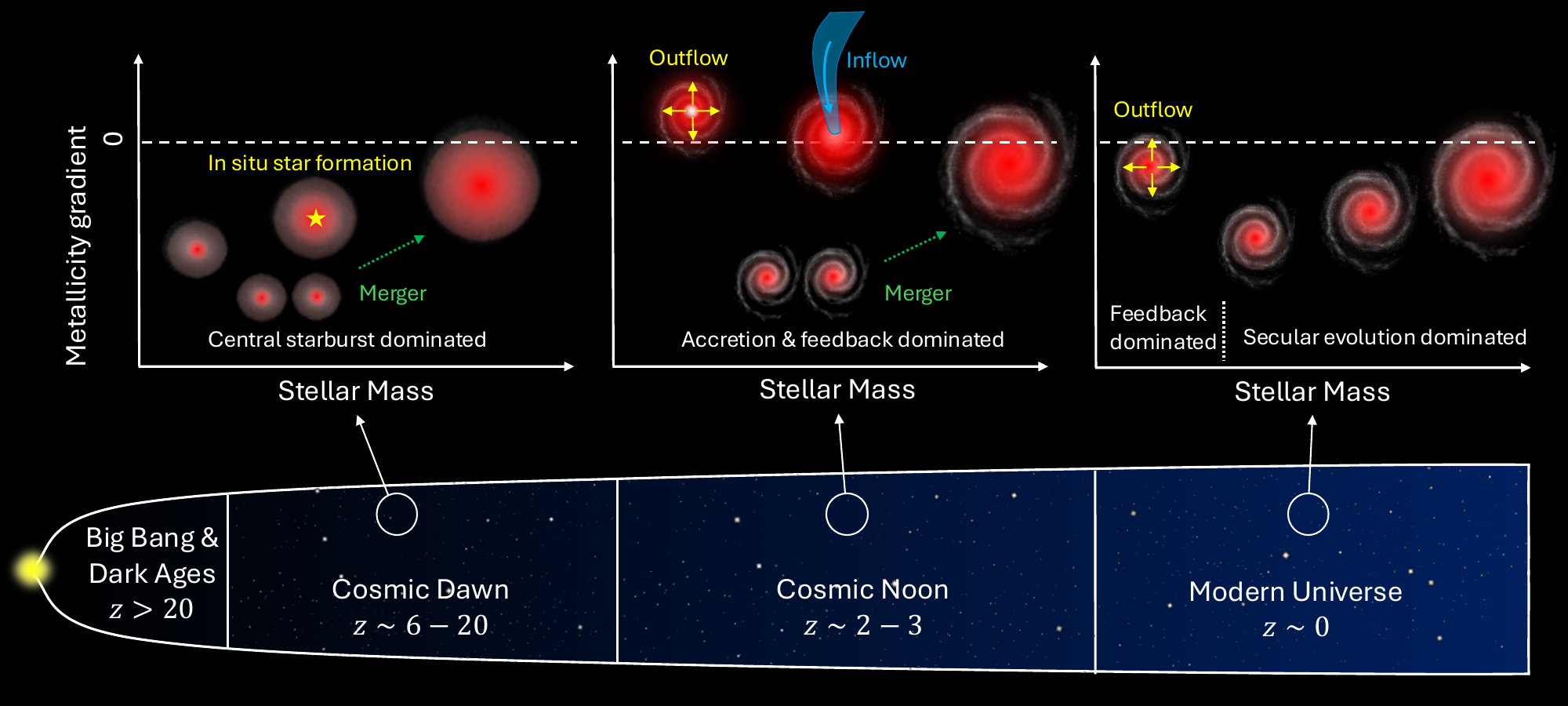}
    \caption{A sketch of how metallicity gradients are thought to evolve as galaxies grow across different cosmic epochs, based on our current knowledge. At cosmic dawn, the galaxies are observed with negative metallicity gradients, and they flatten with increasing stellar masses. This suggests the metal mixing and transport process may be inefficient, and coincides with the inside-out growth scenario with in situ metal production. At cosmic noon, the galaxies are observed with nearly flat metallicity gradients, and they have an anti-correlation with stellar masses. It suggests that metal mixing and transport mechanisms, including gas outflows, inflows, and radial transport, are efficient in redistributing metals in the ISM, which are sensitive to gravitational potentials. Mergers are also expected in dense high-redshift universe, which also contributes to gas mixing and flat gradients.
    In the local Universe, the MZGRs are observed to turn over at certain stellar masses, where higher-mass galaxies show positive MZGRs while lower-mass galaxies show the opposite trend. It suggests feedback-driven outflows may still be efficient for local dwarf galaxies, while the in situ metal production is dominating high-mass galaxies.\label{fig:sketch}}
\end{figure*}
To reproduce galaxies with both sufficiently low metallicities and low gas fractions, the \citetalias{Erb_08} model requires a high outflow rate. An alternative is provided by the FFB model, which predicts lower metallicities ($Z<0.1Z_\odot$) at low gas fractions, as expected by the high star formation efficiency and the dense cold inflows with inefficient mixing. From Fig. \ref{fig:fgas-z-b}, our observed metallicity gradients in the ASPIRE and FRESCO samples are also in agreement with the predictions of the FFB models at redshifts of $z=6$ and $z=8$, respectively. 
The inefficient gas mixing mechanism in the FFB scenario permits the existence of steep gradients at sufficiently high redshift. This simple analytical model demonstrates that steep gradients can be maintained even in the presence of inflows and outflows, provided that central in situ star formation dominates the galaxy formation and metal enrichment. In addition, we have estimated the sSFR maps of our $5<z<9$ samples based on SFR density and stellar mass density maps (Fig. \ref{fig:fgas}). The sSFR profiles can be visualized in Fig. \ref{fig:fgas-z-a}. We find that the sSFR is lower in the central bins ($r\in[0.0,0.8)$ kpc) and increases toward the outer regions ($r\in[1.7,2.5)$ kpc). This provides additional evidence for inside-out growth, echoing the recently observed core–disk structure at $z = 7.4$ \citep{Baker_24}, where the sSFR was found to increase with radius. The central sSFR suppression likely traces early-forming, dense stellar cores, while elevated outer-disk sSFR indicates ongoing gas accretion and star formation in extended disks. These results across our $5<z<9$ sample suggest that inside-out growth may be a dominant pattern for galaxy formation at the epoch of reionization. Consistent with this picture, \citet{Matharu_24} also reported evidence for the inside-out growth by comparing the \Ha\ and stellar continuum profiles at $4.8 < z < 6.5$.

\section{Summary}\label{sec:conclusion}

We summarize the results from the synthesis of our new JWST observations and literature observations in Fig. \ref{fig:sketch}. We find different phases in the redshift evolution of metallicity gradients, with a growing phase from sufficiently high-redshifts $z>5$, which flattens at $z\approx2$ and transitions to a descending phase until the present day. The implications of these three phases are as follows:

\begin{enumerate}
    \item The ascending phase may indicate rapid growth in inside-out mode, supported by continuous replenishment from cold gas accretion. This stage reflects the early buildup of galaxies in a universe dominated by cold gas. The steep gradients align with inside-out growth and limited radial mixing. To explain this, we apply a toy model within the Feedback-Free Starburst scenario. In this framework, dense inflowing gas sustains efficient, localized star formation while the redistribution of metals is inefficient \citep{Dekel_23,Lizhaozhou_23}. Such conditions preserve negative metallicity gradients, reflecting the early stages of galaxy assembly where internal chemical enrichment occurs earlier in the inner regions than in the outer disk. The observed positive correlation between metallicity gradient and stellar mass further supports the notion that massive galaxies grow more efficiently in an inside-out fashion during this period, linking their internal structure directly to their rapid mass accumulation history.
    \item The metallicity gradients flatten over time toward $z\approx2$, suggesting enhanced radial gas flows and mixing mechanisms. This epoch corresponds to the peak of the SFRD, and is driven by a combination of intense cold gas accretion, elevated SFR, and strong stellar/SNe feedback \citep{Dekel_09, Heckman_23,Sanders_23a}. The near-flat gradients observed at this time, along with large galaxy-to-galaxy scatter, reflect the turbulent internal dynamics triggered by gas accretion and feedback, which redistributes metals and flattens chemical abundances. The negative MZGR at this epoch signatures the dominant role of feedback in flattening gradients across stellar masses, especially in lower-mass systems \citep{Ma_17}.
    \item The gradual decline in metallicity gradients observed at $z<2$ coincides with the decrease in cold-mode accretion and a shift toward hot-mode accretion \citep{Voort_11}. This reduces the efficiency of star formation and limits pristine gas inflows, leading to slower chemical evolution. The reduced star formation activity and feedback at later times allow the galactic disks to stabilize and steepen the metallicity gradients. This is reflected by the declining global SFRD and indicates the transition to a more gradual mode of galaxy growth driven by secular evolution. The positive MZGRs for massive galaxies also provide evidence. Although low-mass galaxies still show a negative MZGR \citep{Lit_25}, feedback effects mainly impact young, low-mass systems in the local universe, whereas high-mass galaxies tend to be more evolved and dynamically stable with steeper metallicity gradients.
\end{enumerate}

These findings highlight the intricate connection between galaxy internal star formation and the cosmic-scale galaxy formation across cosmic time. The synchronized evolution pattern between metallicity gradients and cosmic star formation density reveals that gas accretion, star formation, and feedback not only shape the evolutionary pathways of individual galaxies but also collectively drive the large-scale formation of the galaxy population in the universe. Nevertheless, how these various physical processes interact to drive the observed evolution remains inconclusive and calls for further extensive investigation with both observations and simulations.

\begin{acknowledgments}
The authors thank the editors and anonymous referees for their careful reading and constructive comments, which greatly helped improve the manuscript.
This work is based on observations made with the NASA/ESA/CSA James Webb Space Telescope.
% associated with the programs GO-2078 (\url{https://doi.org/10.17909/vt74-kd84}), GO-1895 (\url{https://doi.org/10.17909/gdyc-7g80}), and GO-2079.
The JWST data presented in this article were obtained from the Mikulski Archive for Space Telescopes (\href{http://archive.stsci.edu}{MAST}) at the Space Telescope Science Institute, which is operated by the Association of Universities for Research in Astronomy, Inc., under NASA contract NAS 5-03127 for JWST. The specific observations analyzed can be accessed via GO-2078 (\dataset[doi:10.17909/vt74-kd84]{https://doi.org/10.17909/vt74-kd84},\citealt{aspire}), GO-1895 (\dataset[doi:10.17909/gdyc-7g80]{https://doi.org/10.17909/gdyc-7g80}, \citealt{fresco}), and GO-2079 (\dataset[doi:10.17909/02wx-6j29]{https://doi.org/10.17909/02wx-6j29}, \citealt{ngdeep}).
The data has been processed using public software \textsc{grizli} \citep{Brammer_23} and \texttt{Calwebb} \citep{bushouse_2024_10870758}.
The photometric catalog can be downloaded from the Dawn JWST Archive (DJA). The authors sincerely thank the FRESCO team (PI: Pascal Oesch) and the NGDEEP team (PI: Steven Finkelstein, Casey Papovich, and Nor Pirzkal) for developing their observing programs with a zero-exclusive-access period for the exquisite data.
This work made use of the High-Performance Computing resources at Tsinghua University. This work uses images and photometry catalogs retrieved from the Dawn JWST Archive (DJA). DJA is an initiative of the Cosmic Dawn Center, which is funded by the Danish National Research Foundation under grant No. 140. 

We thank Steven Finkelstein and Raymond Simons for helpful discussions regarding NGDEEP data; Pascal Oesch and Romain Meyer for helpful discussions regarding FRESCO data. Z.H.L acknowledges Dandan Xu and Cheng Li for insightful discussions about the observational and theoretical context; Xiangcheng Ma, Guochao Sun, and Xunda Sun for valuable input about the FIRE simulations; Ayan Acharyya for helpful discussions on FOGGIE simulations and their theoretical interpretation; Alex Garcia for sharing results from EAGLE, Illustris, TNG and SIMBA simulations; Piyush Sharda for helpful discussions with metallicity gradient models; Prerak Grag for help with the photoionization model; Zechang Sun for maintaining High-Performance computers at Tsinghua University; Mengting Ju for sharing results from JWST MSA-3D; Zefeng Li and William Baker for helpful comments.

Z.H.L., Z.C., X.L., and Y.W. are supported by the National Science Foundation of China (grant no. 12073014), the science research grants from the China Manned Space Project with No. CMS CSST-2021-A05, and the Tsinghua University Initiative Scientific Research Program (No. 20223080023). Z.H.L. also acknowledges that this work was conducted in part at the Cosmic Dawn Center with financial support from the center.
X. W. is supported by the China Manned Space Program with grant no. CMS-CSST-2025-A06, the National Natural Science Foundation of China (grant 12373009), the CAS Project for Young Scientists in Basic Research Grant No. YSBR-062, the Fundamental Research Funds for the Central Universities, and the Xiaomi Young Talents Program. 
X.W. also acknowledges work carried out, in part, at the Swinburne University of Technology, sponsored by the ACAMAR visiting fellowship.
Z.Z.L. acknowledges the European Union’s Horizon 2020 programme under the Marie Skłodowska-Curie grant No. 101109759 (“CuspCore”)  and the Israel Science Foundation Grant ISF 861/20.
SEIB is supported by the Deutsche Forschungsgemeinschaft (DFG) under Emmy Noether grant number BO 5771/1-1. F.W. acknowledges support from NSF grant AST-2308258. M.T. acknowledges support from the NWO grant 016.VIDI.189.162 (“ODIN”). S.ZOU acknowledges support from the National Natural Science Foundation of China (No.12303011).

\end{acknowledgments}
\facilities{JWST (NIRCam, NIRISS)}
\software{\textsc{Grizli} \citep{Brammer_23}, \texttt{Calwebb} \citep{bushouse_2024_10870758}, \textsc{Linmix}\citep{Kelly_07}, Astropy \citep{2013ascl.soft04002G}, Scipy \citep{scipy}, $\textsc{Emcee}$ \citep{emcee}, \texttt{BEAGLE} \citep{Chevallard_16}, \texttt{SourceXtractor++} \citep{sepp}, Photutils \citep{larry_bradley_2024_13989456}.}
\bibliography{sample7}{}
\bibliographystyle{aasjournalv7}

\begin{appendix}

\section{Measuring Metallicity gradients in mass bins}\label{sec:massbin}
\begin{figure*}[t!]
	\centering
	\begin{minipage}{1\linewidth}
		\centering
\includegraphics[width=1\linewidth]{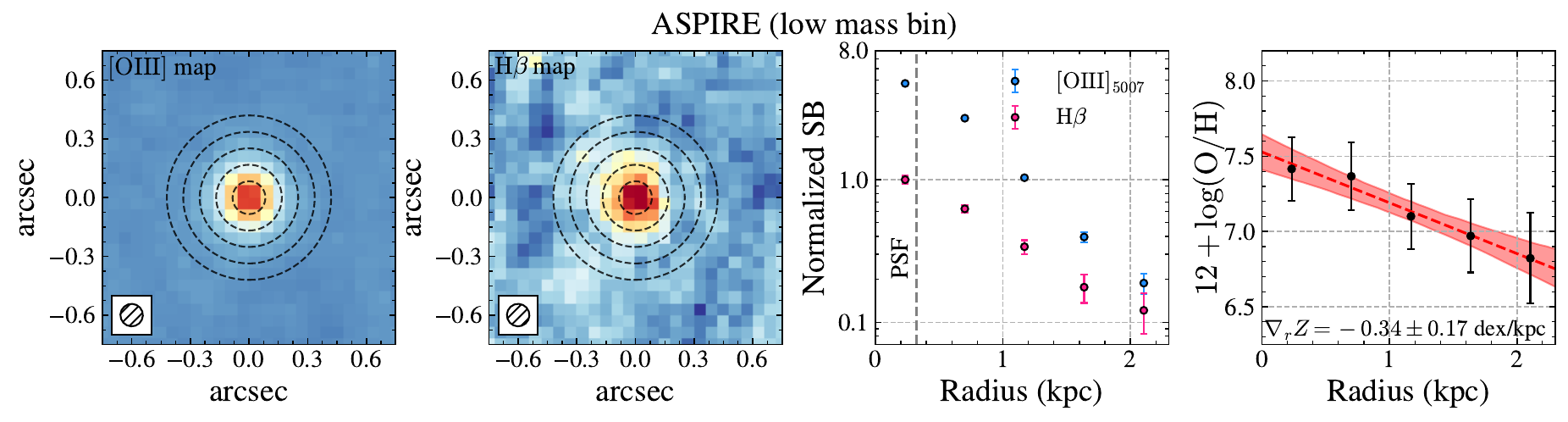}
	\end{minipage}
	\begin{minipage}{1\linewidth}
		\centering
\includegraphics[width=1\linewidth]{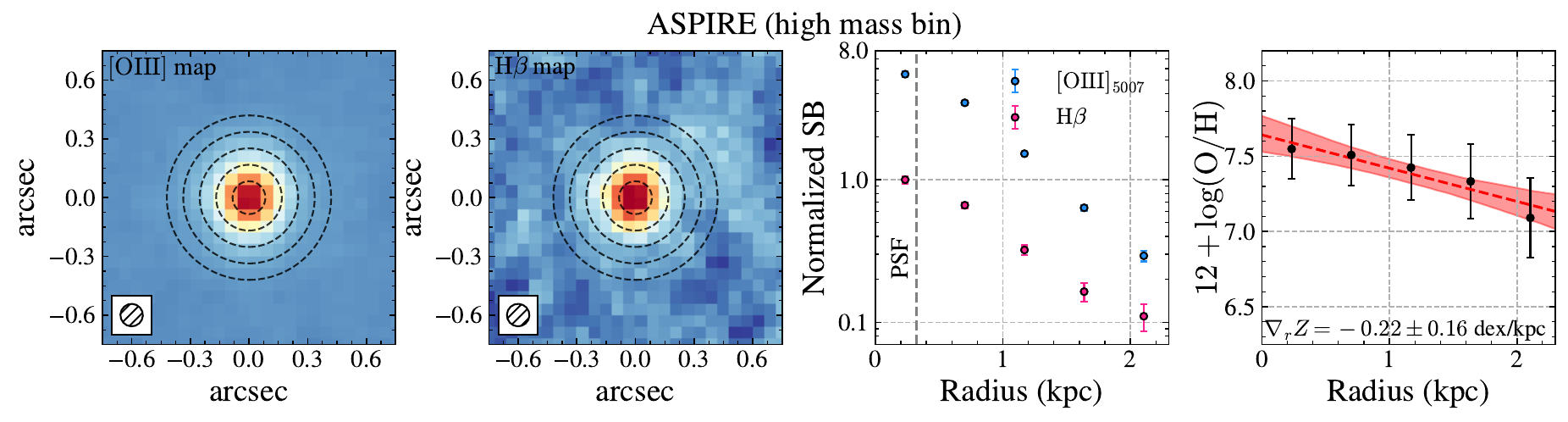}
	\end{minipage}
	\begin{minipage}{1\linewidth}
		\centering
\includegraphics[width=1\linewidth]{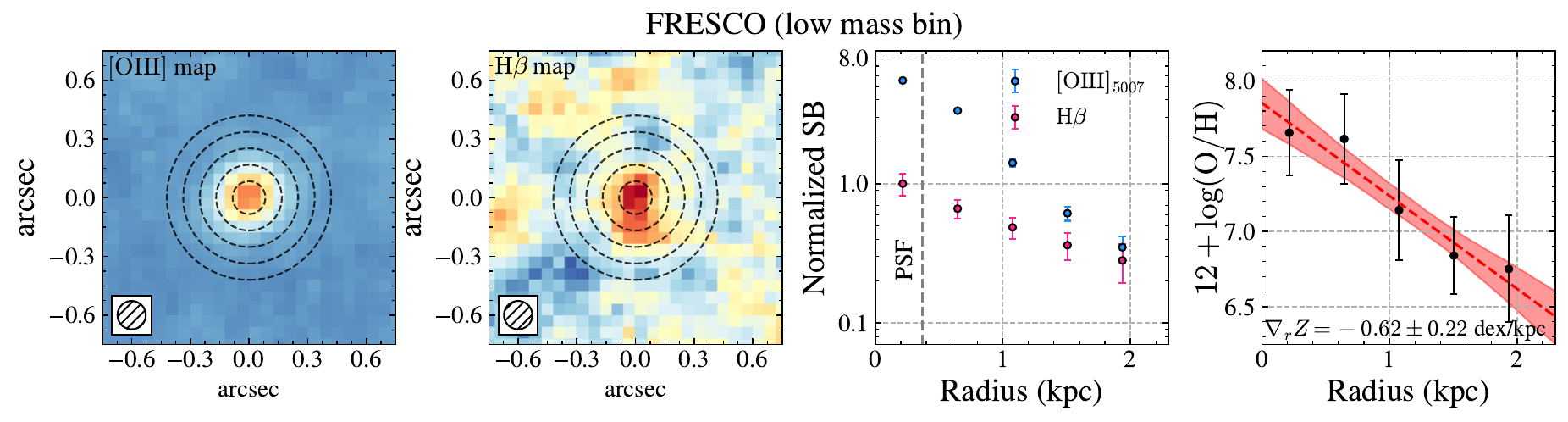}
	\end{minipage}
	\begin{minipage}{1\linewidth}
		\centering
\includegraphics[width=1\linewidth]{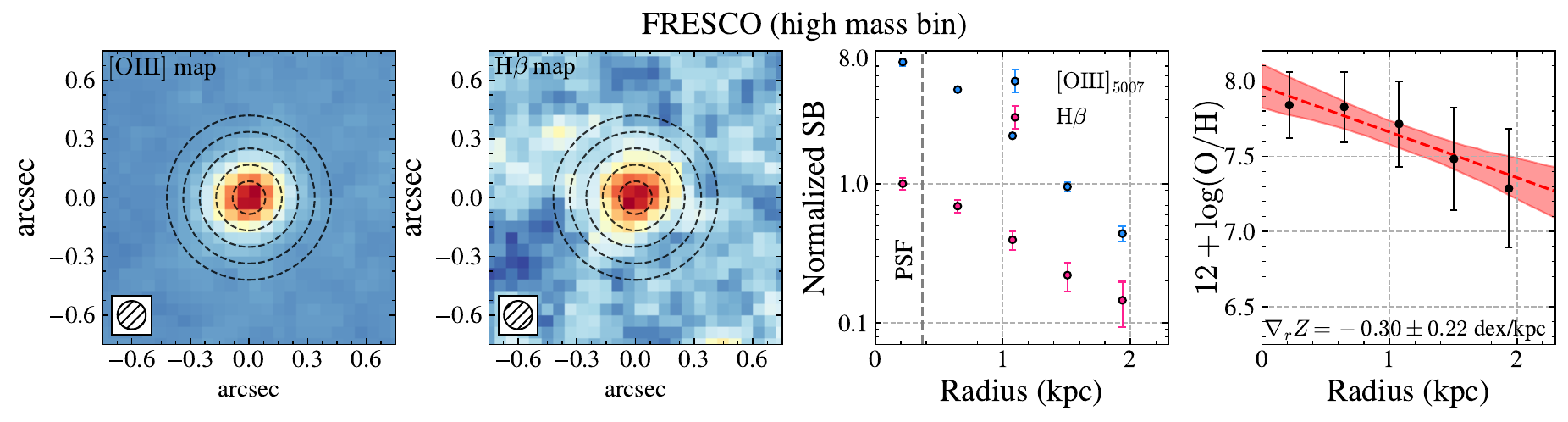}
        \vspace*{-2em}
	\end{minipage}
 		\caption{Stacking results in different mass bins. First two rows: results from ASPIRE galaxies in low-mass bin $(7.18<\log(M_*/M_\odot)<8.09)$ and high mass bin $(8.09<\log(M_*/M_\odot)<9.00)$. Last two rows: results from FRESCO galaxies in low-mass bin $(7.30<\log(M_*/M_\odot)<7.95)$ and high mass bin $(7.95<\log(M_*/M_\odot)<8.70)$.
    \label{fig:massbin}}
    
\end{figure*}

\begin{figure*}
    \centering
    \plotone{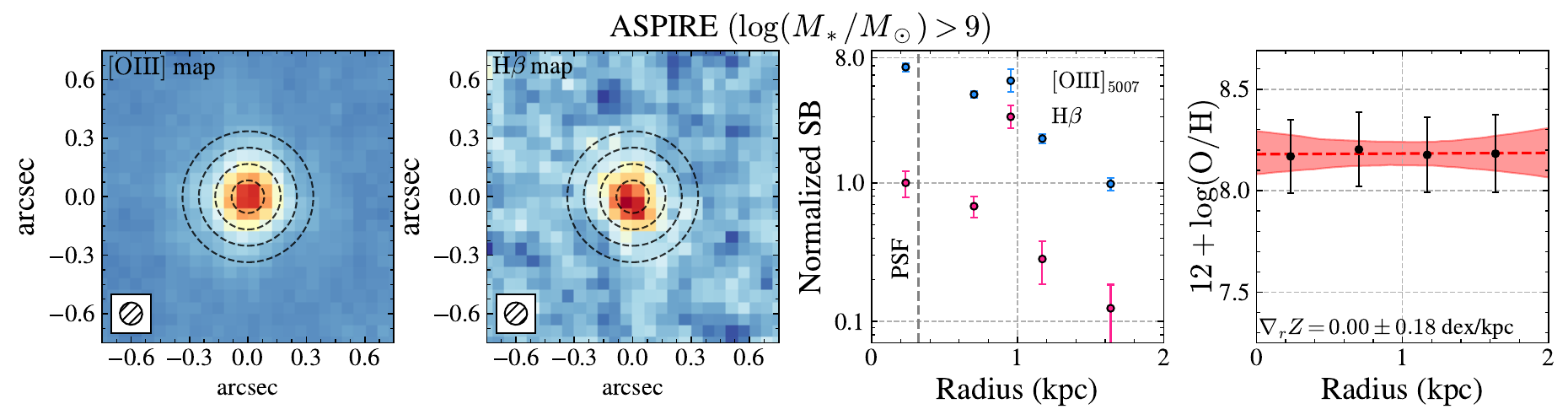}
    \caption{The same as Fig. \ref{fig:massbin}, but for galaxies with highest stellar masses ($\log(M_*/M_\odot)>9.00$) in ASPIRE sample. The metallicities are estimated using the upper branch solution of the R3 relation in \citet{Sanders_23}.}
    \label{fig:mass-ge9}
\end{figure*}

\begin{figure*}[ht!]
	\centering
	% \plotone{NGDEEP_stacking_lowmass.pdf}
    \includegraphics[width=1\linewidth]{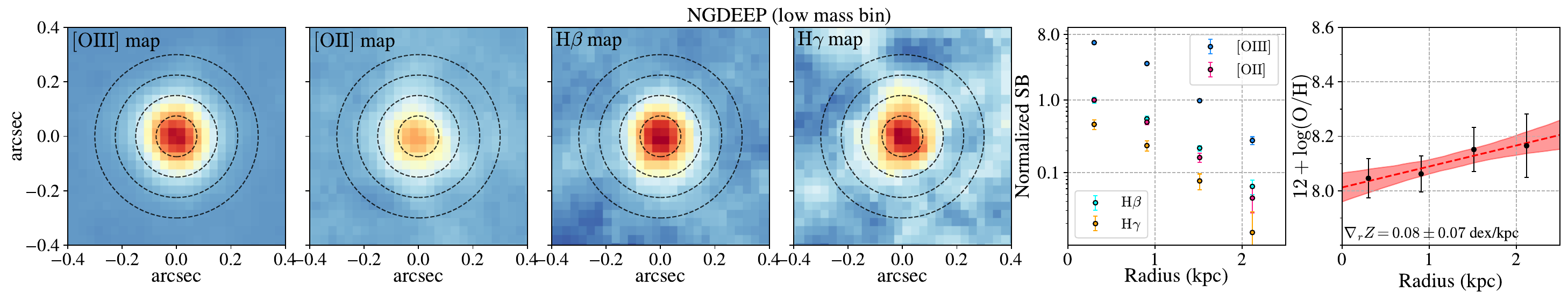}
    \includegraphics[width=1\linewidth]{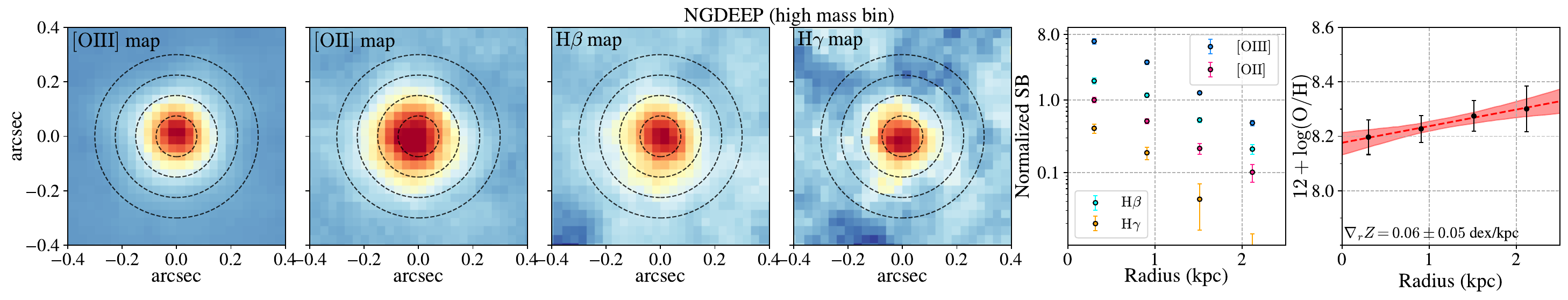}
	% \plotone{NGDEEP_stacking_highmass.pdf}
 	\caption{Stacking results from NGDEEP galaxies in low-mass bin $(6.87<\log(M_*/M_\odot)<8.44)$ and high mass  bin $(8.44<\log(M_*/M_\odot)<9.81)$.
    \label{fig:ngdeep-mass-bin}}
    
\end{figure*}

The galaxies grow in different sizes and masses, which are correlated through the mass-size relation \citep{Shibuya_15, Langeroodi_23}. The metallicity gradients may be different for galaxies of different sizes. To investigate the impact of galaxy masses on the metallicity gradient, we divided the stacking into different mass bins at each redshift. 
% The mass distributions of ASPIRE and FRESCO samples are shown in Fig. \ref{fig:mzr_pred}.
For the ASPIRE sample, we utilized three mass bins: the most massive bin includes galaxies with $\log(M_*/M_\odot)>9.00$, while the remaining lower-mass galaxies ($\log(M_*/M_\odot)<9.00$) were evenly split into two bins: a low-mass bin $(7.18<\log(M_*/M_\odot)<8.09)$ and a high-mass bin $(8.09<\log(M_*/M_\odot)<9.00)$. The FRESCO sample is less massive, and we evenly divide them into low mass bin $(7.30<\log(M_*/M_\odot)<7.95)$ and high mass bin $(7.95<\log(M_*/M_\odot)<8.70)$. For comparison between individual and stacked measurements, we also separate the NGDEEP sample into low mass bin $(6.87<\log(M_*/M_\odot)<8.44)$ and high mass bin $(8.44<\log(M_*/M_\odot)<9.81)$.
Within each mass bin, the stellar mass range spans approximately $\sim1$ dex, corresponding to a change of $R_e\sim0.1-0.2$ kpc as inferred from the mass-size relation \citep{Langeroodi_23}. We applied the same median stacking method as used for galaxies in redshift bins. We applied the same R3 method \citep{Sanders_23} to measure metallicity gradients for high-mass bins and low-mass bins in ASPIRE and FRESCO samples, shown in Fig.~\ref{fig:massbin}. However, since the bin with $\log(M_*/M_\odot)>9.00$ is massive enough that MZR predicts these galaxies have higher metallicity, which is out of the lower branch range of the R3 relation, we use the upper branch instead to measure their metallicity gradient, shown in Fig.~\ref{fig:mass-ge9}.  For the NGDEEP sample, we use the same method with both R3 and R2 calibrations \citep{Bian_18} to jointly constrain metallicity as for individual galaxies, shown in Fig.~\ref{fig:ngdeep-mass-bin}. The stacks in mass bins provide a more representative characterization of the galaxy population with similar masses and sizes. From the stacked map of NGDEEP sample, we measure the MZGR slope of $\rm-0.019~dex~kpc^{-1}$ using the two stacked points at low and high masses, which is consistent with the slope of $\rm -0.020~dex~kpc^{-1}$ obtained from a linear regression on individual measurements (Section \ref{sec:mzgr}).

\section{Metallicity gradients in units of effective radius}\label{sec:Re}

As galaxies intrinsically have smaller sizes at higher redshift \citep{Morishita_24}, we expect steeper gradients for small galaxies if the changes in metallicity are the same. Galaxy size is also correlated with galaxy stellar mass, with massive galaxies having larger effective radii.
Thus, the metallicity gradient measured with respect to effective radius $\rm R_e$ should provide a quantity that is less dependent on stellar mass and size. 

\begin{figure*}[ht!]
	\centering
\includegraphics[width=0.9\linewidth]{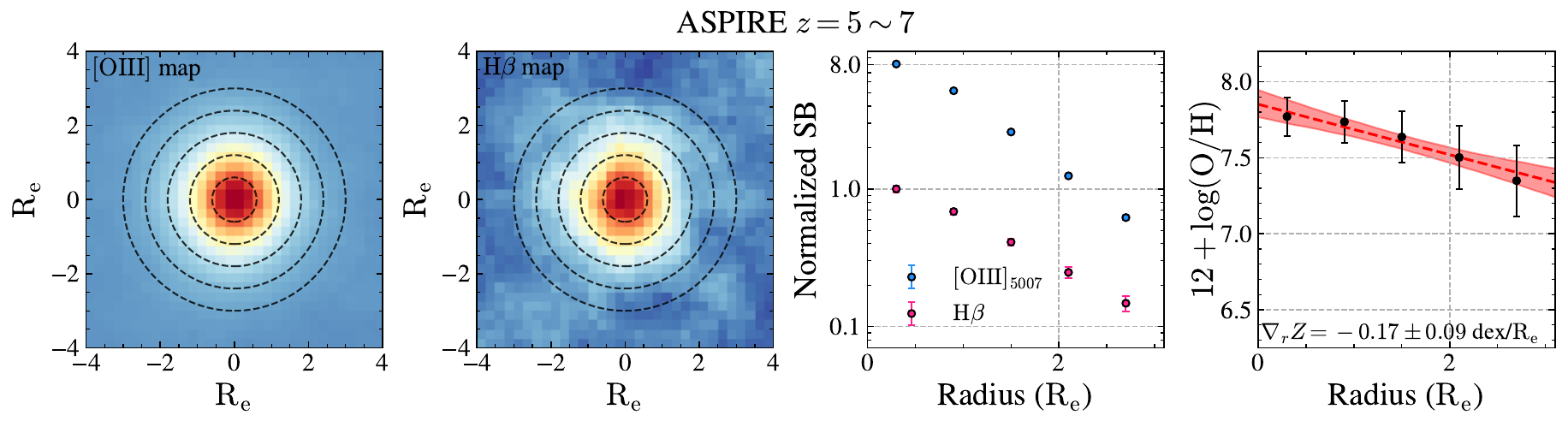}
\includegraphics[width=0.9\linewidth]{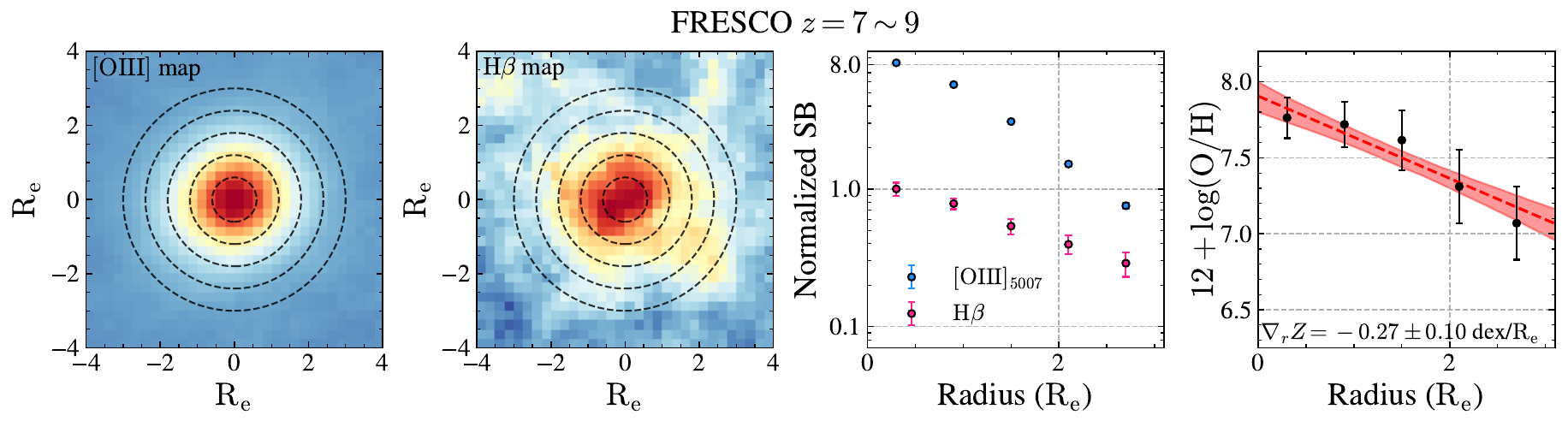}
		\caption{Stacking results in scale of effective radius ($\mathrm{R_e}$). 
    \label{fig:re-stacking}}
\end{figure*}

% We fit S\'{e}rsic profiles convolved with PSFs on direct images using \texttt{PetroFit}\citep{Geda_2022}. The PSFs in each band are modeled by \texttt{WebbPSF}\citep{Perrin_14}. For the ASPIRE and FRESCO sample, we measure $\rm R_e$ on stacked images in the F200W and F210M bands, respectively. 
We use a s\'ersic model convolved with PSF to fit galaxy morphology with \textsc{Galfit}. With fitted $\rm R_e$, we resample the emission maps to the pixel scale of $0.25~\rm R_e$ with flux conservation. Then we stack the galaxy emission maps using the median stacking method, before measuring the gradients on the stacks. In Fig. \ref{fig:re-stacking}, we show the stacked results from the ASPIRE and FRESCO samples. We still observe steep negative gradients in units of $\rm dex~R_e^{-1}$. This test eliminates possible bias from the variation of galaxy sizes.

\begin{figure*}[t]
 \centering
 % \plotone{zgrad_re-z.pdf}
 \includegraphics[width=0.6\textwidth]{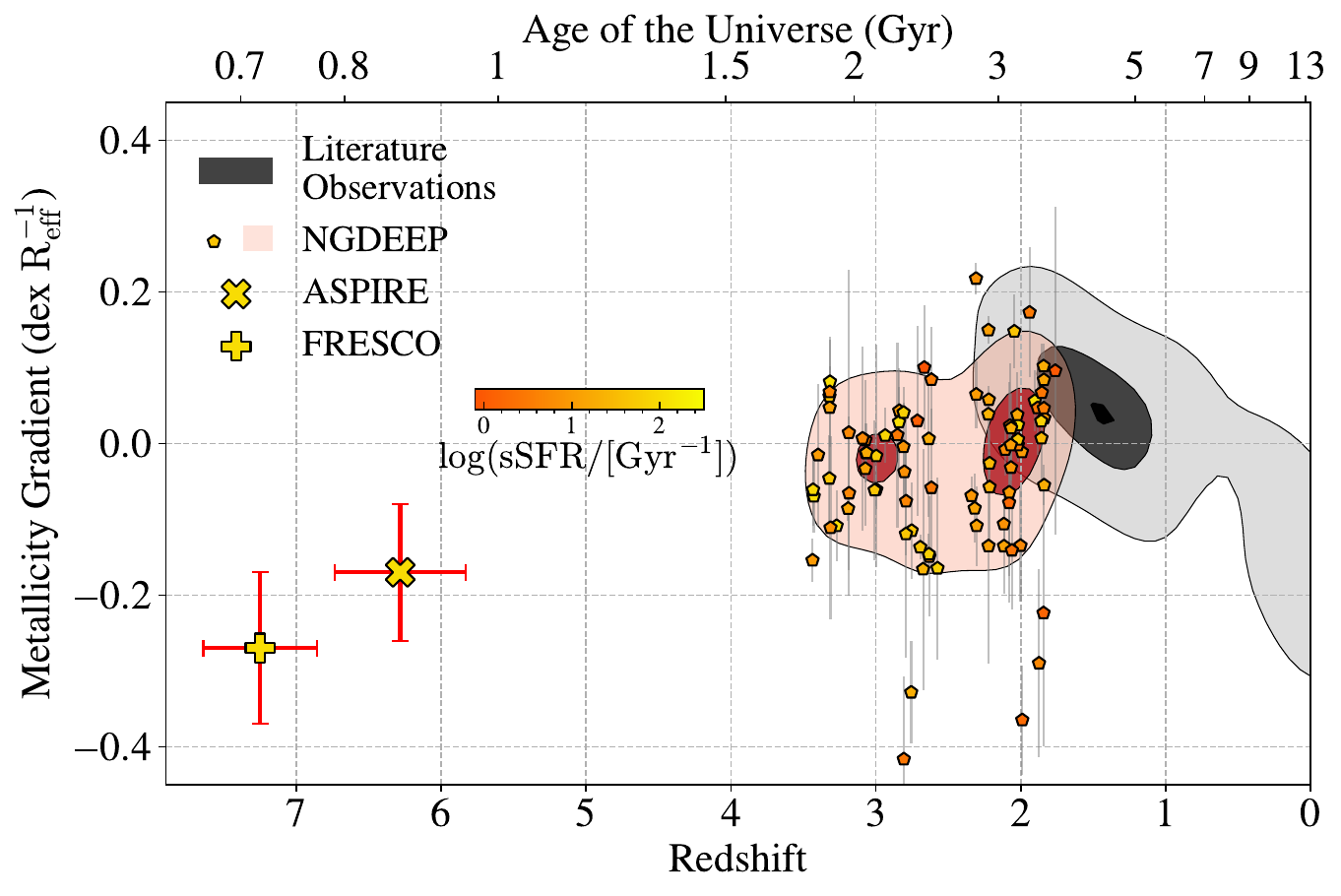}
 \caption{Metallicity gradients measured with respect to $\rm R_{eff}$. The symbols are the same as in Fig. \ref{fig:zgrad-z}. The gray shadow encompasses literature observations \citep{Carton_18, Franchetto_21, Simons_21}.
 \label{fig:zgrad_re-z}}
\end{figure*}

The effective radii for the NGDEEP sample are measured on the F115W image. The metallicity gradients rescaled by $\rm R_e$ are shown in Fig. \ref{fig:zgrad_re-z}. We also include literature observations at high redshift \citep{Carton_18, Simons_21}, and local observations from MaNGA \citep{Franchetto_21}, which provide metallicity gradients in the same units for comparison. The ascending phase at $3.5<z<1.75$ remains in the NGDEEP galaxies. The gradients at $z>5$ are still steeper than galaxy populations at lower redshift despite their intrinsically smaller sizes. The redshift evolution of metallicity gradients measured with $\rm R_e$ is consistent with that measured with kpc.

\section{Gas fraction estimation}\label{sec:gas}

The gas fraction is useful information when comparing our observations with analytical models shown in Fig.~4. 
In this section, we present our gas fraction estimation as follows. 
We estimate gas density by inverting the Kennicutt–Schmidt law \citep[KS law,][]{Schmidt_59, Kennicutt_89}:

\begin{equation}
    \Sigma_{\rm SFR}=10^{-12}\kappa_s\Sigma_{\rm gas}^{1.4},\label{eq:ks}
\end{equation}
where $\Sigma_{\rm SFR}$ is in $M_\odot\rm yr^{-1}\ kpc^{-2}$, and $\Sigma_{\rm gas}$ is in $M_\odot\rm\ kpc^{-2}$. $\kappa_s$ is the burstiness parameter representing the deviation from the original KS relation \citep{Ferrara_19}. There is evidence that the KS relation is not universally applicable and may undergo an upward shift at higher redshifts \citep{Pallottini_22, Markov_22}. For example, \citet{Daddi_10} found that the normalization of $\Sigma_{\rm SFR}$ for highly star-forming galaxies is shifted by $\sim0.9$ dex from the conventional KS relation.
Thus, the highly star-forming galaxies are expected to have $\kappa_s>1$. Thus, we adopt $\kappa_s=10$ for the galaxy sample at $z\approx1-3$, and a higher value of $\kappa_s=20$ for galaxies at $z\approx6-7$, due to burstier star formation suggested by recent measurements \citep{Markov_22,Vallini_21,Vallini_24}. Since $\kappa_s$ for individual galaxies is vastly unconstrained, we add $50\%$ uncertainty to the values we adopted to have a conservative estimation of the gas fraction.

We adopt the Balmer ratio $\Ha/\Hb=2.86$ in Case B recombination, where we estimate the SFR
by using \Hb luminosity and the following calibration \citep{Kennicutt_98}:
\begin{equation}
    \mathrm{SFR}=13.16\times10^{-42}\frac{L_{\Hb}}{\rm erg\ s^{-1}} (M_\odot\ \rm yr^{-1}).
\end{equation}
Since we cannot derive spatially resolved dust attenuation without other Balmer series, we cannot directly correct the \Hb\ flux for dust. Recent studies of galaxies within a similar mass range at $z>5$ have been found with a certain amount of dust, either through SED modeling \citep{Matthee_23, Tacchella_23}, or Balmer decrement \citep{Sandles_23}. The typical dust is expected to be $\rm E(B-V)\sim0.1$ corresponding to the uncertainty $35\%$ for the SFR derived from the uncorrected flux \Hb. Since this uncertainty is not negligible, we add one-sided $35\%$ uncertainties to SFR maps. While for the NGDEEP sample, the \Hb fluxes have been corrected by dust using the Balmer decrement between \Ha, \Hb, and \Hg (equation \ref{eq:mcmc}).

The $\sum_{\rm star}$ is obtained by spatially resolved SED fitting to each pixel. 
We match the F115W and F200W with the PSF to the F356W band in the ASPIRE sample, and we match all the photometric bands with the PSF
to F444W in the FRESCO sample. After the PSF matching, both the images and emission maps have the same PSF. We use a median stack for all cutouts in units of luminosity and convert back to flux density using the median redshifts $z=6.28$ and $z=7.24$ for ASPIRE and FRESCO, respectively. 
We then apply the SED fitting to each pixel, assuming the fixed redshifts $z=6.28$ and $z=7.24$ for the ASPIRE and FRESCO samples, respectively. 

\begin{figure*}[ht!]
 \centering
 \includegraphics[width=\textwidth]{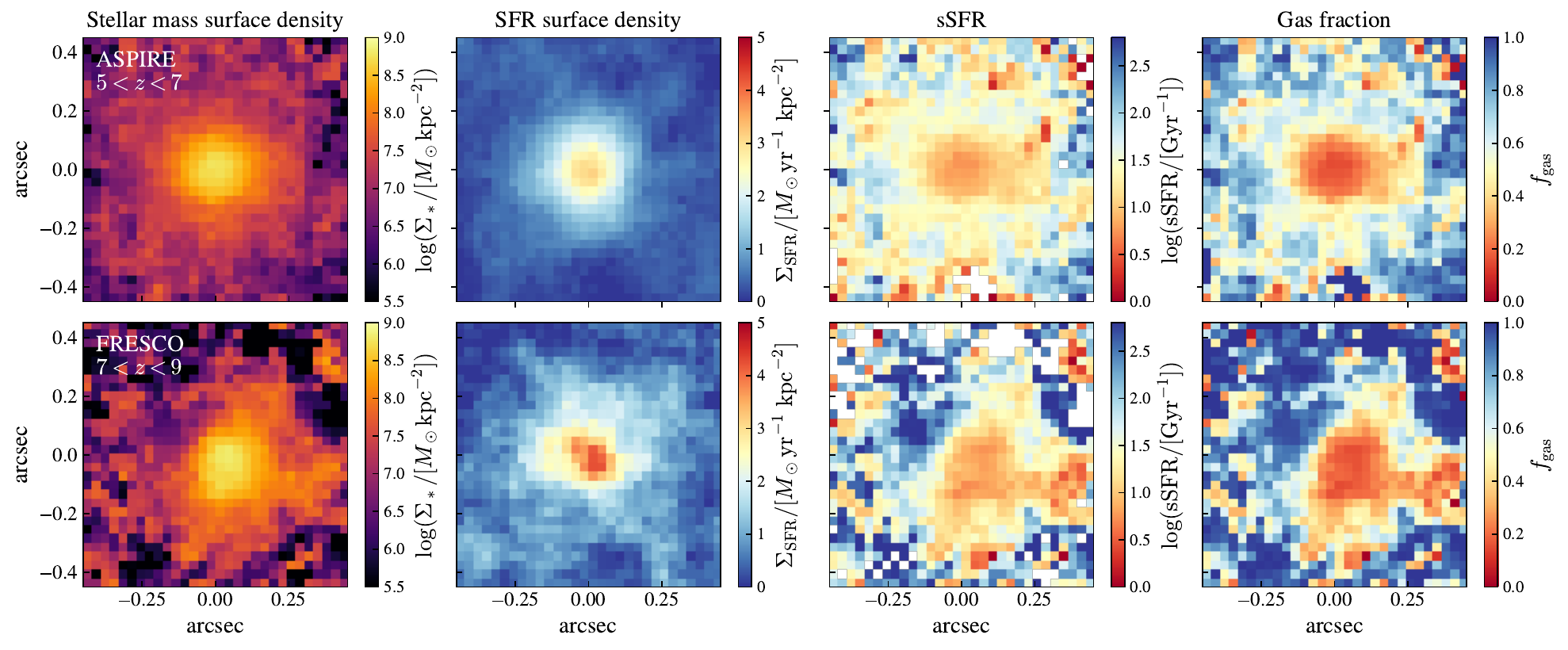}
 \caption{
From left to right: stellar mass surface density, SFR surface density, specific star formation rate (sSFR), and gas fraction of ASPIRE (upper) and FRESCO (lower) samples.
 \label{fig:fgas}}
\end{figure*}

The gas surface density $\Sigma_{\rm gas}$ can be derived by the inverse KS law in Eq. \ref{eq:ks}. The gas fraction is then expressed as:
\begin{equation}
    f_{\rm gas}=\Sigma_{\rm gas}/(\Sigma_{\rm gas}+\Sigma_{\rm star}).
\end{equation}
The stellar mass surface density map, the SFR surface density map, and the derived gas fraction map are shown in Fig. \ref{fig:fgas}. The gas fraction we derived here has been used in Fig. \ref{fig:fgas-z-a}. 

\section{Fit redshift evolution of metallicity gradients}\label{sec:zfit}
\begin{figure*}
    \centering
    \includegraphics[width=0.65\linewidth]{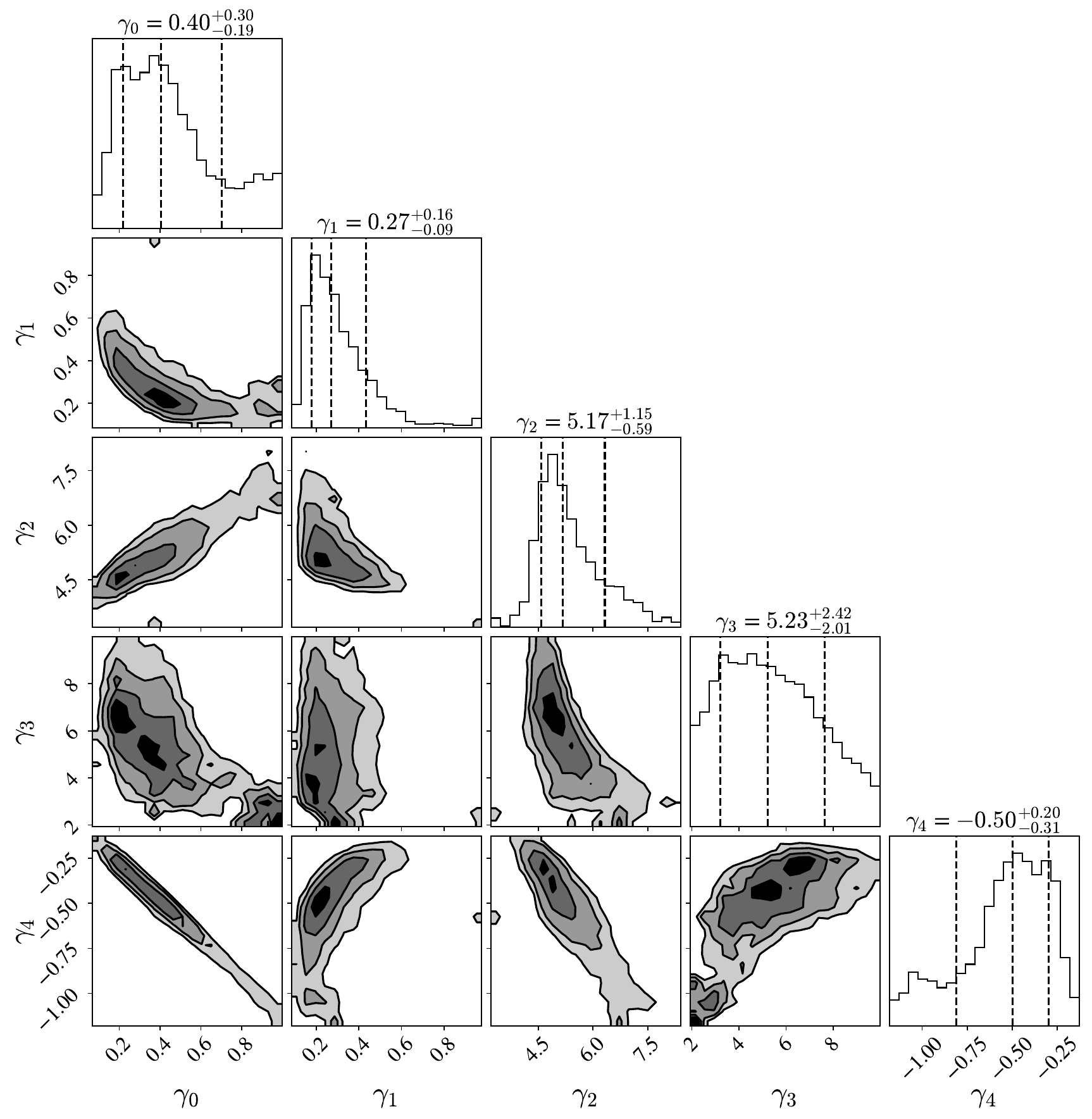}
    \caption{Posterior distribution of $\gamma_0,~\gamma_1,~\gamma_2,~\gamma_3$, and $\gamma_4$ in Equation \ref{eq:zgrad-eq} from MCMC sampling. The values at the top of each column are the medians with $1\sigma$ interval.}
    \label{fig:mcmc}
\end{figure*}

We perform MCMC sampling of the multidimensional parameter space with the package \textsc{Emcee} \citep{2013PASP..125..306F} to fit the double-power-law model of Eq (\ref{eq:zgrad-eq}).
We apply a flat prior for the following parameters: $\gamma_0\sim\mathcal{U}(-10,10)$,  $\gamma_1\sim\mathcal{U}(-1,1)$, $\gamma_2\sim\mathcal{U}(0,10)$, $\gamma_3\sim\mathcal{U}(-10,10)$, and $\gamma_4\sim\mathcal{U}(-10,10)$. The sampling is performed with 32 walkers, 5000 iterations each, and with a burn-in period $n=1000$. The posterior probability distribution for the parameters is shown in Fig. \ref{fig:mcmc}. The median model and $1\sigma$ prediction interval are shown in Fig. \ref{fig:zgrad-z} and Fig. \ref{fig:zgrad-z-sim}.

\section{Systematics using different emission-line calibrations}\label{sec:syst}

To verify that our results are not significantly altered by different metallicity diagnostics, we compare our fiducial measurements using \citet{Bian_18} in the NGDEEP sample with other popular calibrations based on different samples and methods \citep{Maiolino_08,Curti_20b,Nakajima_22}.
The calibrations in \citet{Maiolino_08} combine the direct electron temperature measurements from the Sloan Digital Sky Survey (SDSS) in the low-metallicity 
% ($\oh\lesssim8.35$) 
branch \citep{Nagao_06} and the predictions of a photoionization model in the high-metallicity branch \citep{Kewley_02}. The calibrations in \citet{Curti_20b} are derived from a set of individual low-metallicity galaxies together with stacks of high-metallicity galaxies in the redshift range $0.027 < z < 0.25$, where auroral lines are detected in composite spectra, where the metallicities are self-consistently measured via the $\rm T_e$ method for both the high and low metallicity range. The calibrations in \citet{Nakajima_22} are derived from more metal-poor galaxies with auroral lines detected and selected from HSC-SSP and SDSS catalog and follow-up observations \citep{Kojima_20,Isobe_22}, in addition to high metallicity galaxies used in \citet{Curti_20b}. 

\begin{figure}[t!]
 \centering
 \includegraphics[width=0.5\textwidth]{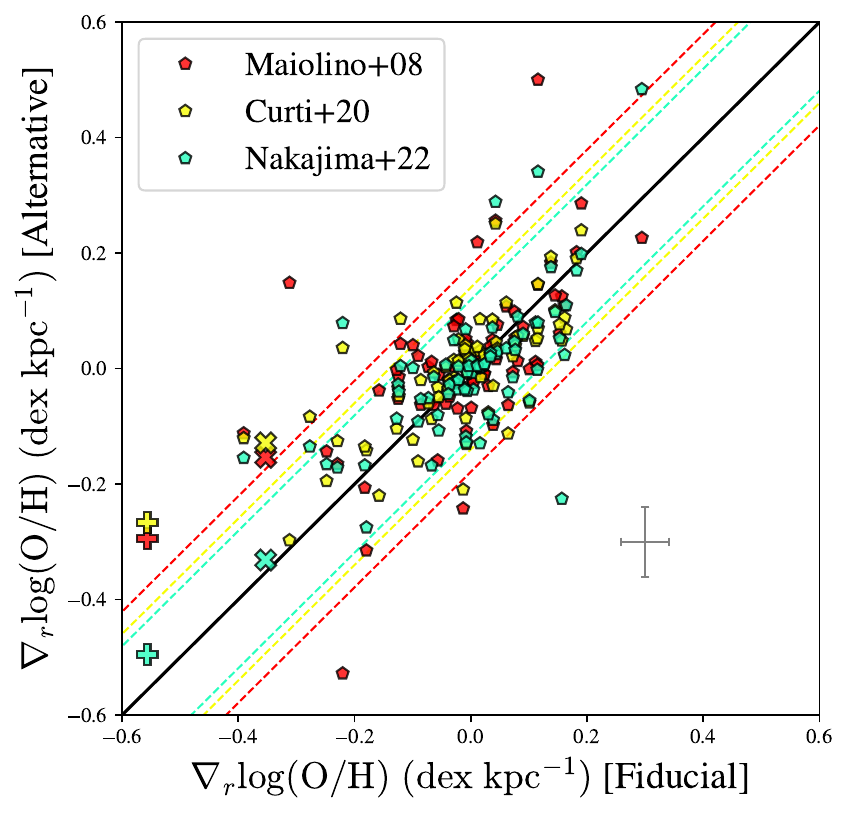}
 \caption{Comparisons between different strong-line calibrations for metallicity gradients. NGDEEP, ASPIRE, and FRESCO measurements are marked with pentagons, ``$\times$" and ``$+$" respectively.
 The x-axis is the fiducial measurements with \citet{Bian_18} for NGDEEP galaxies and \citet{Sanders_23} for ASPIRE and FRESCO galaxies, and the y-axis is the measurements using alternative calibrations \citep{Maiolino_08, Curti_20b, Nakajima_22}. The diagonal black line shows equality, and the dashed lines show the standard deviation around equality for each calibration in the corresponding color. 
 % Different calibrations yield a 1-$\sigma$ of $\lesssim0.1\rm dex/kpc$ systematic uncertainties in metallicity gradient measurement. 
 \label{fig:calib_comp}}
\end{figure}

Following the same procedures, we derive the metallicity maps, measure the metallicity gradients for each galaxy and the stacks, and compare those with the fiducial measurements. The comparisons are shown in Fig.~\ref{fig:calib_comp}. We find no significant bias using either of the three alternative calibrations, with all of the scatters around equality being within $1\sigma$ measurement uncertainties $\lesssim0.1~\rm dex~kpc^{-1}$. The different calibrations for two $z>5$ stacks also introduces small uncertainties $\lesssim0.1~\rm dex~kpc^{-1}$. Thus, the choice of metallicity calibrations does not alter our results of the evolution of the metallicity gradient.

\section{Integrated metallicity of $z>5$ sample}\label{integ}

\begin{figure*}[t!]
    \gridline{\fig{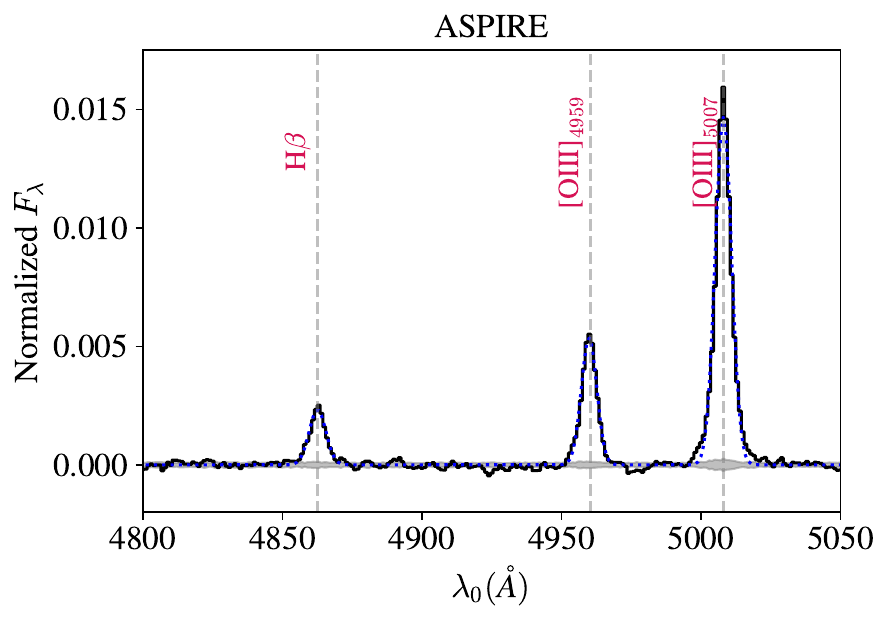}{0.5\textwidth}{}\fig{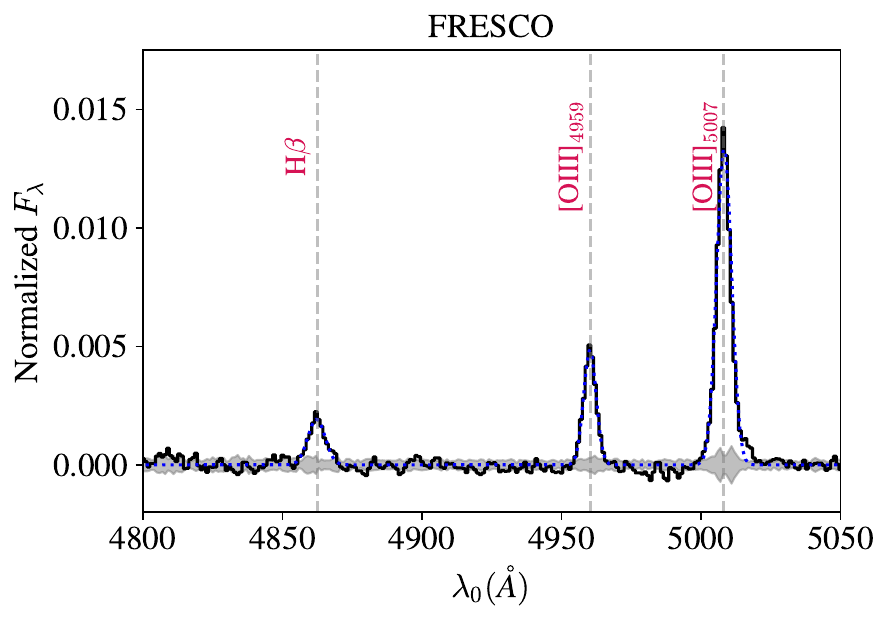}{0.5\textwidth}{}}
    \caption{Median stacked 1D rest-frame spectra of galaxies in ASPIRE and FRESCO. The fluxes are normalized by \OIII$_{4959+5007}$ flux to avoid excessive weighting towards bright sources. The gray shadow represents the spectra uncertainty. The blue dotted lines are the best-fit Gaussian models.\label{fig:spec1d}}	
\end{figure*}

Here we measure the integrated metallicity of our stacks in Fig.~\ref{fig:zgrad-stacking}.
We resample our 1D spectra to rest-frame on a common 1 \AA wavelength grid with flux preserved using \texttt{spectres} \citep{Carnall_17}. Following \citet{Wang_22a}, to avoid the excessive weighting towards bright sources with stronger line fluxes, we normalized each spectrum by its measured \OIII flux. We take the median value of the normalized spectra at each wavelength grid, and the uncertainty is estimated by measuring the standard deviation from 1000 bootstrap realizations of the sample. The median stacked 1D rest-frame spectra of ASPIRE and FRESCO are shown in Fig. \ref{fig:spec1d}.
We fit \OIII and \Hb line fluxes using Gaussian profiles. We do not set line-ratio constraints between $\OIII_{5007}$ and $\OIII_{4959}$. Assuming the lower branch solution, the metallicities are measured using R3 calibrations from \citet{Sanders_23}, the same as we use for metallicity gradient measurements. We also compare the metallicities from different calibrations in \citet{Nakajima_22}.
We list the integrated line fluxes and metallicities in Table \ref{tab:line_info}. We note that \citet{Nakajima_22} calibration gives $\sim0.1-0.2$ dex high metallicities. More detailed analysis of integrated metallicity and mass-metallicity relation of the ASPIRE and NGDEEP sample is presented in \citet{Li_25} and X. He et al. submitted.

\begin{deluxetable}{ccc}
\tablecaption{Measurements from stacked 1D spectra in ASPIRE and FRESCO.\label{tab:line_info}}
\tablewidth{0pt}
\tablehead{
\colhead{Property} & \colhead{ASPIRE} & \colhead{FRESCO}
}
\startdata
$\log(R3)$ & $0.79\pm0.02$ & $0.78\pm0.04$\\
$^\textrm{a}F_{\OIII_{4959}}$ & $34.17\pm0.70$ & $30.18\pm1.64$ \\
$^\textrm{a}F_{H_\beta}$ & $16.10\pm0.60$ & $16.51\pm1.45$ \\
$^\textrm{b}12+\log(\mathrm{O/H})_{\mathrm{Sanders+23}}$ & $7.61^{+0.07}_{-0.05}$ & $7.57^{+0.18}_{-0.10}$\\
$^\textrm{c}12+\log(\mathrm{O/H})_{\mathrm{Nakajima+22}}$ & $7.78^{+0.09}_{-0.06}$ & $7.74^{+0.16}_{-0.10}$\\
\enddata
\tablecomments{$^\textrm{a}$Normalized as $F_{\OIII_{5007}}=100$\\
$^\textrm{b}$With \citet{Sanders_23} calibration assuming lower branch solution.\\
$^\textrm{c}$With \citet{Nakajima_22} calibration assuming lower branch solution.}
\end{deluxetable}

\section{Uncertainties in gradient measurements}\label{sec:uncertain}

To quantify how uncertainties in emission may impact our measurements in the NGDEEP sample, we do a mock test in the same manner as in \citet{Li_22}. We fit a normal model to the distribution of effective radius $R_e$, and log-normal models to axis ratio $b/a$, and $\rm SNR_{\OIII}$ in the NGDEEP sample and randomly generate mock galaxies with 2D S\'ersic surface brightness profiles given the distribution of $R_e$, $b/a$ and $\rm SNR_{\OIII}$. 
We apply a flat gradient to all mock galaxies and use the same method discussed before to reconstruct the metallicity gradients. 
We measure the scatter of 200 mock galaxies and find the intrinsic scatter in the measurement to be $\approx 0.09$ dex kpc$^{-1}$. 

\begin{figure*}[ht!]
    \begin{minipage}{0.5\linewidth}
    \plotone{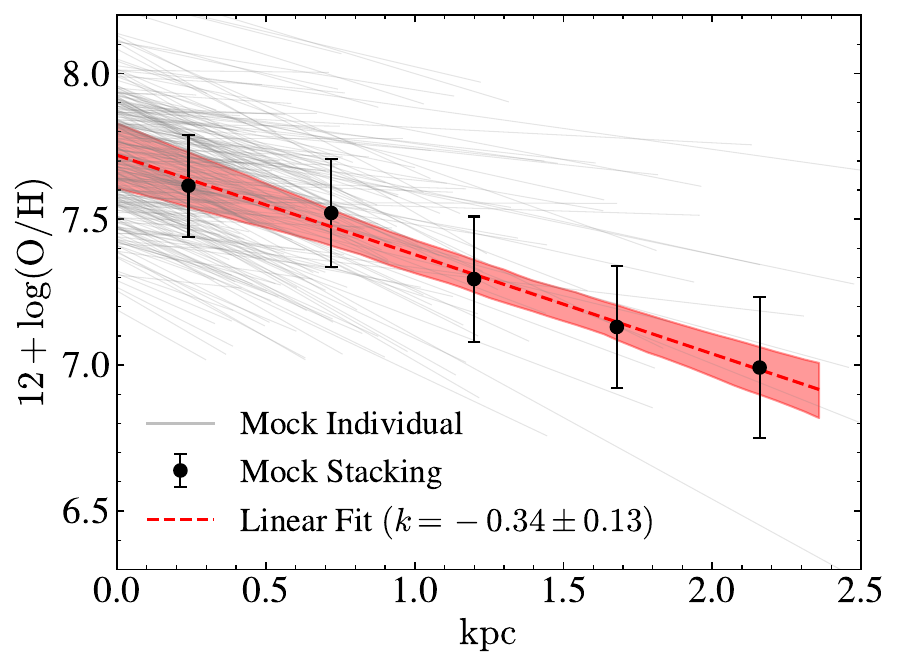}
    \end{minipage}
    \begin{minipage}{0.5\linewidth}
    \plotone{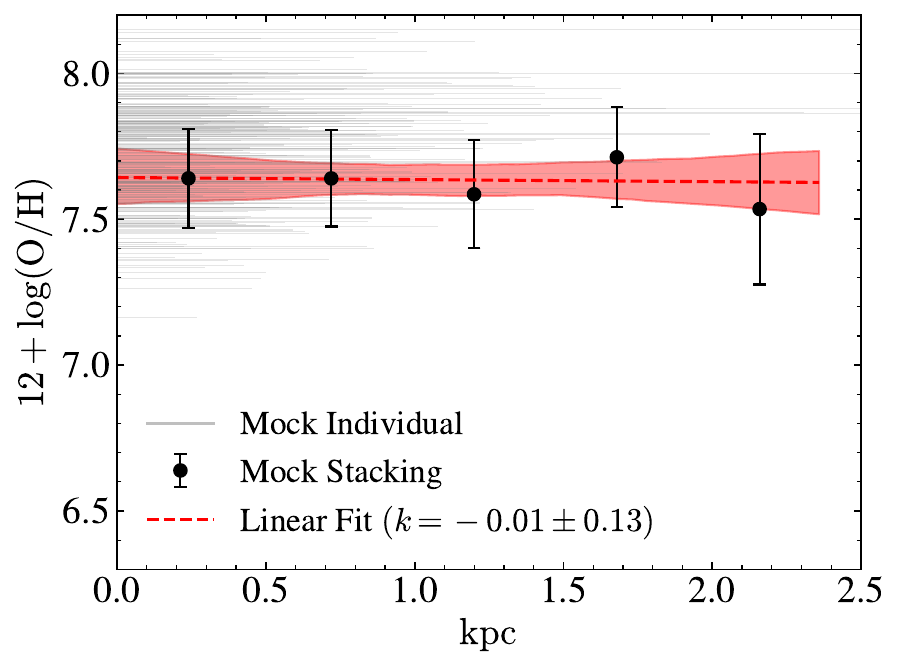}
    \end{minipage}
    
    \caption{Metallicity gradients in the mock test. In the left panel, the metallicity gradients of mock galaxies are randomly chosen from a uniform distribution $\mathcal{U}\left(-0.6,0\right)$, while all the mock galaxies in the right panel have a zero gradient. Gray lines represent individual galaxies, with length representing twice the effective radius of the galaxy. Black points are measured from the stacked emission maps,  with error bars showing the $1\sigma$ uncertainty. The red dashed line shows the linear fitting to the measured points, and the red shadowed region is its $1\sigma$ confidence interval. \label{fig:zgrad_mock}}
\end{figure*}

Since we used the stacking method for galaxies at $z>5$ in ASPIRE and FRESCO, we also used the mock data set to test whether it reflects the median metallicity gradients by stacking emission maps. 
We start by randomly generating 300 galaxies with stellar masses following a log-norm distribution, which is fitted to our SED results. We then assign the effective radius and metallicity of each mock galaxy using the mass-size relation \citep{Langeroodi_23} and the mass-metallicity relation \citep{Sarkar_24} measured at similar redshifts. We also considered the intrinsic scatter of these relations, so we added Gaussian noise to $\rm R_e$ and \oh, with $\sigma_{\log(\rm R_e/kpc)}=0.25$ and $\sigma_{\oh}=0.16$ \citep{Langeroodi_23, Sarkar_24}.  We fit a normal distribution to the observed axis ratios and then randomly assigned the axis ratios to the S\'ersic model of the emission maps using the normal distribution. Additionally, we found that $\rm SNR_{\OIII}$ is not strongly correlated with other physical parameters, so we randomly assigned $\rm SNR_{\OIII}$ for these galaxies using a log-normal distribution that best represents the observed $\rm SNR_{\OIII}$ distribution. Finally, we construct two different mock sets, the one with metallicity gradients randomly chosen from a uniform distribution $\mathcal{U}\left(-0.6,0\right)$, and the other with all zero gradients.
We measure the median metallicity gradient from the stacked maps \OIII and \Hb. 
The inferred metallicity gradients are $k=-0.34\pm0.13$ for the negative gradient dataset and $k=-0.01\pm0.13$ for the flat gradient dataset, as shown in Fig. \ref{fig:zgrad_mock}. 
We have found that the gradient measured from the stacked emission maps can effectively represent the median gradients of the mock sample. Therefore, the gradients observed for galaxies at $z\approx6-7$ are strong representations of the populations.

\section{The impact on the metallicity of the variation of the ionization parameter (U)}\label{sec:ion}

The observed line ratios depend on both metallicity and the ionization parameter (U, the ratio of the number of ionizing photons to gas number density). Galaxies at $z\sim2-9$ typically have values of $\rm \log(U)$ varying from $-3.5$ to $-1.5$ \citep{Strom_18,Reddy_23,Tang_23,Trump_23}. The varied $\rm \log(U)$ can lead to a line-ratio gradient, even if there is no gradient in metallicity.

\begin{figure*}[!ht]
  \gridline{
    \fig{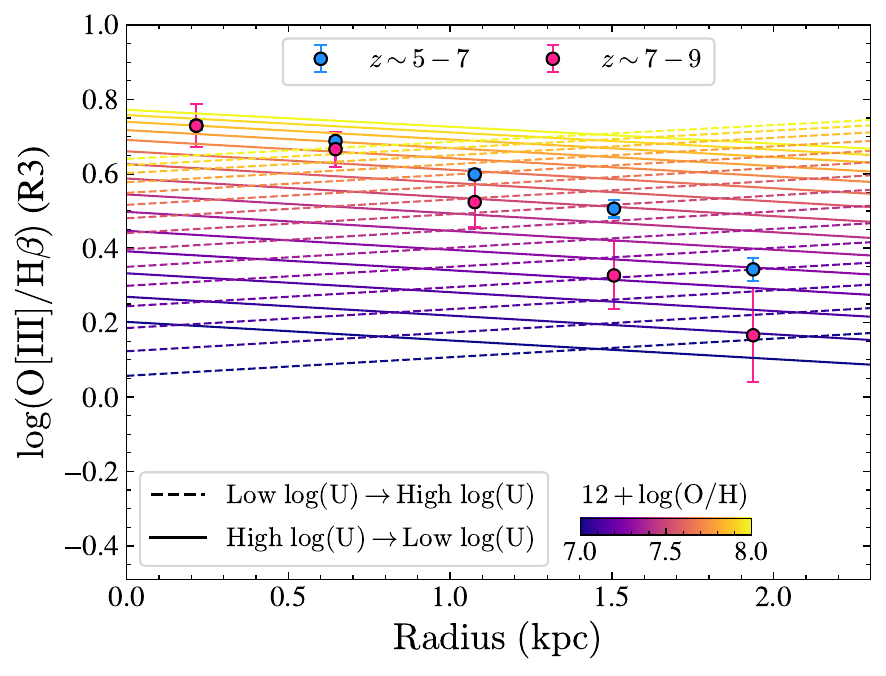}{0.49\textwidth}{\phantomsubcaption{(a)}\label{fig:logu-a}}
    \fig{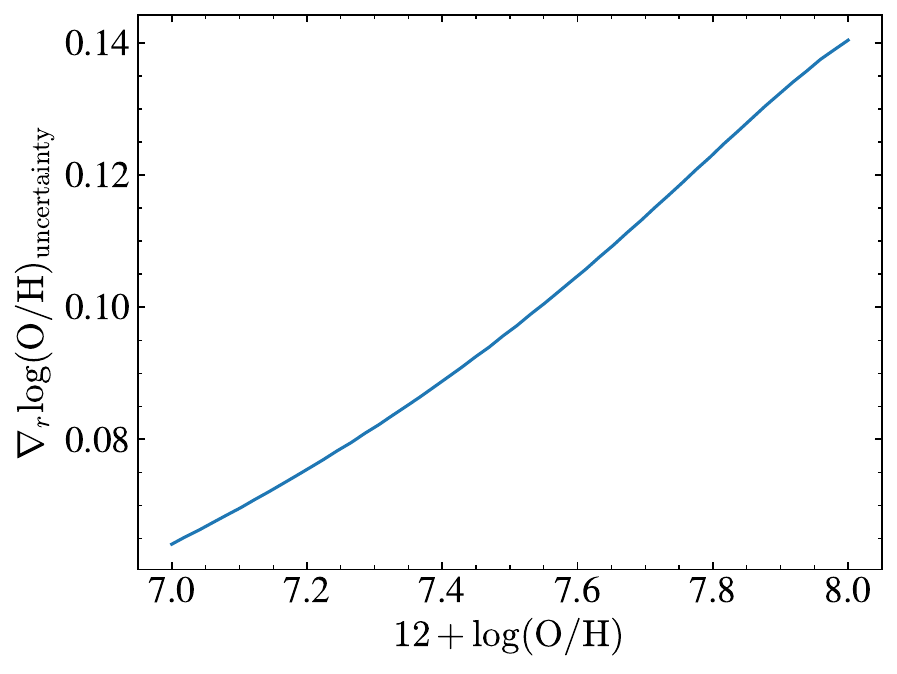}{0.49\textwidth}{\phantomsubcaption{(b)}\label{fig:logu-b}}
  }
\caption{(a) The observed R3 ratio in two redshift bins (red and blue circles), and the modeled R3 ratio from \citet{Garg_23}. 
The models are color-coded with different metallicities in the range $12+\rm\log(O/H)=7$--$8$. 
We show that the ionization parameter, $\rm U$, varies from high in the center to low in the outskirts (solid line), and from low in the center to high in the outskirts (dashed line). 
(b) The uncertainty of the metallicity gradient due to the variation of ionization parameters.
}
\label{fig:logu}
\end{figure*}

To quantify the effects of varied $\rm \log(U)$ on measured metallicity gradients, we examine the dependence of the observed line ratio on $\rm U$ using the recent photoionization model provided by \citet{Garg_23}. We applied their two models that simulate galaxies at $z\approx5$, one with low ionization (median $\rm \log(U)=-2.3$) and the other with high ionization (median $\rm \log(U)=-0.86$). 
In Fig. \ref{fig:logu-a}, we vary $\rm \log(U)$ from the center to the outskirts, transitioning either from low to high or from high to low. 
In Fig. \ref{fig:logu-b}, we observe that different $\rm \log(U)$ can produce the maximum $\approx 0.1-0.2$ dex in the R3 ratio. While our observed R3 ratios vary by $\approx 0.5$ dex, larger than can be explained by the ionization parameter. As a result, different metallicities are required to reproduce the observed R3 ratios.
% and the metallicity does not change by more than $\approx 0.2$ dex in the possible range of $\rm \log(U)$.

To further quantify the uncertainty induced by varied $\rm \log(U)$. We consider the measurement of a galaxy out to a radius $R=2$ kpc (as in our observations). Assuming the galaxy with uniform metallicity, but with an ionization parameter $\rm \log(U)$ varying with radius, we expect to observe varied R3 ratios. Then, we use the calibration from \citet{Sanders_23} to derive metallicity from such observed line ratios, we obtain a spurious non-zero metallicity gradient -- either positive or negative -- corresponding to positive or negative gradients of $\rm \log(U)$, respectively. In the right panel of Fig. \ref{fig:logu}, we show such uncertainty as a function of galaxy metallicity $Z_{\mathrm{gal}}$. 
We see that the error increases with metallicity. For typical galaxies in our sample with $Z_{\mathrm{gal}}\sim7.6$, the induced uncertainty is $\approx 0.1\mathrm{\ dex~kpc^{-1}}$, with the maximum variation of $\rm \log(U)$ from $\log(\mathrm{U})=-2.3$ to $-0.86$ \citep{Garg_23}. Compared with the metallicity gradient of $\nabla Z\approx -0.4~\mathrm{dex~kpc^{-1}}$ in our JWST observations, the maximum uncertainty induced by $\rm \log(U)$ falls within the range of the statistical error. Supporting this, \citet{Poetrodjojo_18} found no significant variation in ionization parameters with ionization parameter gradients $\lesssim0.05~\mathrm{dex~R_e^{-1}}$.
If the gradient continues to $4~\rm R_e\sim2~kpc$, the variation in $\rm \log(U)$ is only $\lesssim0.2$ dex, much less than the extreme variation we assumed $\sim1.44$ dex.
Thus, possible variation in $\rm U$ does not dominate the variation of line ratios (also see \citealt{Cresci_10}), and we expect lower uncertainty $< 0.1\mathrm{\ dex~kpc^{-1}}$ from varied $\rm \log(U)$.
% assuming the same metallicity calibration across galaxies does not alter the results. 

\section{AGN contamination}\label{sec:agn}
\begin{figure}[t!]
\centering
\includegraphics[width=0.5\textwidth]{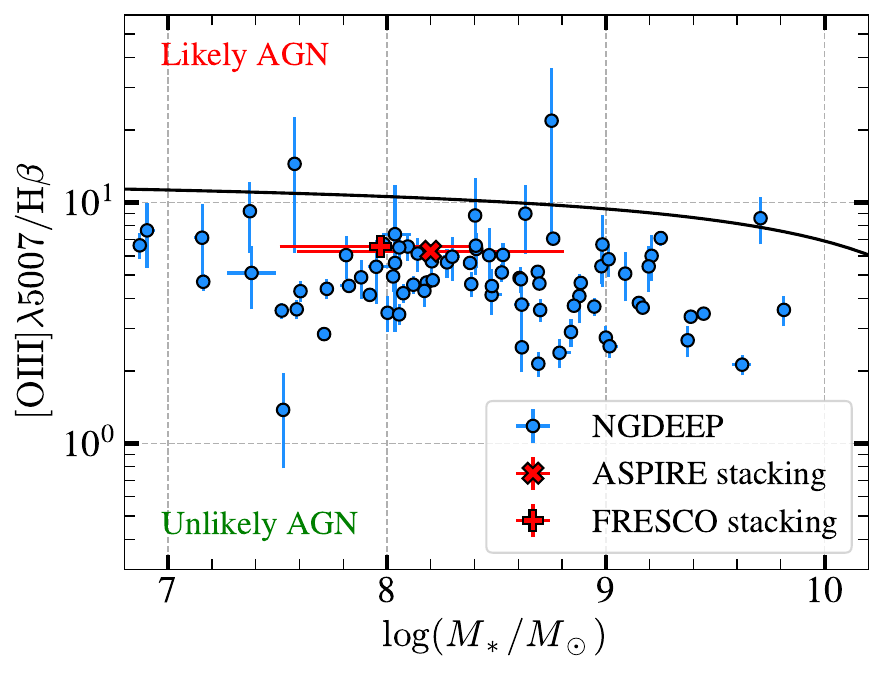}
\caption{The mass-excitation diagram for our sample galaxies.
The blue points show individual measurements in NGDEEP, and the red cross represents the median stacked result in ASPIRE and FRESCO. The solid lines are the demarcation scheme \citep{Coil_15}, where the points below the line are unlikely to be AGNs. 
Our sample galaxies in NGDEEP at $z\approx1-3$ are safely excluded from being AGNs, and the ASPIRE and FRESCO stackings at $z\approx6-7$ are also negligibly contaminated by possible AGNs.}
\label{fig:mex}
\end{figure}

Since galaxies with AGNs are powered by additional AGN ionization, the standard metallicity calibrations \citep{Bian_18, Sanders_23} are not suitable for AGN-contaminated sources.
As such, we remove sources with possible AGN contamination. 
We use the mass-excitation diagram \citep{Coil_15} to separate AGNs from star-forming galaxies. In Fig. \ref{fig:mex}, we have removed the sources $\ge 2\sigma$ from the demarcation. 
As shown, all of our sample galaxies are classified as typical star-forming galaxies.

Nevertheless, we also note that AGN may be more common in high-$z$ galaxies. \citet{Harikane_23, Maiolino_23} find red AGN with broad line features take a fraction as high as $10-20\%$ at $z>5$.
\citet{Harikane_23, Maiolino_23} suggest an offset for high-$z$ AGN in the BPT diagram (and also in the mass-excitation diagram). 
% In Fig. \ref{fig:mex}, we also show the points of type-1 AGN at $z>4$ from \citet{Harikane_23, Maiolino_23}. The reported high-$z$ type-1 AGN may be mixed with a small number of galaxies in our sample from this figure. 
\citet{Maiolino_23} provided several possibilities for high-$z$ AGN offset in BPT diagram. One is that the narrow line emission is dominated by star formation, instead of the AGN narrow line region (NLR). \citet{Maiolino_24a} further suggested a high covering factor of the broad line region (BLR), prohibiting photons from escaping to produce NLR emissions. 
\begin{figure*}[t!]
\centering
	\includegraphics[width=0.7\linewidth]{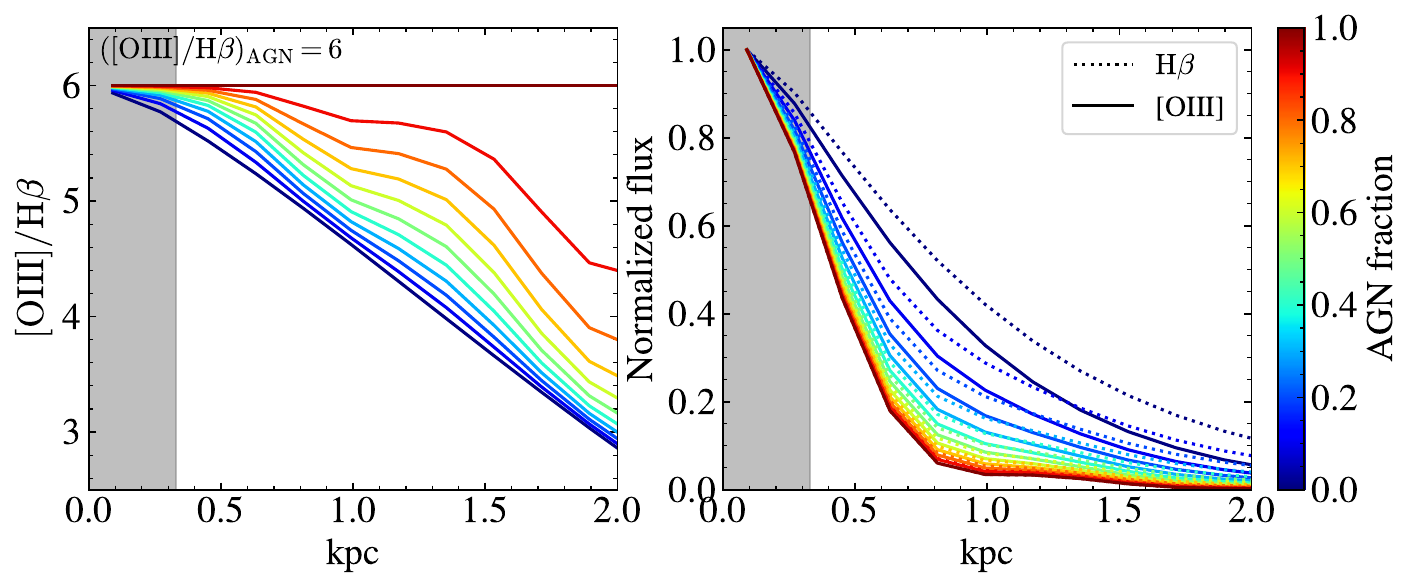}
	\includegraphics[width=0.7\linewidth]{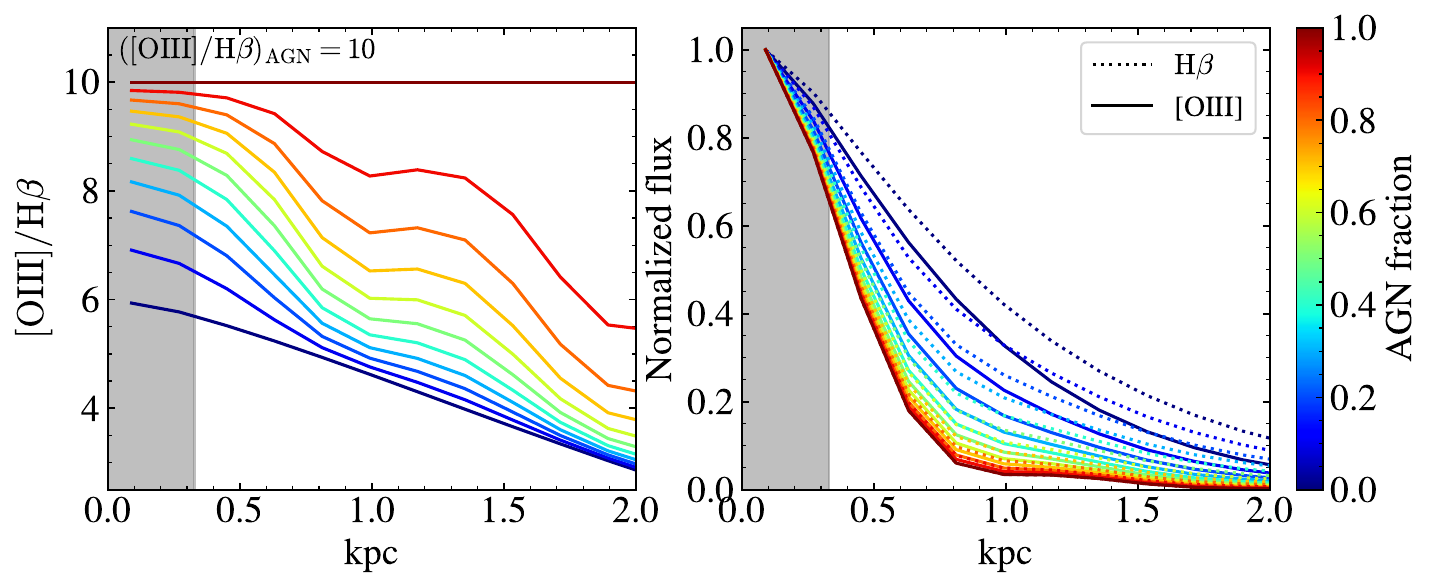}
 \caption{The \OIII/\Hb ratios (left) and their normalized profiles (right) with varying AGN fractions. The profiles are normalized by their central peak flux. In the top row, we assume a ratio of $(\OIII/\Hb)_{\rm AGN}=6$, indicative of AGN emission comparable to typical star-forming galaxies at $z>5$. In the bottom row, we consider an enhanced \OIII emission from AGN, with a ratio of $(\OIII/\Hb)_{\rm AGN}=10$.}
 \label{fig:agn_mock}
\end{figure*}
An alternative explanation is that the NLR of high-$z$ AGN is characterized by low metallicities \citep{Harikane_23, Maiolino_23}. With decreasing metallicity in NLR, the emission line ratios shift from the local AGN locus to be mixed with the star-forming locus.
AGN may be confused with low-mass, low-metallicity star-forming-dominated galaxies. \citet{Izotov_08} show that AGN non-thermal radiation should contribute less than $10\%$ to total ionizing radiation in metal-poor ($\oh<8$) AGN host galaxies, to reproduce the shift in BPT diagram. In this case, the narrow line emission from black holes should still be much weaker ($<10\%$) than emissions from stellar radiation. From these models, AGN contamination in our sample should be small, and the \OIII fluxes should come predominantly from star formation. 

Type-2 AGN, however, has been observed to be similar or more prevalent than type-1 AGN at high redshift. A similar issue has been raised by \citet{Scholtz_24}  that type-2 AGN have offsets in the BPT diagram, and it is very difficult to distinguish from star-forming galaxies using the standard BPT diagram unless high ionization lines are detected. \citet{Scholtz_24} selected type-2 AGN from a sample of galaxies with JWST/NIRSpec observations using high ionization lines in combination with UV transitions. While they found that type-2 AGN have weaker \NII/\Ha ratios, they have a similar \OIII/\Hb ratio to star-forming galaxies, and they measured the \OIII/\Hb ratio to be $4.55\pm0.15$ and $4.70\pm0.14$ from stacked spectra of type-2 AGN and star-forming galaxies, respectively. 
Although it is difficult to distinguish type-2 AGN solely based on \OIII/\Hb ratios, nevertheless, the central AGN should not significantly change the line ratios, as type-2 AGN contribute to a very similar amount of \OIII and \Hb emission as star-forming galaxies do.

To further quantify the impact of AGN on emission maps at $z>5$, we have conducted a set of mock tests. We define the AGN fraction as the AGN contribution of \OIII luminosity to the total \OIII luminosity of the galaxy. As AGN are point sources, their emissions are detected as PSF components in emission maps. We construct a mock galaxy at $z\sim6$ with a s\'ersic profile, and we add a central AGN component as a NIRCam F356W PSF. We assume two cases, the one with high \OIII/\Hb ratio as high as $(\OIII/\Hb)_{\rm AGN}=10$, as is expected at lower redshift \citep{Juneau_14}, and another with similar ratio as star-forming galaxy $(\OIII/\Hb)_{\rm AGN}=6$, as is expected by recent AGN observations at $z>4$ \citep{Scholtz_24}. In Fig. \ref{fig:agn_mock}, we show the \OIII, \Hb line, and line ratio profiles in galaxies with different AGN fractions. We find that an obvious sign of AGN in emission maps is a bump in line ratio profiles at $r=1.5$ kpc, induced by the wing of the PSF. We can also observe that when $(\OIII/\Hb)_{\rm AGN}=10$, higher than the line ratio from star formation, the AGN component contributes to a steeper \OIII/\Hb profile as a result of boosting \OIII emission in the center. However, if the AGN emission $(\OIII/\Hb)_{\rm AGN}=6$ is comparable to the emission from galaxies (e.g., as revealed in \citealt{Scholtz_24}), the change of \OIII/\Hb slope is quite small if AGN fraction is below $\sim50\%$. \citet{Maiolino_23, Maiolino_24a} suggested that AGN NLR emission is subdominant, possibly due to either the high cover fraction in BLR or the low metallicity nature of NLR. The AGN component should contribute to a small fraction of the total \OIII emissions from the host galaxy. \citet{Curti_24} also found no significant change in MZR when including AGN candidates, assuming the metallicity calibration is still valid with the contribution of AGN ionization. As a result, our measured metallicity gradient based on the \OIII/\Hb is less likely to be biased by AGN contamination. 

Similar concerns are raised by \citet{Trump_11}. The positive metallicity gradient observed at $z \sim 2$ is associated with more extended \Hb\ emission, resulting in a negative gradient of the R3 ratio. Since the metallicities fall on the upper branch of the R3 calibration, this corresponds to a positive gradient in metallicity. \citet{Trump_11} found more extended \Hb emission at $z\sim2$. They attributed this to the consequence of enhanced \OIII emission boosted by the central AGN. Based on our earlier discussion, if AGN are unlikely to dominate the flux profile, the observed trend can also be interpreted as a result of positive metallicity gradients, consistent with our results at $z \sim 2$.

\section{Halo mass-redshift distribution and analytic predictions}\label{sec:halo}
\begin{figure}[t]
    \centering
    \includegraphics[width=0.5\textwidth]{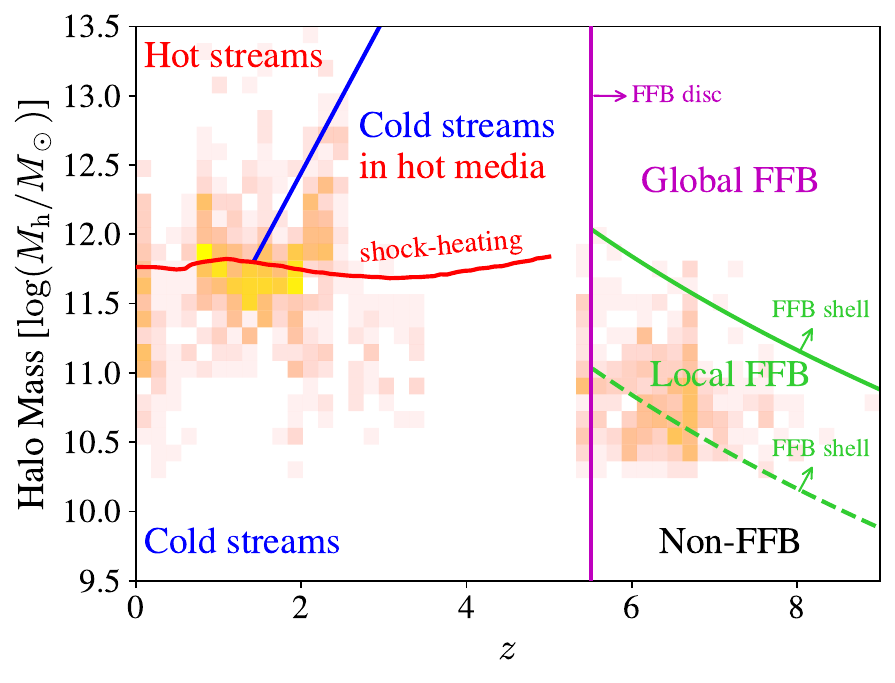}
    \caption{The 2D-histogram of halo mass and redshift distribution of all sample galaxies. The regimes predicted by analytic models of cold accretion, hot accretion \citep{Dekel_06,Dekel_09}, and FFB \citep{Dekel_23} are demarcated by lines in different colors.
    }
    \label{fig:halo_mass}
\end{figure}

We estimate the halo masses using the empirical stellar-to-halo mass relation in \citet{Shuntov_22}. It is worth noting that their highest redshift bin extends only to $z=5.5$, whereas we extended this relation for our $5<z<9$ sample. We should be cautious of this halo mass estimation, which may deviate at higher redshifts. As this is a qualitative comparison, we are less sensitive to accurate halo masses. In Fig. \ref{fig:halo_mass}, we show the halo mass vs. redshift for the whole galaxy sample, in comparison with the analytic predictions of cold/hot accretion \citep{Dekel_06,Dekel_09}, and FFB \citep{Dekel_23}. We find that at $z\gtrsim2$, most of the sample galaxies are capable of sustaining cold streams. While at lower redshift $z\lesssim2$, the inflow gas is expected to be shock-heated for massive galaxies, and only low mass galaxies can be fed with cold streams. This suggests that cold-mode accretion plays a significant role at $z \sim 2$, contributing to flat metallicity gradients, while hot-mode accretion becomes dominant at lower redshifts, leading to negative gradients.

At higher redshift $z\gtrsim6$, most of the galaxies can fall within the FFB regime. In massive halos above the green solid line, mean gas density in the ISM is expected to reach the FFB threshold to boost global star formation in FFB mode. 
The green dashed line shows the model with higher density contrast \citep[$c=10$, Eq. 37 in][]{Dekel_23} between FFB clouds and the mean overdensity.
For less massive galaxies between green solid and dashed lines, the gas clouds can reach the FFB threshold when their contraction reaches a higher density than the mean gas density, so that the FFB can happen locally. If the gas can further collapse into higher density, we would expect a lower halo mass threshold for the occurrence of local FFB. As a result, local FFB might happen in most of our sample galaxies, and supports the steep negative gradients in the FFB scenario. This qualitative comparison hints at the transition of different gas accretion and star formation scenarios across redshifts, supporting our interpretation in the main text.

\section{Degeneracies in simulations}\label{sec:deg_sim}

Simulations have shown that the radial distribution of metals is shaped by factors such as gas-phase structure, feedback mechanisms, metal yields, and diffusion in simulations \citep{Roca_24}.  
Below, we briefly discuss the impact of these factors.

{\bf (i) The gas-phase structure.} MUGS, FIRE, and FOGGIE all allow explicitly the multi-phase ISM, while the ISM gas in EAGLE and TNG50 relies on the effective equation of state (eEOS, e.g., \citet{Springel_03}), resulting in different levels of metal mixing. 
The application of eEOS in TNG50 is not able to explicitly model small-scale (i.e., $< 100$ pc) structures that pressurize the ISM, such as turbulence \citep{Hemler_21}. The unresolved small-scale turbulence may serve to radially mix chemically enriched gas and flatten the metallicity gradients \citep{Sharda_21b}. Moreover, they are not affected by the continuous burst of feedback, and they go through more steady star formation, which consequently develops a steeper gradient \citep{Hemler_21}. However, with continuous feedback, it can erase the metallicity gradient in simulations with multi-phase ISM (such as FIRE \citep{Ma_17}).

{\bf (ii) Feedback strategy.} The different implementations of feedback (e.g., stellar and AGN feedback) in different simulations yield vastly different halo baryon fractions (Fig.~9 of \citet{Crain_23}, in particular at the low-mass end, influencing the mixing and evacuation of gas to different extents. 
TNG50 simulations include the feedback of both stars and AGN \citep{Hemler_21}, but their sub-grid prescriptions and parameterized winds result in smooth star formation and feedback. FIRE does not have AGN feedback, but their explicitly modeled strong stellar and radiative feedback will also play a significant role in the evacuation of gas from the circumgalactic medium (CGM) \citep{Ma_17,Sunxd_24}. 
For the supernova feedback, previous simulations have shown that a factor of $\approx 2.5\times$ more powerful supernova feedback could lead to $\approx 0.2$ dex kpc$^{-1}$ higher metallicity gradient at $z\geq 1.5$ \citep{Gibson_13}. However, the supernova feedback prescription in FOGGIE simulations is insufficient and unable to expel enough metal-rich gas \citep{Acharyya_25}.
Different modes of feedback can also cause different extents of gas transport. \citet{Shimizu_19} found that kinetic-only feedback can more easily enrich the ISM and CGM than thermal-only feedback, leading to more metal-enriched outskirts and a shallower gradient. They also found that turn-off cooling also influences SNe feedback efficiency. Thus, how we treat feedback and how we include different fractions of feedback models lead to different extents of metal enrichment.

{\bf (iii) The metal yield and diffusion.} How the mass of stars converts to the mass of metals in the ISM and how the metal mixes with the ISM in simulations also influence the galaxy metallicity gradient. 
These two processes are described by the parameters of metal yield (the ratio between the metal mass and the stellar mass) and the yield reduction factor ($\phi_{y}$, the fraction of metals mixed with the ISM). FIRE, MUGS, and FOGGIE use three different metal yields, with FIRE using Type II Supernova (SNII), Type Ia Supernova (SNIa), and winds as sources (yields from \citet{Woosley_95}, \citet{Iwamoto_99}, and \citet{Izzard_04}, respectively), MUGS only accounts for SNII \citep{Raiteri_96} and SNIa \citep{Thielemann_86}, and FOGGIE uses a custom recipe for the metal enrichment. Different metal yield influences the absolute metals produced in galaxies, and, for example, can imprint on the normalization of mass-metallicity relations \citep[e.g.][]{Ma_16,Marszewski_24}. While the absolute amount of metal is less important on the metallicity gradient, as the gradient relies on the relative abundance of metal at different radii.
For the yield reduction factor, \citet{Sharda_21b} has shown that the galaxy metallicity gradient is inversely correlated with $\phi_{y}$.
% where $\nabla_{r} log(\mathcal{Z})$ could decline from $0.0$ dex kpc$^{-1}$ to $-0.2$ dex kpc$^{-1}$ as $\phi_{y}$ increase from 0.1 to 1.0 for galaxy with $M_{\star}\approx 10^{9} \ M_{\odot}$ at $z=2$ \citep{Sharda_21b,Roca_24}. 
In the FOGGIE simulation, they found the scatter of metallicity gradients may be related to this yield reduction factor \citep{Acharyya_25}. Note that the gas-phase metallicity gradient measures the radial change of the oxygen abundance relative to hydrogen in the ionized gas, and it is less prone to the different values of stellar yield assumed in the models. 
A factor of 10$\times$ increase in the yield reduction factor can only bring about $\lesssim 0.2$ dex change in gradient slopes according to analytical models \citep{Sharda_21b}. 
Thus, different cosmological simulation results can be compared here for qualitative comparison.

These factors all influence the distribution of gas-phase metallicity in galaxies to different extents. Although the degeneracies inherent in these different simulation architectures complicate the ability to isolate individual effects based on limited observation constraints, understanding the relative importance of these factors helps improve our interpretation of galaxy formation and evolution processes. It is hard to quantify the exact mechanisms that drive the metallicity gradients we observed; we generally prefer stronger feedback at $z\sim2$ and less efficient feedback at higher redshift $z>5$.
This nuanced understanding is essential for future studies aiming to reconcile simulation outcomes with observational data. 

\section{Measured quantities of individual galaxies.}
{\setlength{\tabcolsep}{2pt}
\renewcommand*{\arraystretch}{1.2}
{
\begin{longtable*}{ccccccccc}
\caption{Measured quantities of individual galaxies in our sample.}\\
\hline
\hline
Field & ID & ra & dec & redshift & $\log(M_*/M_\odot)$ & SFR & $12+\log(\rm O/H)$& $\nabla_r\log(\rm O/H)$\\
- & - & [deg] & [deg] &- & - & $M_\odot~\text{yr}^{-1}$ & - & [$\rm dex~kpc^{-1}$] \\
\hline
\endfirsthead
\hline
\hline
Field & ID & ra & dec & redshift & $\log(M_*/M_\odot)$ & SFR & $12+\log(\rm O/H)$& $\nabla_r\log(\rm O/H)$\\
- & - & [deg] & [deg] &- & - & $M_\odot~\text{yr}^{-1}$ & - & [$\rm dex~kpc^{-1}$] \\
\hline
\endhead
\hline
\endfoot
% ASPIRE & Stack $(N=284)$ & - & - & $6.28^{+0.38}_{-0.66}$ & $8.19^{+0.69}_{-0.55}$ & $9.03^{+13.82}_{-5.02}$ & {\red $7.61^{+0.07}_{-0.05}$} & $-0.34\pm0.12$ \\
% FRESCO & Stack $(N=42)$ & - & - & $7.24^{+0.42}_{-0.24}$ & $7.97^{+0.41}_{-0.41}$ & $7.27^{+8.67}_{-3.98}$ & {\red $7.57^{+0.18}_{-0.10}$} & $-0.46\pm0.16$ \\
ASPIRE & Zgrad-1 & 315.583439 & -14.976233 & 6.66 & $8.41_{-0.06}^{+0.10}$ & $25.67_{-3.23}^{+6.76}$ & $7.73_{-0.18}^{+0.14}$ & $-0.18\pm0.26$ \\
ASPIRE & Zgrad-2 & 137.723621 & -4.222954 & 6.64 & $8.24_{-0.03}^{+0.08}$ & $17.26_{-1.06}^{+3.53}$ & $7.75_{-0.17}^{+0.12}$ & $-0.16\pm0.26$ \\
ASPIRE & Zgrad-3 & 137.700518 & -4.272600 & 6.21 & $8.22_{-0.05}^{+0.04}$ & $16.67_{-1.70}^{+1.62}$ & $7.72_{-0.20}^{+0.15}$ & $-0.25\pm0.27$ \\
ASPIRE & Zgrad-4 & 46.318882 & -31.839629 & 6.67 & $8.55_{-0.09}^{+0.09}$ & $33.19_{-5.98}^{+6.64}$ & $7.61_{-0.21}^{+0.21}$ & $-0.15\pm0.28$ \\
ASPIRE & Zgrad-5 & 17.475148 & -30.798016 & 6.71 & $8.35_{-0.06}^{+0.09}$ & $22.16_{-2.75}^{+4.98}$ & $7.69_{-0.19}^{+0.16}$ & $-0.10\pm0.39$ \\
ASPIRE & Zgrad-6 & 140.957012 & 4.049241 & 6.39 & $8.04_{-0.02}^{+0.03}$ & $10.98_{-0.60}^{+0.69}$ & $7.31_{-0.17}^{+0.28}$ & $-0.45\pm0.31$ \\
NGDEEP & 00060 & 53.150753 & -27.804927 & 2.04 & $7.52_{-0.02}^{+0.02}$ & $1.36_{-0.06}^{+0.05}$ & $7.87_{-0.05}^{+0.06}$ & $0.14\pm0.05$ \\
NGDEEP & 00096 & 53.151582 & -27.803721 & 2.12 & $8.41_{-0.02}^{+0.03}$ & $1.83_{-0.13}^{+0.10}$ & $8.29_{-0.05}^{+0.05}$ & $-0.07\pm0.02$ \\
NGDEEP & 00173 & 53.148643 & -27.801635 & 2.12 & $8.10_{-0.05}^{+0.03}$ & $1.52_{-0.07}^{+0.09}$ & $7.98_{-0.04}^{+0.04}$ & $-0.09\pm0.04$ \\
NGDEEP & 00197 & 53.153745 & -27.801119 & 2.02 & $7.16_{-0.03}^{+0.03}$ & $1.34_{-0.07}^{+0.05}$ & $7.87_{-0.04}^{+0.05}$ & $0.05\pm0.04$ \\
NGDEEP & 00228 & 53.150438 & -27.800622 & 2.62 & $9.00_{-0.02}^{+0.02}$ & $3.26_{-0.20}^{+0.22}$ & $8.50_{-0.08}^{+0.06}$ & $0.08\pm0.07$ \\
NGDEEP & 00242 & 53.160060 & -27.800122 & 2.22 & $8.00_{-0.01}^{+0.01}$ & $1.11_{-0.04}^{+0.04}$ & $7.93_{-0.07}^{+0.07}$ & $-0.11\pm0.13$ \\
NGDEEP & 00310 & 53.156367 & -27.799048 & 2.67 & $9.37_{-0.01}^{+0.01}$ & $1.29_{-0.06}^{+0.07}$ & $8.52_{-0.05}^{+0.04}$ & $0.07\pm0.05$ \\
NGDEEP & 00317 & 53.145614 & -27.798979 & 2.62 & $9.62_{-0.04}^{+0.04}$ & $11.65_{-0.82}^{+0.99}$ & $8.61_{-0.04}^{+0.04}$ & $-0.03\pm0.05$ \\
NGDEEP & 00407 & 53.171345 & -27.797899 & 1.99 & $8.84_{-0.02}^{+0.02}$ & $1.23_{-0.13}^{+0.13}$ & $8.47_{-0.05}^{+0.04}$ & $-0.15\pm0.03$ \\
NGDEEP & 00413 & 53.152521 & -27.797787 & 3.00 & $7.58_{-0.03}^{+0.04}$ & $2.08_{-0.11}^{+0.08}$ & $8.05_{-0.07}^{+0.07}$ & $-0.01\pm0.13$ \\
NGDEEP & 00436 & 53.170460 & -27.797382 & 2.08 & $9.02_{-0.03}^{+0.04}$ & $1.72_{-0.12}^{+0.12}$ & $8.42_{-0.07}^{+0.05}$ & $-0.03\pm0.04$ \\
NGDEEP & 00466 & 53.168830 & -27.796994 & 1.99 & $8.61_{-0.01}^{+0.01}$ & $4.08_{-0.24}^{+0.19}$ & $8.05_{-0.05}^{+0.05}$ & $-0.04\pm0.02$ \\
NGDEEP & 00508 & 53.173491 & -27.796419 & 2.75 & $7.38_{-0.11}^{+0.11}$ & $0.99_{-0.08}^{+0.09}$ & $8.01_{-0.10}^{+0.12}$ & $-0.19\pm0.10$ \\
NGDEEP & 00536 & 53.148336 & -27.796032 & 2.02 & $8.61_{-0.03}^{+0.03}$ & $1.56_{-0.13}^{+0.11}$ & $8.34_{-0.05}^{+0.05}$ & $-0.00\pm0.03$ \\
NGDEEP & 00618 & 53.146372 & -27.795289 & 2.63 & $8.69_{-0.01}^{+0.01}$ & $2.38_{-0.11}^{+0.13}$ & $8.59_{-0.05}^{+0.04}$ & $-0.13\pm0.07$ \\
NGDEEP & 00619 & 53.143575 & -27.795297 & 2.02 & $8.18_{-0.01}^{+0.02}$ & $7.02_{-0.32}^{+0.33}$ & $8.09_{-0.06}^{+0.07}$ & $0.01\pm0.01$ \\
NGDEEP & 00683 & 53.148356 & -27.794324 & 2.63 & $7.73_{-0.02}^{+0.02}$ & $3.21_{-0.11}^{+0.13}$ & $8.14_{-0.11}^{+0.11}$ & $-0.10\pm0.02$ \\
NGDEEP & 00729 & 53.172439 & -27.793993 & 2.80 & $8.95_{-0.01}^{+0.02}$ & $5.88_{-0.11}^{+0.12}$ & $8.39_{-0.07}^{+0.06}$ & $-0.03\pm0.03$ \\
NGDEEP & 00745 & 53.152584 & -27.793907 & 3.07 & $9.71_{-0.01}^{+0.01}$ & $15.44_{-0.39}^{+0.59}$ & $8.34_{-0.03}^{+0.03}$ & $0.00\pm0.02$ \\
NGDEEP & 00760 & 53.165077 & -27.793719 & 2.34 & $8.89_{-0.01}^{+0.01}$ & $6.19_{-0.15}^{+0.16}$ & $8.35_{-0.04}^{+0.04}$ & $-0.04\pm0.01$ \\
NGDEEP & 00782 & 53.147407 & -27.793410 & 2.79 & $7.81_{-0.02}^{+0.02}$ & $3.08_{-0.29}^{+0.15}$ & $8.12_{-0.08}^{+0.09}$ & $-0.08\pm0.11$ \\
NGDEEP & 00784 & 53.170795 & -27.793366 & 2.67 & $7.71_{-0.02}^{+0.02}$ & $2.79_{-0.19}^{+0.25}$ & $7.93_{-0.07}^{+0.09}$ & $0.59\pm0.05$ \\
NGDEEP & 00789 & 53.172714 & -27.793316 & 2.81 & $8.88_{-0.02}^{+0.02}$ & $2.08_{-0.31}^{+0.32}$ & $8.36_{-0.07}^{+0.07}$ & $-0.32\pm0.08$ \\
NGDEEP & 00860 & 53.153127 & -27.792447 & 3.32 & $8.63_{-0.02}^{+0.02}$ & $4.95_{-0.23}^{+0.18}$ & $8.17_{-0.05}^{+0.06}$ & $0.05\pm0.09$ \\
NGDEEP & 00868 & 53.151284 & -27.792470 & 1.84 & $8.85_{-0.03}^{+0.03}$ & $5.95_{-0.20}^{+0.21}$ & $8.28_{-0.04}^{+0.04}$ & $0.04\pm0.01$ \\
NGDEEP & 00917 & 53.147644 & -27.791848 & 1.87 & $7.95_{-0.04}^{+0.07}$ & $0.45_{-0.02}^{+0.02}$ & $8.41_{-0.06}^{+0.05}$ & $-0.26\pm0.11$ \\
NGDEEP & 00928 & 53.146782 & -27.791739 & 2.79 & $8.40_{-0.02}^{+0.03}$ & $1.16_{-0.08}^{+0.09}$ & $7.94_{-0.10}^{+0.10}$ & $-0.13\pm0.12$ \\
NGDEEP & 00996 & 53.178202 & -27.790838 & 1.76 & $8.62_{-0.02}^{+0.02}$ & $0.41_{-0.02}^{+0.02}$ & $8.19_{-0.11}^{+0.08}$ & $0.08\pm0.18$ \\
NGDEEP & 01014 & 53.153116 & -27.790565 & 3.32 & $8.14_{-0.02}^{+0.02}$ & $5.61_{-0.45}^{+0.25}$ & $7.96_{-0.08}^{+0.07}$ & $0.11\pm0.04$ \\
NGDEEP & 01066 & 53.167405 & -27.790198 & 2.76 & $8.27_{-0.02}^{+0.02}$ & $3.63_{-0.15}^{+0.23}$ & $8.16_{-0.08}^{+0.08}$ & $-0.24\pm0.05$ \\
NGDEEP & 01145 & 53.171118 & -27.789204 & 2.84 & $6.90_{-0.02}^{+0.04}$ & $0.81_{-0.04}^{+0.07}$ & $7.93_{-0.08}^{+0.08}$ & $0.08\pm0.19$ \\
NGDEEP & 01173 & 53.141930 & -27.788913 & 2.71 & $9.09_{-0.03}^{+0.03}$ & $1.48_{-0.08}^{+0.09}$ & $8.28_{-0.07}^{+0.08}$ & $0.03\pm0.13$ \\
NGDEEP & 01222 & 53.169688 & -27.788158 & 3.18 & $9.21_{-0.01}^{+0.01}$ & $7.56_{-0.25}^{+0.25}$ & $8.29_{-0.04}^{+0.05}$ & $-0.04\pm0.06$ \\
NGDEEP & 01223 & 53.146252 & -27.788128 & 2.00 & $8.30_{-0.04}^{+0.04}$ & $1.00_{-0.04}^{+0.04}$ & $8.20_{-0.04}^{+0.04}$ & $-0.07\pm0.04$ \\
NGDEEP & 01227 & 53.143415 & -27.788081 & 3.40 & $8.75_{-0.01}^{+0.01}$ & $5.86_{-0.17}^{+0.19}$ & $8.05_{-0.05}^{+0.05}$ & $-0.02\pm0.11$ \\
NGDEEP & 01233 & 53.152356 & -27.788058 & 2.31 & $8.39_{-0.02}^{+0.01}$ & $2.52_{-0.10}^{+0.08}$ & $8.23_{-0.06}^{+0.06}$ & $0.06\pm0.04$ \\
NGDEEP & 01237 & 53.163104 & -27.787936 & 2.22 & $7.37_{-0.02}^{+0.03}$ & $0.57_{-0.04}^{+0.06}$ & $8.08_{-0.05}^{+0.05}$ & $0.06\pm0.06$ \\
NGDEEP & 01255 & 53.174319 & -27.787763 & 2.07 & $8.20_{-0.02}^{+0.02}$ & $1.43_{-0.02}^{+0.03}$ & $8.09_{-0.05}^{+0.06}$ & $-0.02\pm0.03$ \\
NGDEEP & 01291 & 53.177252 & -27.787383 & 3.00 & $9.81_{-0.01}^{+0.01}$ & $4.68_{-0.24}^{+0.25}$ & $8.41_{-0.06}^{+0.06}$ & $-0.10\pm0.04$ \\
NGDEEP & 01346 & 53.180819 & -27.787398 & 1.94 & $8.93_{-0.02}^{+0.04}$ & $2.54_{-0.17}^{+0.18}$ & $8.39_{-0.04}^{+0.03}$ & $0.05\pm0.02$ \\
NGDEEP & 01385 & 53.180809 & -27.786328 & 2.69 & $8.76_{-0.01}^{+0.01}$ & $23.20_{-0.64}^{+0.43}$ & $8.13_{-0.04}^{+0.04}$ & $-0.09\pm0.01$ \\
NGDEEP & 01482 & 53.169359 & -27.785060 & 1.90 & $6.87_{-0.01}^{+0.01}$ & $0.75_{-0.02}^{+0.03}$ & $7.71_{-0.10}^{+0.08}$ & $0.07\pm0.05$ \\
NGDEEP & 01489 & 53.149093 & -27.785142 & 2.06 & $9.07_{-0.02}^{+0.02}$ & $2.79_{-0.20}^{+0.20}$ & $8.39_{-0.05}^{+0.05}$ & $-0.04\pm0.02$ \\
NGDEEP & 01494 & 53.178699 & -27.785025 & 2.11 & $8.70_{-0.02}^{+0.03}$ & $2.95_{-0.18}^{+0.20}$ & $8.38_{-0.05}^{+0.05}$ & $-0.00\pm0.02$ \\
NGDEEP & 01544 & 53.142345 & -27.784448 & 2.94 & $8.04_{-0.02}^{+0.02}$ & $8.36_{-0.26}^{+0.25}$ & $8.09_{-0.12}^{+0.10}$ & $0.01\pm0.03$ \\
NGDEEP & 01583 & 53.144136 & -27.783940 & 2.81 & $8.48_{-0.03}^{+0.05}$ & $2.28_{-0.24}^{+0.29}$ & $8.26_{-0.10}^{+0.09}$ & $-0.00\pm0.08$ \\
NGDEEP & 01633 & 53.157671 & -27.783403 & 2.63 & $8.17_{-0.02}^{+0.02}$ & $2.59_{-0.12}^{+0.17}$ & $8.22_{-0.12}^{+0.10}$ & $0.00\pm0.06$ \\
NGDEEP & 01634 & 53.182599 & -27.783504 & 2.07 & $8.47_{-0.02}^{+0.02}$ & $1.63_{-0.10}^{+0.10}$ & $8.28_{-0.05}^{+0.05}$ & $0.01\pm0.03$ \\
NGDEEP & 01635 & 53.182278 & -27.783395 & 2.07 & $8.54_{-0.02}^{+0.02}$ & $3.45_{-0.13}^{+0.13}$ & $8.33_{-0.05}^{+0.04}$ & $-0.00\pm0.02$ \\
NGDEEP & 01641 & 53.167902 & -27.783264 & 2.08 & $7.88_{-0.03}^{+0.03}$ & $0.70_{-0.03}^{+0.03}$ & $8.20_{-0.06}^{+0.06}$ & $-0.04\pm0.08$ \\
NGDEEP & 01751 & 53.167504 & -27.781909 & 2.07 & $8.12_{-0.01}^{+0.01}$ & $2.36_{-0.23}^{+0.26}$ & $7.87_{-0.04}^{+0.05}$ & $0.01\pm0.03$ \\
NGDEEP & 01774 & 53.165166 & -27.781715 & 2.22 & $8.69_{-0.01}^{+0.01}$ & $5.49_{-0.33}^{+0.34}$ & $8.15_{-0.05}^{+0.05}$ & $0.08\pm0.01$ \\
NGDEEP & 01786 & 53.151504 & -27.781429 & 2.30 & $8.41_{-0.02}^{+0.02}$ & $3.05_{-0.11}^{+0.15}$ & $8.12_{-0.05}^{+0.05}$ & $-0.07\pm0.04$ \\
NGDEEP & 01865 & 53.154815 & -27.780452 & 2.57 & $7.53_{-0.02}^{+0.02}$ & $2.19_{-0.23}^{+0.26}$ & $8.10_{-0.11}^{+0.12}$ & $-0.12\pm0.09$ \\
NGDEEP & 01874 & 53.157122 & -27.780335 & 3.01 & $7.94_{-0.01}^{+0.02}$ & $3.63_{-0.14}^{+0.11}$ & $8.05_{-0.07}^{+0.09}$ & $-0.05\pm0.08$ \\
NGDEEP & 01893 & 53.152903 & -27.780194 & 1.85 & $9.06_{-0.02}^{+0.02}$ & $3.20_{-0.13}^{+0.12}$ & $8.39_{-0.04}^{+0.04}$ & $0.02\pm0.02$ \\
NGDEEP & 01918 & 53.150506 & -27.779833 & 3.07 & $8.14_{-0.03}^{+0.03}$ & $1.93_{-0.16}^{+0.23}$ & $8.08_{-0.10}^{+0.10}$ & $-0.01\pm0.09$ \\
NGDEEP & 01966 & 53.157887 & -27.779274 & 1.84 & $8.62_{-0.03}^{+0.04}$ & $0.61_{-0.06}^{+0.05}$ & $8.32_{-0.06}^{+0.05}$ & $-0.10\pm0.08$ \\
NGDEEP & 01971 & 53.163741 & -27.779153 & 3.27 & $7.60_{-0.03}^{+0.02}$ & $4.04_{-0.26}^{+0.22}$ & $8.02_{-0.08}^{+0.08}$ & $-0.08\pm0.03$ \\
NGDEEP & 01986 & 53.176312 & -27.778978 & 3.19 & $9.15_{-0.01}^{+0.01}$ & $20.62_{-0.68}^{+0.71}$ & $8.39_{-0.06}^{+0.05}$ & $-0.10\pm0.02$ \\
NGDEEP & 02030 & 53.148379 & -27.778614 & 3.32 & $7.59_{-0.01}^{+0.01}$ & $3.89_{-0.11}^{+0.08}$ & $7.84_{-0.06}^{+0.07}$ & $0.22\pm0.05$ \\
NGDEEP & 02033 & 53.146435 & -27.778375 & 1.86 & $7.92_{-0.01}^{+0.01}$ & $5.18_{-0.24}^{+0.32}$ & $7.92_{-0.03}^{+0.04}$ & $0.06\pm0.02$ \\
NGDEEP & 02059 & 53.146068 & -27.778118 & 1.86 & $7.16_{-0.04}^{+0.03}$ & $0.40_{-0.05}^{+0.07}$ & $8.01_{-0.12}^{+0.09}$ & $0.02\pm0.13$ \\
NGDEEP & 02167 & 53.159582 & -27.776803 & 3.43 & $7.99_{-0.01}^{+0.01}$ & $9.84_{-0.23}^{+0.25}$ & $7.96_{-0.04}^{+0.05}$ & $-0.05\pm0.04$ \\
NGDEEP & 02180 & 53.144354 & -27.776543 & 2.32 & $8.06_{-0.02}^{+0.02}$ & $1.56_{-0.09}^{+0.07}$ & $8.14_{-0.08}^{+0.08}$ & $-0.15\pm0.08$ \\
NGDEEP & 02297 & 53.139253 & -27.774915 & 1.84 & $7.83_{-0.04}^{+0.03}$ & $1.20_{-0.04}^{+0.03}$ & $8.06_{-0.04}^{+0.05}$ & $-0.07\pm0.03$ \\
NGDEEP & 02304 & 53.167609 & -27.774783 & 3.06 & $8.04_{-0.03}^{+0.02}$ & $1.93_{-0.17}^{+0.19}$ & $8.08_{-0.09}^{+0.11}$ & $-0.02\pm0.06$ \\
NGDEEP & 02396 & 53.173541 & -27.773773 & 1.88 & $8.48_{-0.03}^{+0.02}$ & $0.80_{-0.06}^{+0.08}$ & $8.39_{-0.05}^{+0.04}$ & $0.05\pm0.07$ \\
NGDEEP & 02441 & 53.174137 & -27.773047 & 3.44 & $8.98_{-0.03}^{+0.03}$ & $8.89_{-0.32}^{+0.39}$ & $8.27_{-0.05}^{+0.06}$ & $-0.22\pm0.04$ \\
NGDEEP & 02446 & 53.148715 & -27.773010 & 3.31 & $9.20_{-0.01}^{+0.01}$ & $10.61_{-0.27}^{+0.30}$ & $8.28_{-0.06}^{+0.06}$ & $-0.05\pm0.06$ \\
NGDEEP & 02458 & 53.164201 & -27.772965 & 2.84 & $9.14_{-0.02}^{+0.03}$ & $12.63_{-0.38}^{+0.43}$ & $8.26_{-0.04}^{+0.04}$ & $0.04\pm0.03$ \\
NGDEEP & 02477 & 53.169766 & -27.772581 & 3.32 & $8.74_{-0.03}^{+0.02}$ & $2.59_{-0.12}^{+0.15}$ & $8.06_{-0.09}^{+0.10}$ & $0.06\pm0.07$ \\
NGDEEP & 02482 & 53.152918 & -27.772575 & 1.84 & $8.70_{-0.01}^{+0.01}$ & $9.12_{-0.46}^{+0.21}$ & $8.15_{-0.05}^{+0.06}$ & $0.09\pm0.01$ \\
NGDEEP & 02511 & 53.138215 & -27.772110 & 2.81 & $8.38_{-0.01}^{+0.01}$ & $14.78_{-0.19}^{+0.22}$ & $8.05_{-0.05}^{+0.08}$ & $0.05\pm0.02$ \\
NGDEEP & 02558 & 53.154498 & -27.771493 & 2.22 & $9.39_{-0.01}^{+0.02}$ & $32.22_{-1.05}^{+0.49}$ & $8.39_{-0.10}^{+0.06}$ & $0.08\pm0.01$ \\
NGDEEP & 02564 & 53.140975 & -27.771279 & 2.22 & $7.58_{-0.03}^{+0.03}$ & $0.78_{-0.04}^{+0.05}$ & $7.89_{-0.06}^{+0.06}$ & $-0.08\pm0.11$ \\
NGDEEP & 02578 & 53.144689 & -27.771181 & 3.07 & $9.25_{-0.02}^{+0.02}$ & $22.67_{-0.43}^{+0.41}$ & $8.27_{-0.03}^{+0.03}$ & $-0.03\pm0.01$ \\
NGDEEP & 02581 & 53.160020 & -27.771086 & 1.84 & $8.07_{-0.02}^{+0.02}$ & $1.64_{-0.06}^{+0.06}$ & $8.30_{-0.06}^{+0.06}$ & $0.05\pm0.03$ \\
NGDEEP & 02610 & 53.161804 & -27.770720 & 3.32 & $8.52_{-0.03}^{+0.04}$ & $14.16_{-0.45}^{+0.42}$ & $8.22_{-0.08}^{+0.08}$ & $-0.02\pm0.02$ \\
NGDEEP & 02649 & 53.152318 & -27.770169 & 1.84 & $9.45_{-0.02}^{+0.02}$ & $10.07_{-0.36}^{+0.35}$ & $8.35_{-0.04}^{+0.04}$ & $0.02\pm0.01$ \\
NGDEEP & 02685 & 53.173339 & -27.769147 & 2.85 & $8.80_{-0.02}^{+0.02}$ & $1.95_{-0.12}^{+0.11}$ & $8.14_{-0.08}^{+0.08}$ & $0.01\pm0.10$ \\
NGDEEP & 02724 & 53.169952 & -27.768432 & 3.09 & $9.01_{-0.01}^{+0.01}$ & $5.13_{-0.11}^{+0.12}$ & $8.14_{-0.07}^{+0.08}$ & $0.00\pm0.05$ \\
NGDEEP & 02743 & 53.165934 & -27.767980 & 3.43 & $8.06_{-0.02}^{+0.02}$ & $9.06_{-0.47}^{+0.81}$ & $8.08_{-0.07}^{+0.07}$ & $-0.06\pm0.04$ \\
NGDEEP & 02745 & 53.156388 & -27.767910 & 2.02 & $8.03_{-0.01}^{+0.01}$ & $1.61_{-0.07}^{+0.06}$ & $7.91_{-0.08}^{+0.08}$ & $0.04\pm0.06$ \\
NGDEEP & 02770 & 53.153851 & -27.767378 & 2.31 & $9.17_{-0.01}^{+0.00}$ & $9.82_{-0.13}^{+0.23}$ & $8.42_{-0.04}^{+0.04}$ & $0.11\pm0.01$ \\
NGDEEP & 02798 & 53.171741 & -27.766728 & 3.19 & $8.75_{-0.02}^{+0.02}$ & $2.78_{-0.21}^{+0.27}$ & $8.12_{-0.12}^{+0.11}$ & $0.01\pm0.13$ \\
NGDEEP & 02823 & 53.160498 & -27.766138 & 2.67 & $8.07_{-0.02}^{+0.02}$ & $2.73_{-0.21}^{+0.24}$ & $8.14_{-0.10}^{+0.09}$ & $-0.10\pm0.09$ \\
NGDEEP & 02897 & 53.172172 & -27.763745 & 2.21 & $8.21_{-0.02}^{+0.03}$ & $4.65_{-0.32}^{+0.26}$ & $7.96_{-0.03}^{+0.04}$ & $-0.03\pm0.02$ \\
\label{tab:info}
\end{longtable*}
}
}
% \clearpage

\section{Spectra and metallicity gradients of individual galaxies}

\begin{figure*}[h!]
	\begin{minipage}{1\linewidth}
		\centering
            \includegraphics[width=1\linewidth]{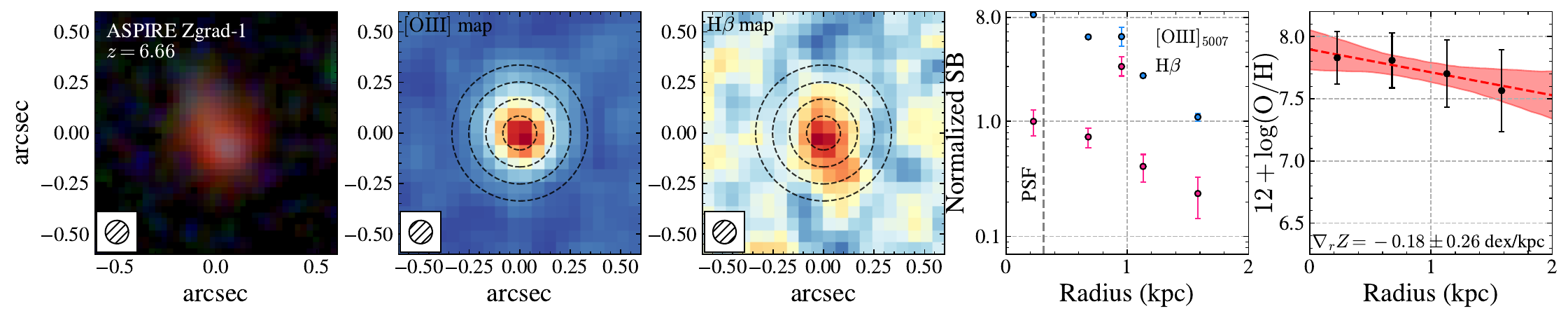}
	\end{minipage}
 	\begin{minipage}{1\linewidth}
		\centering
            \includegraphics[width=1\linewidth]{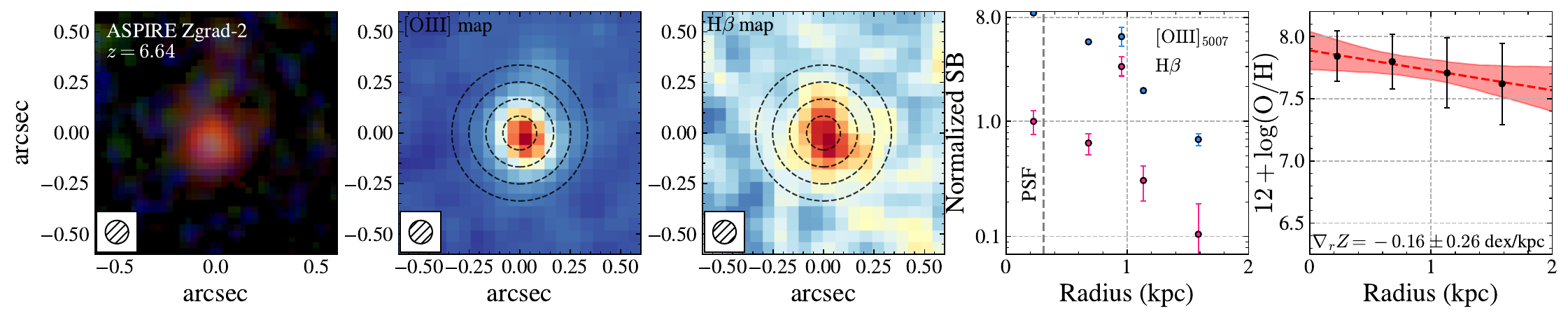}
	\end{minipage}
 	\begin{minipage}{1\linewidth}
		\centering
            \includegraphics[width=1\linewidth]{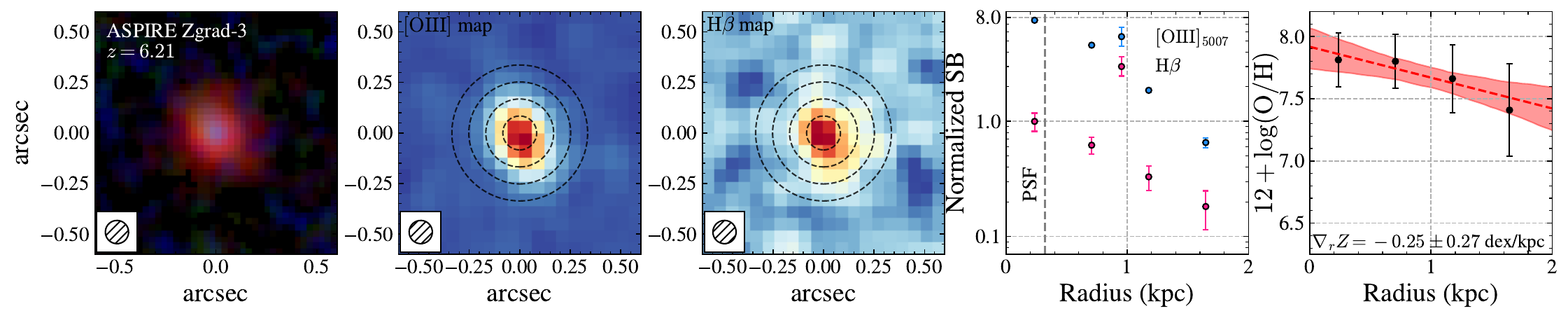}
	\end{minipage}
  	\begin{minipage}{1\linewidth}
		\centering
            \includegraphics[width=1\linewidth]{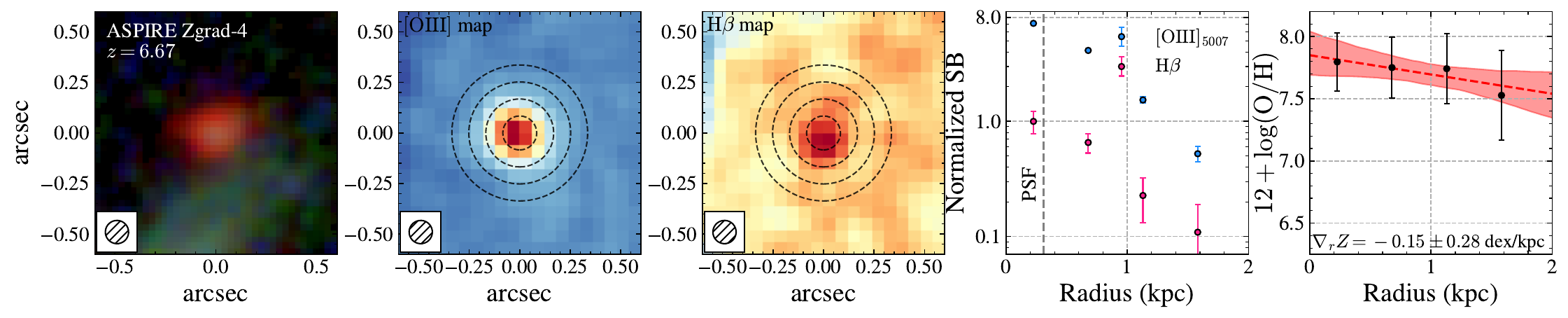}
	\end{minipage}
	\begin{minipage}{1\linewidth}
		\centering
            \includegraphics[width=1\linewidth]{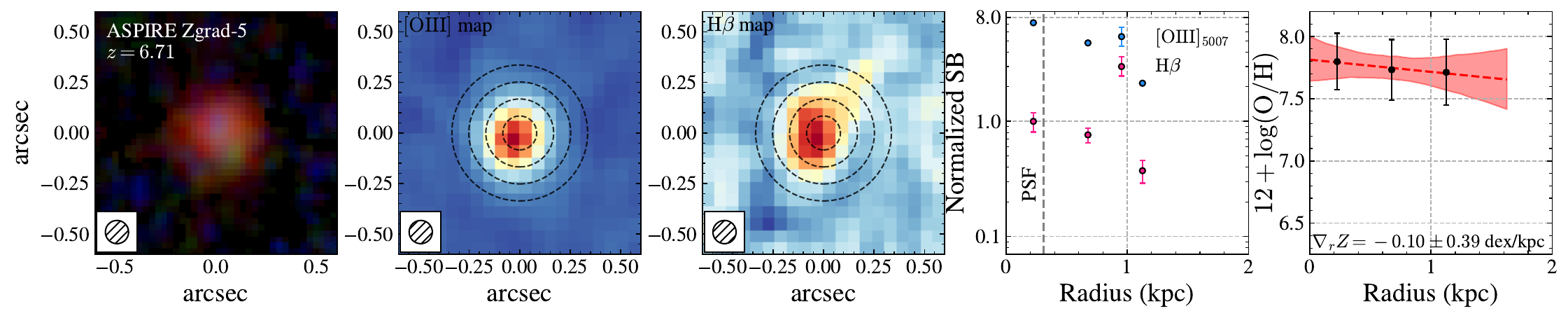}
	\end{minipage}
  	\begin{minipage}{1\linewidth}
		\centering
            \includegraphics[width=1\linewidth]{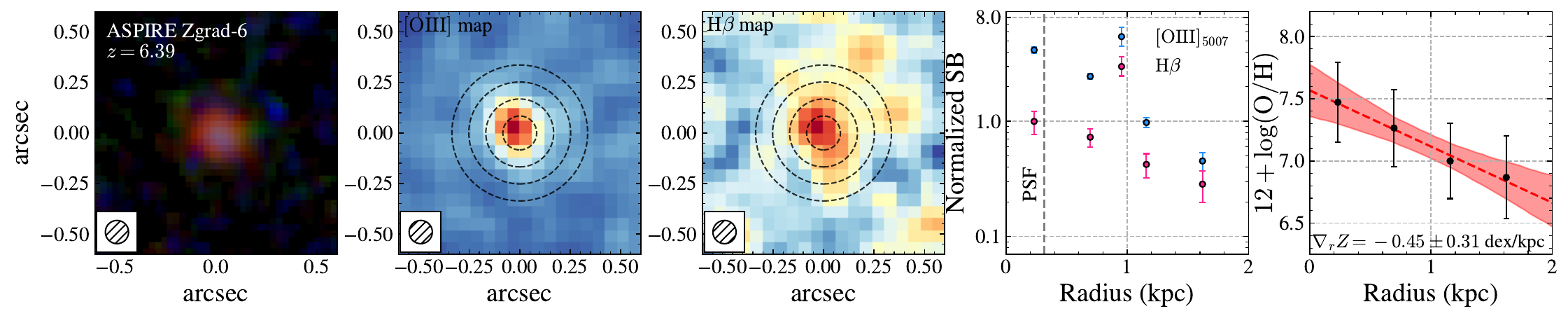}
	\end{minipage}
\caption{\textbf{metallicity gradient measurements of individual galaxies in ASPIRE.} The right four panels are the same as Fig. \ref{fig:zgrad-stacking}, but for individual galaxies with the highest SNR. The first panel shows the false-color JWST NIRCam image (with F115W, F200W, and F356W) centered on each source.\label{fig:aspire_ind}}
\end{figure*}

\begin{figure*}[ht!]
    \begin{minipage}{0.5\linewidth}
    \includegraphics[width=\textwidth]{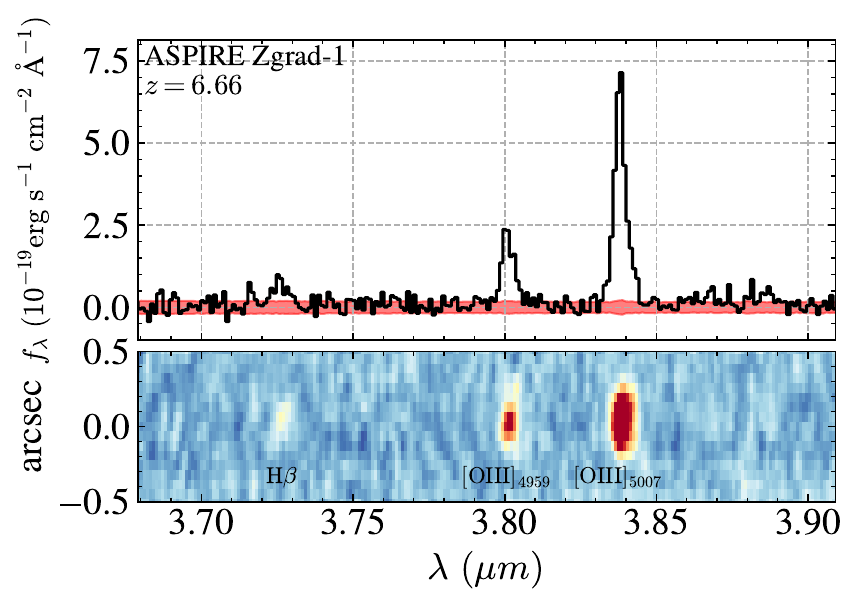}
    \end{minipage}
    \begin{minipage}{0.5\linewidth}
    \includegraphics[width=\textwidth]{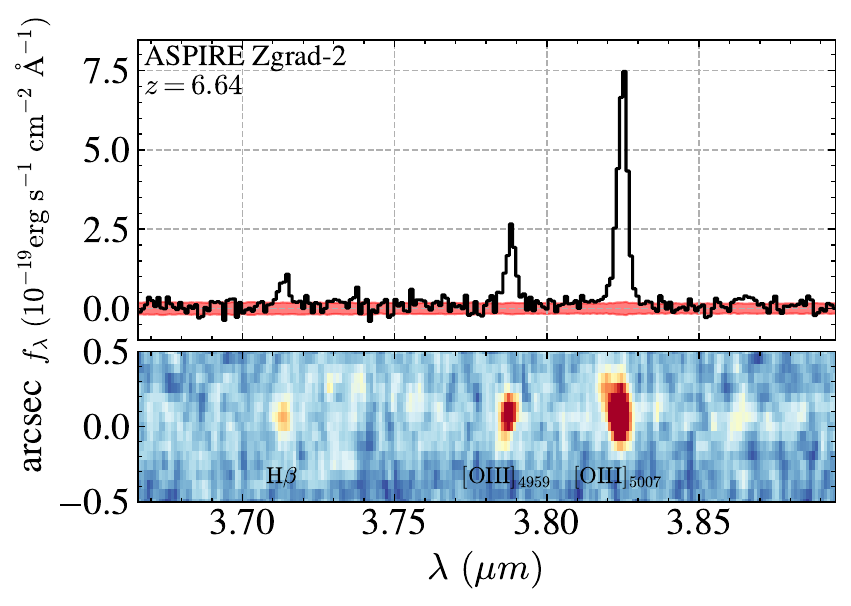}
    \end{minipage}
    \begin{minipage}{0.5\linewidth}
    \includegraphics[width=\textwidth]{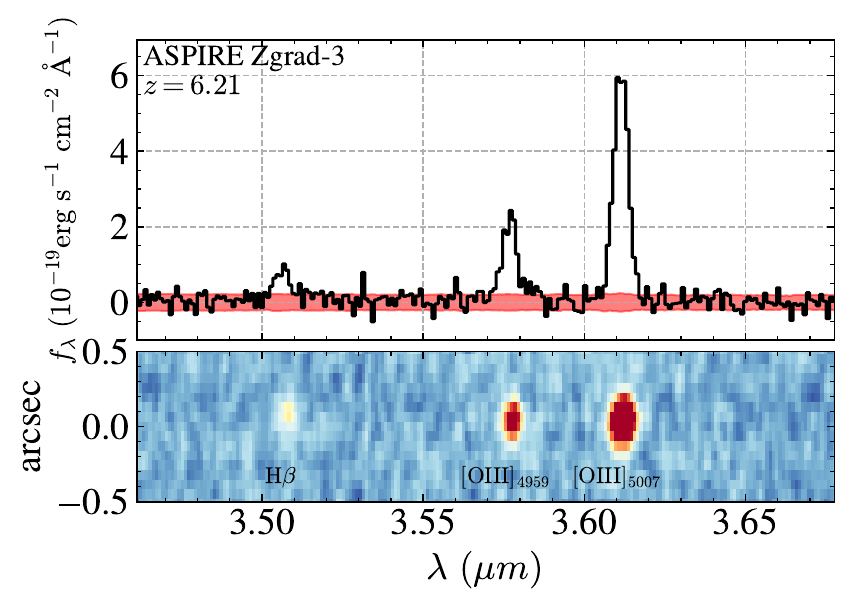}
    \end{minipage}
    \begin{minipage}{0.5\linewidth}
    \includegraphics[width=\textwidth]{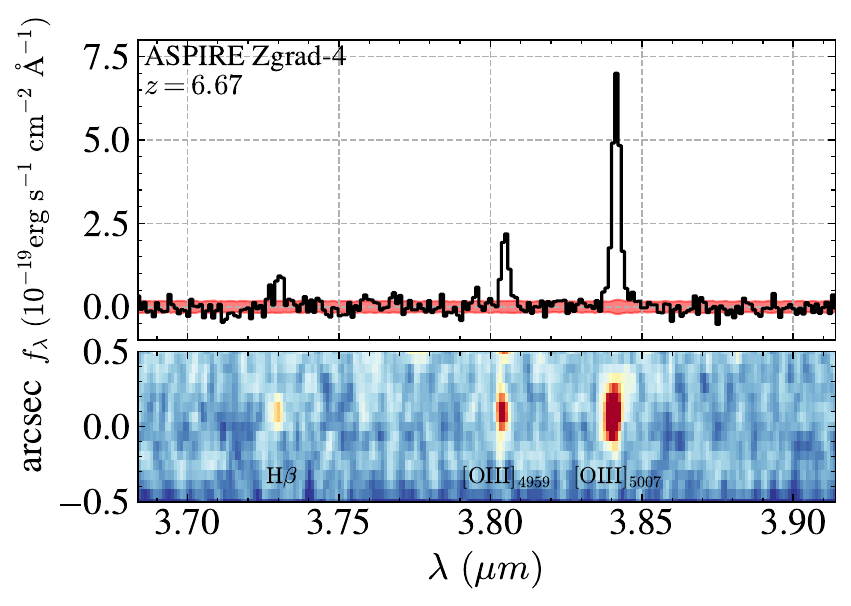}
    \end{minipage}
    \begin{minipage}{0.5\linewidth}
    \includegraphics[width=\textwidth]{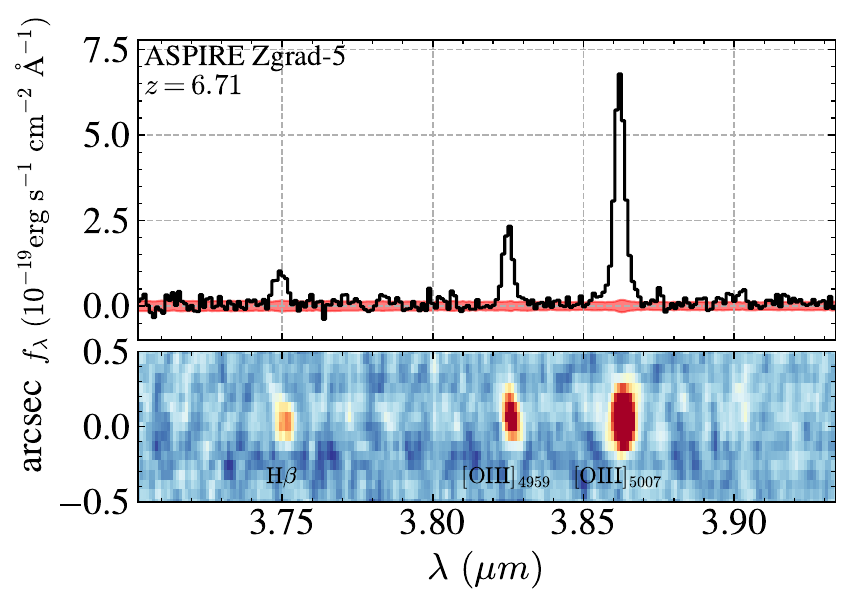}
    \end{minipage}
    \begin{minipage}{0.5\linewidth}
    \includegraphics[width=\textwidth]{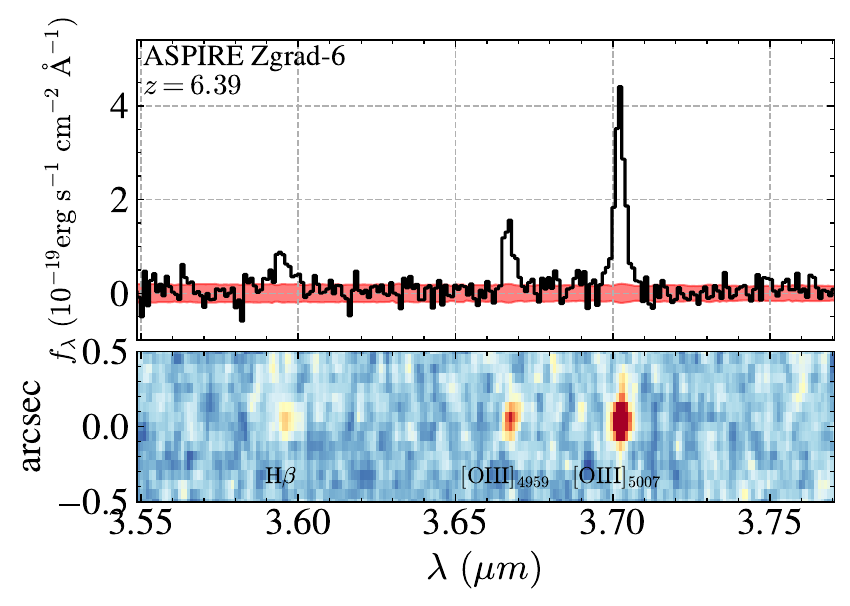}
    \end{minipage}
    \vspace*{-1em}
    \caption{\small \textbf{The 1D and 2D NIRCam F356W grism spectra of each source in Fig.\ref{fig:aspire_ind}.} \label{fig:aspire_spec}}
\end{figure*}

Here we provide figures of spectra and metallicity gradient measurements for all individual galaxies in our sample, including 6 galaxies in the ASPIRE sample (Fig.~\ref{fig:aspire_ind},~\ref{fig:aspire_spec}), and 88 galaxies in the NGDEEP sample (Fig. \ref{fig:ngdeep-ind}).
% \addtocounter{figure}{-1}
% \restoregeometry

% \newpage

\newpage

\begin{figure*}[t!]
\centering
\includegraphics[width=1\linewidth]{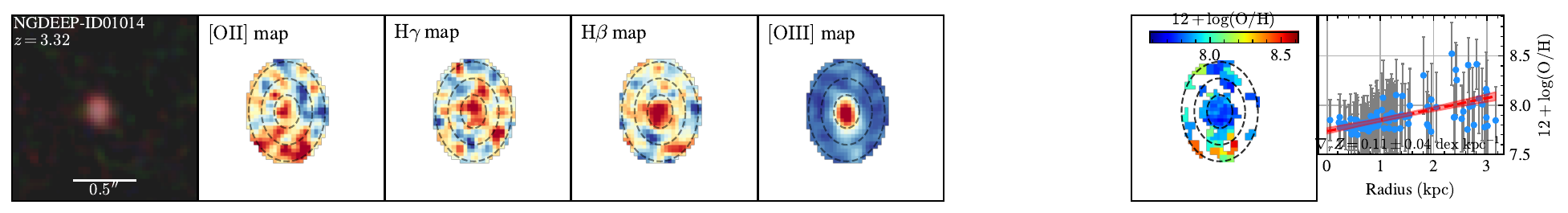}
\includegraphics[width=1\linewidth]{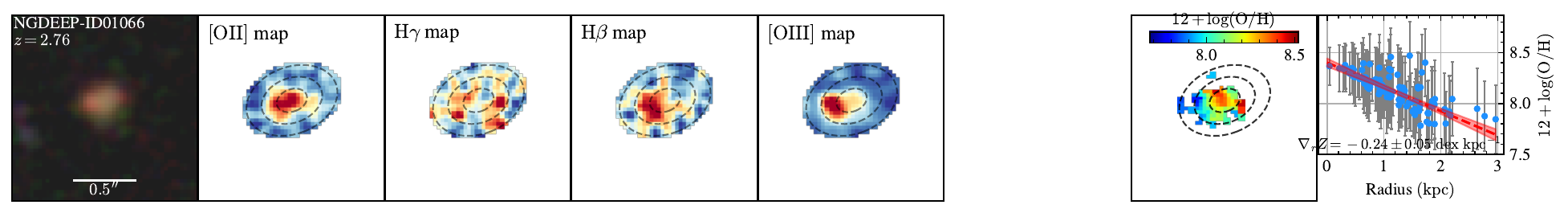}
\includegraphics[width=1\linewidth]{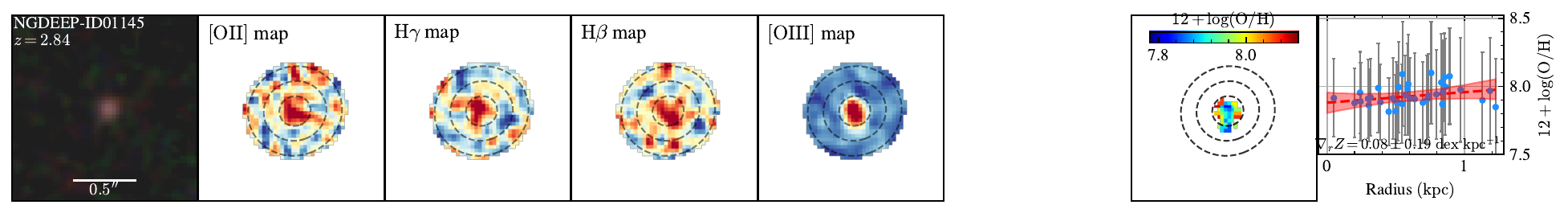}
\includegraphics[width=1\linewidth]{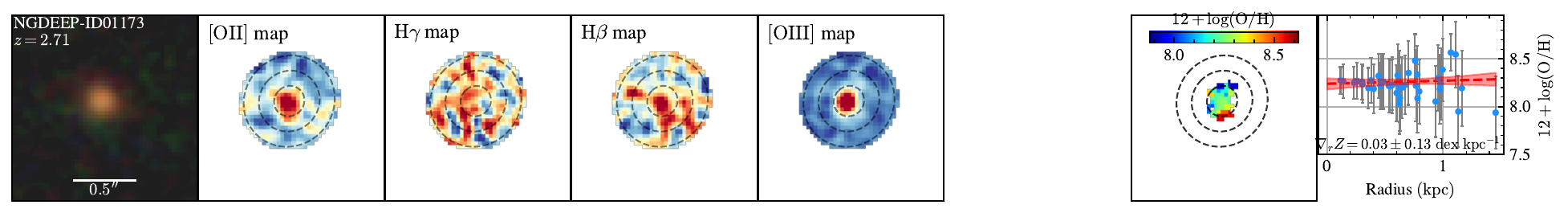}
\includegraphics[width=1\linewidth]{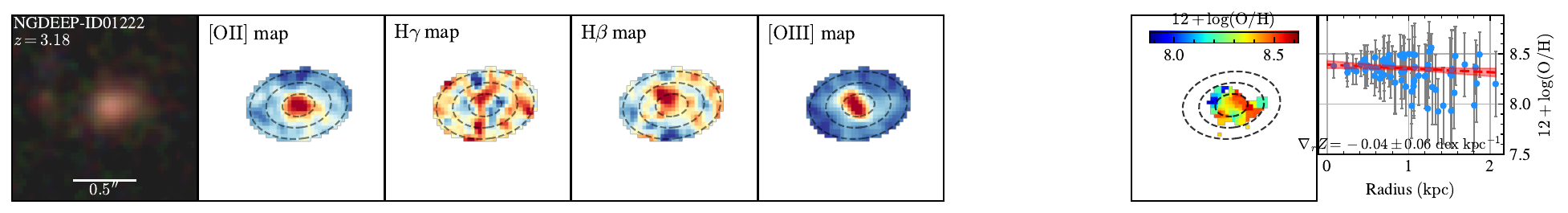}
\includegraphics[width=1\linewidth]{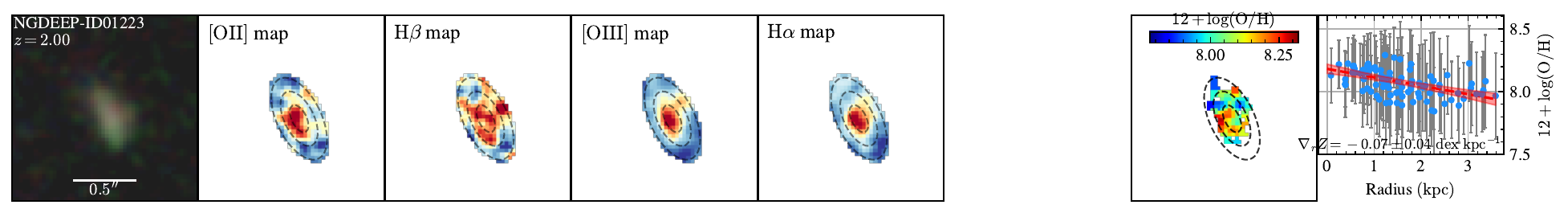}
\includegraphics[width=1\linewidth]{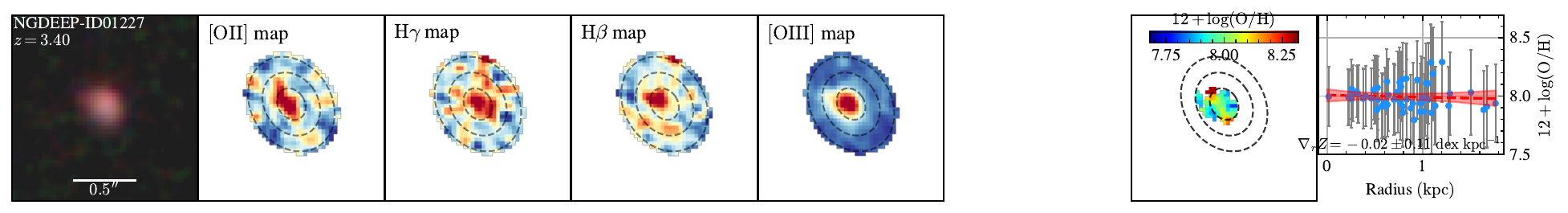}
\includegraphics[width=1\linewidth]{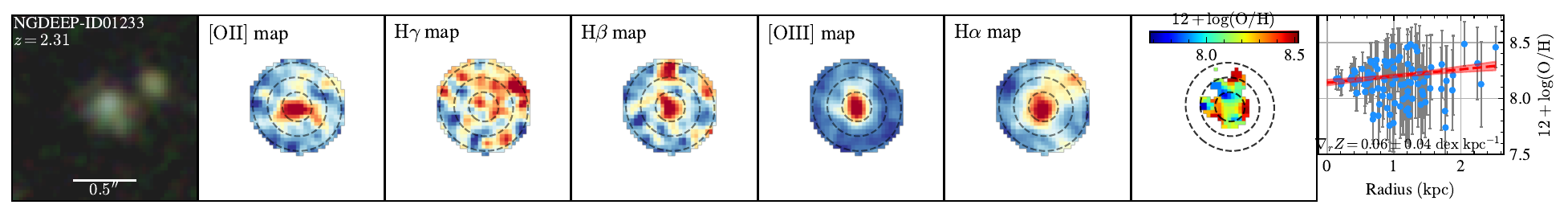}
\includegraphics[width=1\linewidth]{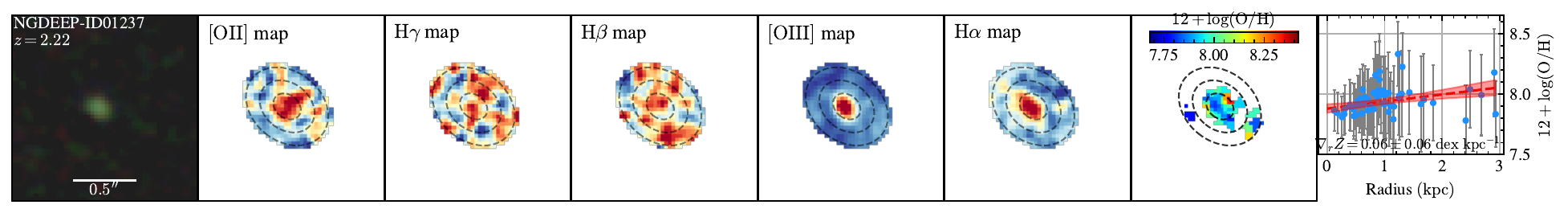}
\caption{\textbf{The false-color image, line maps, metallicity maps, and metallicity gradients for each source in the NGDEEP sample.}}\label{fig:ngdeep-ind}
\end{figure*}
\addtocounter{figure}{-1}

% 第 2 页
\begin{figure*}[t!]
\centering
\includegraphics[width=1\linewidth]{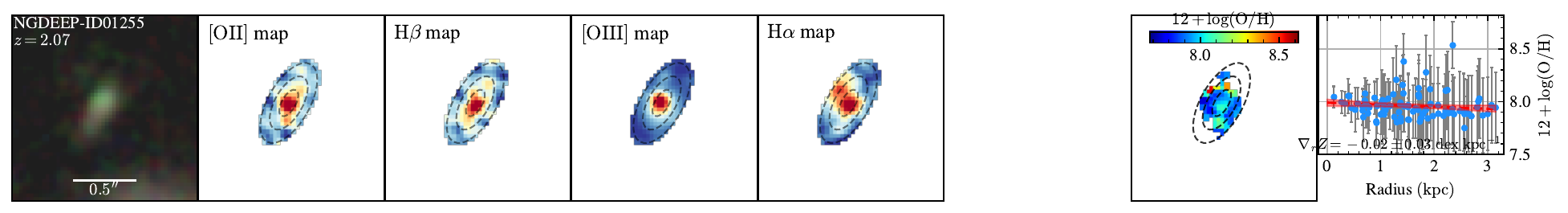}
\includegraphics[width=1\linewidth]{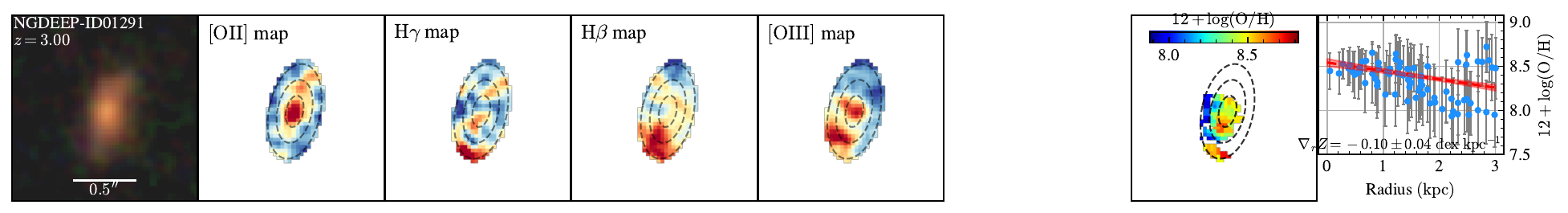}
\includegraphics[width=1\linewidth]{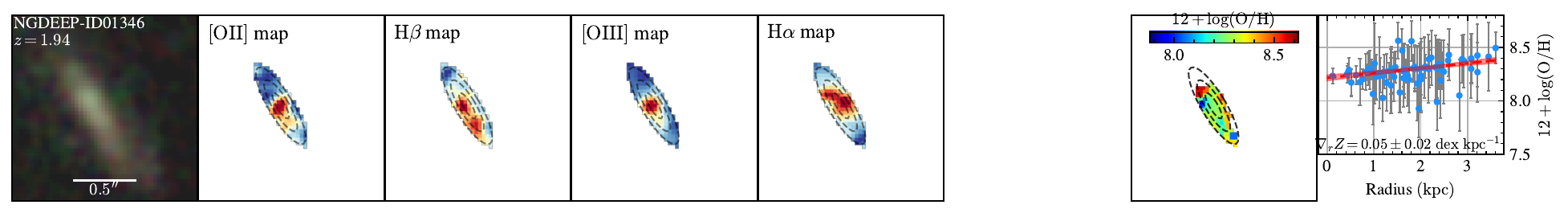}
\includegraphics[width=1\linewidth]{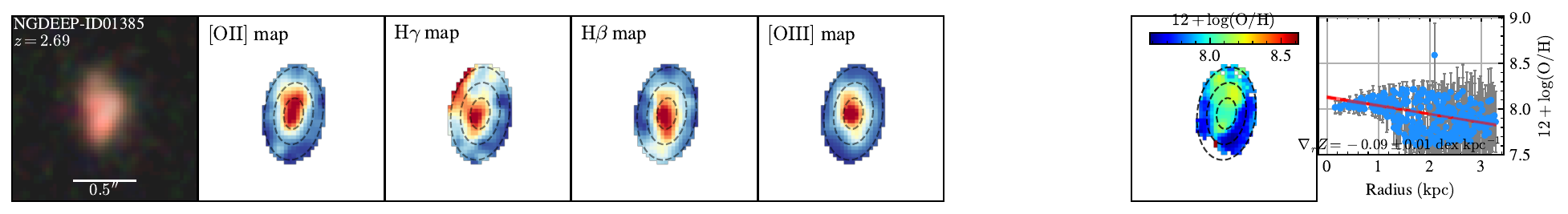}
\includegraphics[width=1\linewidth]{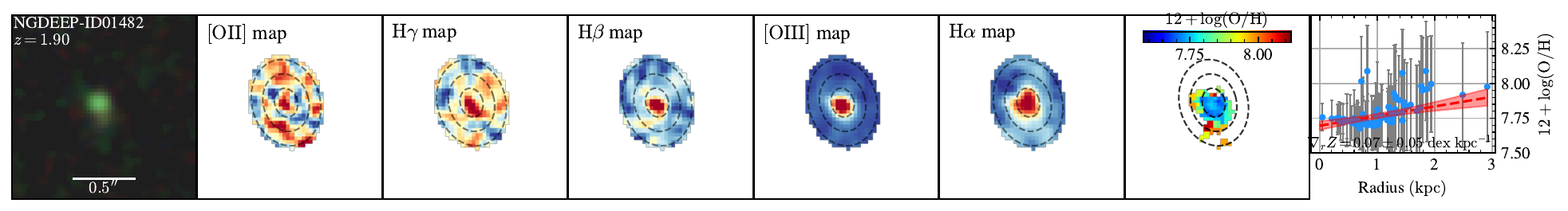}
\includegraphics[width=1\linewidth]{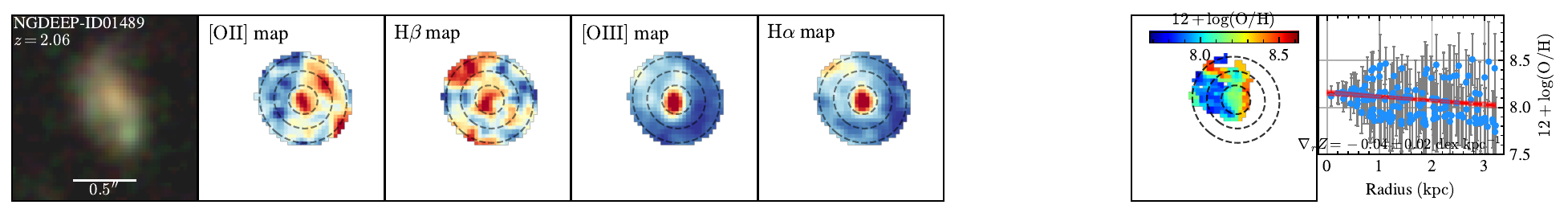}
\includegraphics[width=1\linewidth]{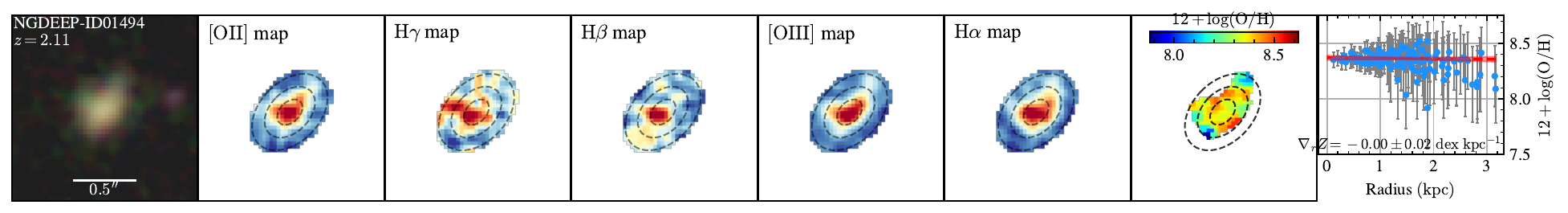}
\includegraphics[width=1\linewidth]{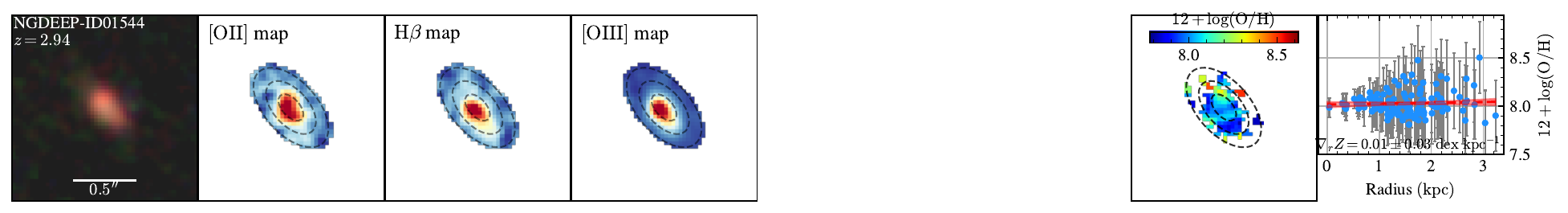}
\includegraphics[width=1\linewidth]{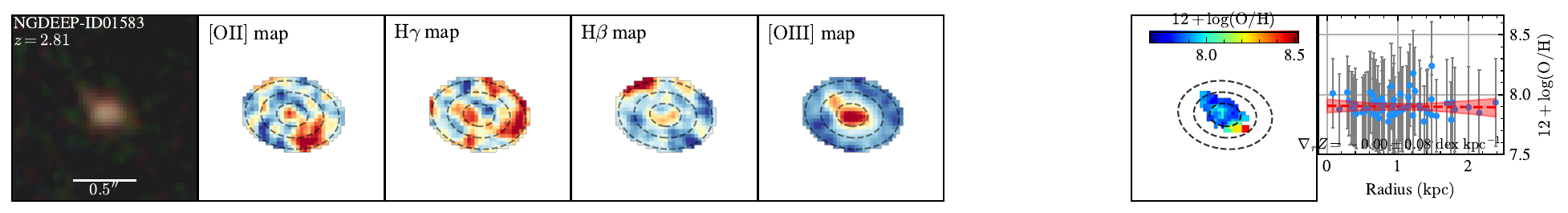}
\caption{Continued.}
\end{figure*}
\addtocounter{figure}{-1}
% 第 3 页
\begin{figure*}[t!]
\centering
\includegraphics[width=1\linewidth]{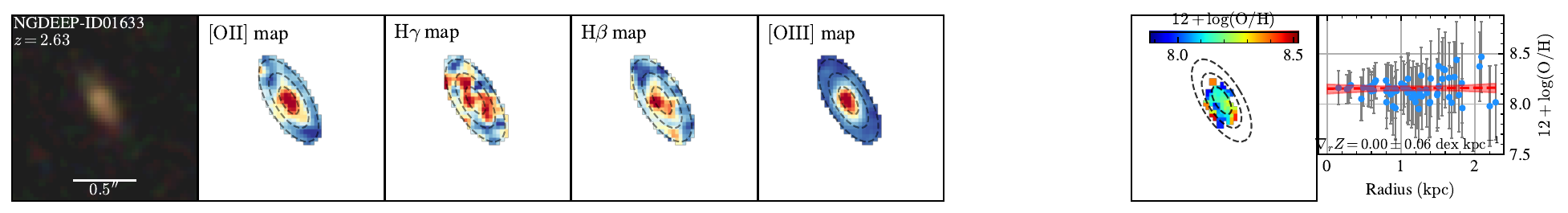}
\includegraphics[width=1\linewidth]{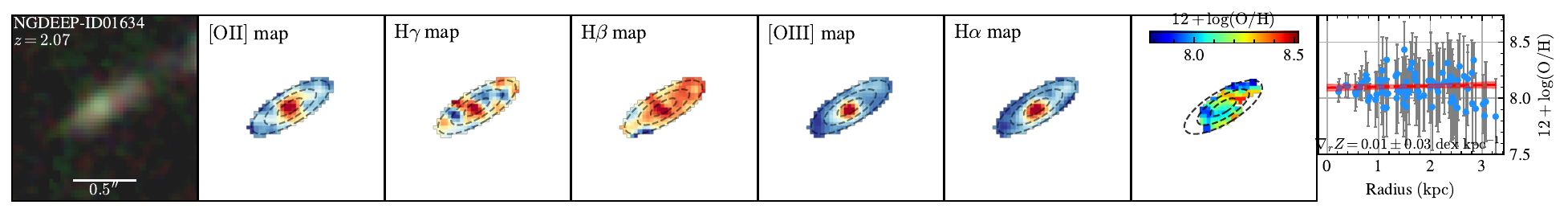}
\includegraphics[width=1\linewidth]{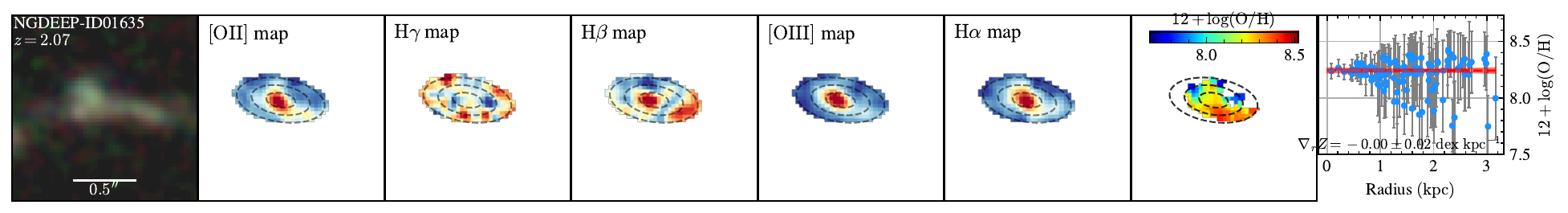}
\includegraphics[width=1\linewidth]{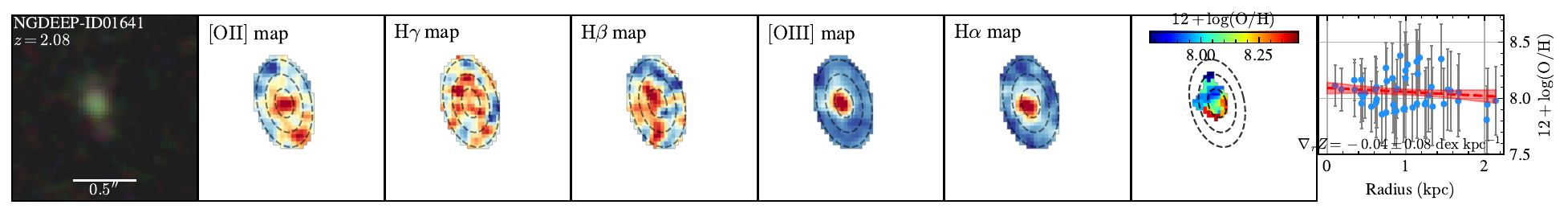}
\includegraphics[width=1\linewidth]{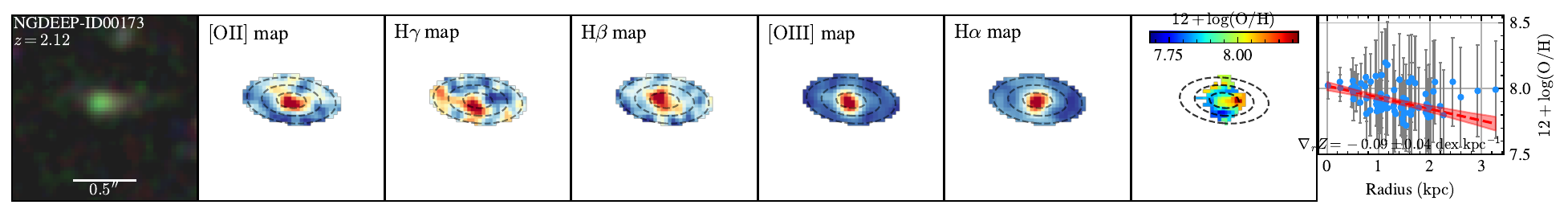}
\includegraphics[width=1\linewidth]{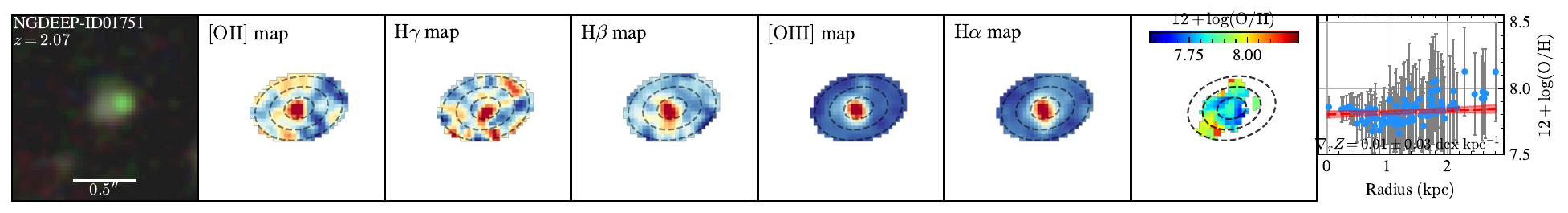}
\includegraphics[width=1\linewidth]{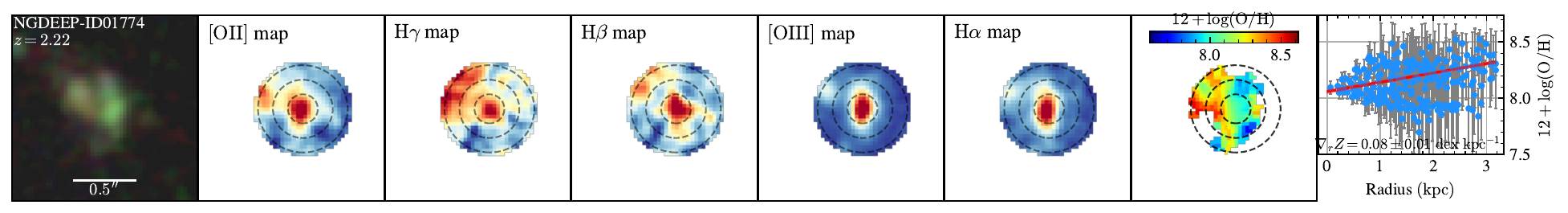}
\includegraphics[width=1\linewidth]{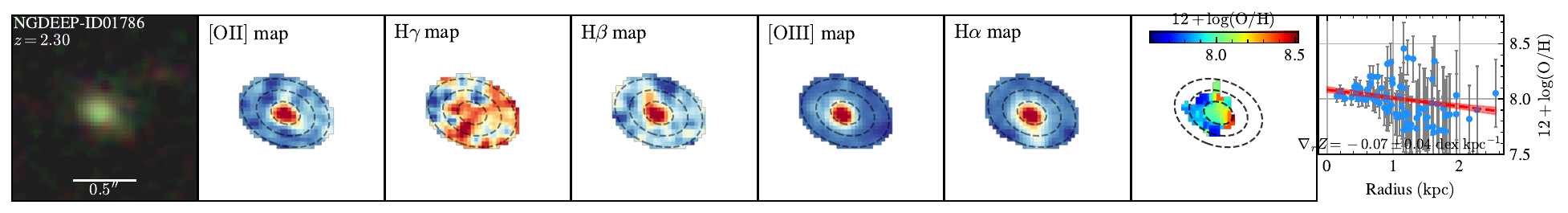}
\includegraphics[width=1\linewidth]{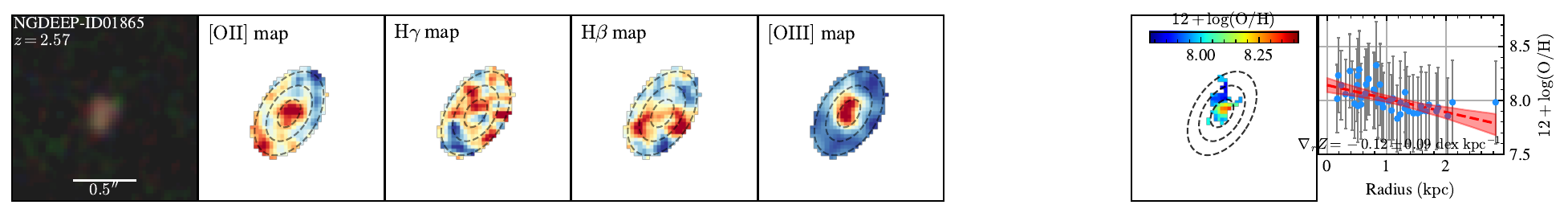}
\caption{Continued.}
\end{figure*}
\addtocounter{figure}{-1}
% 第 4 页
\begin{figure*}[t!]
\centering
\includegraphics[width=1\linewidth]{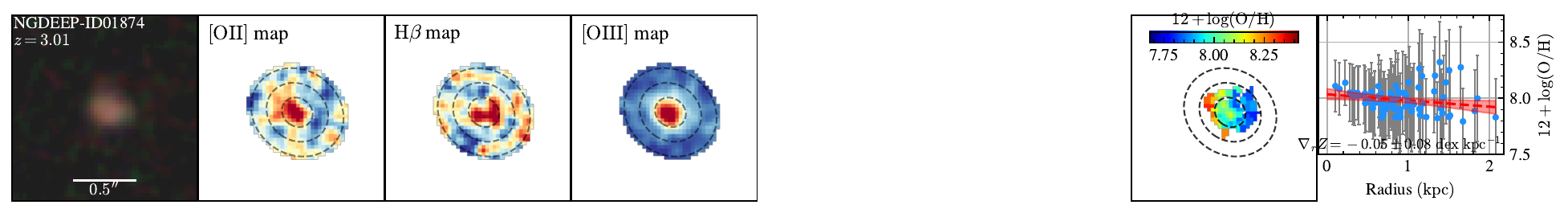}
\includegraphics[width=1\linewidth]{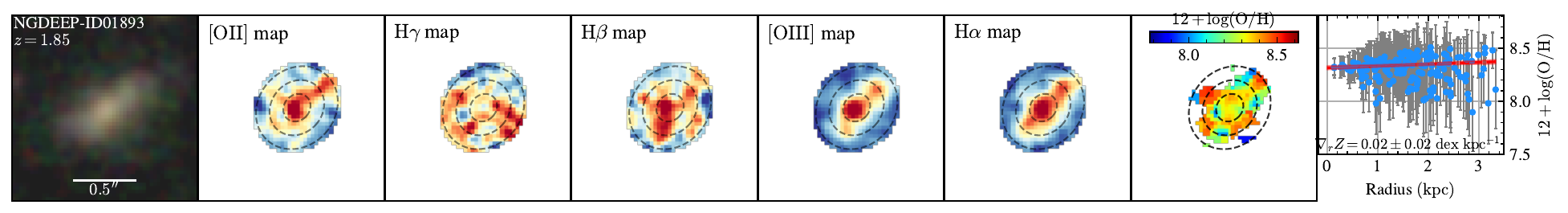}
\includegraphics[width=1\linewidth]{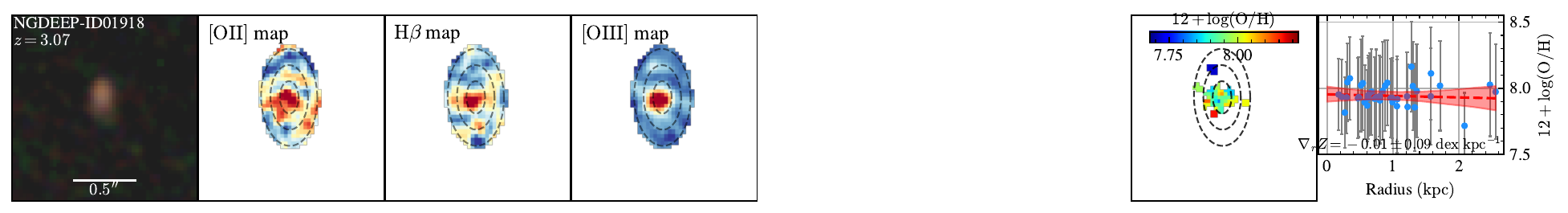}
\includegraphics[width=1\linewidth]{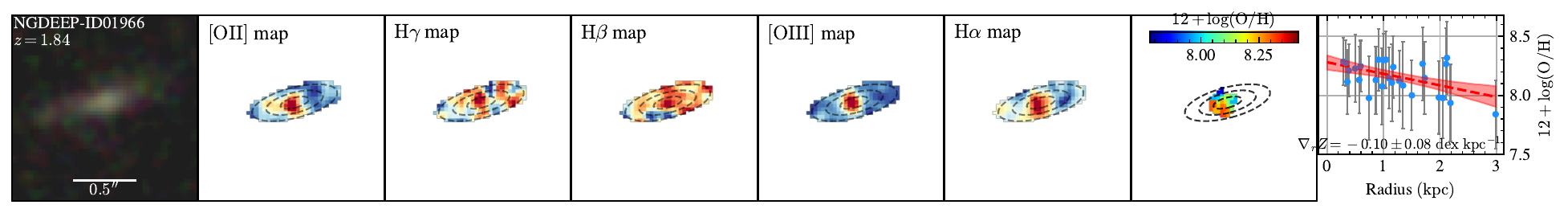}
\includegraphics[width=1\linewidth]{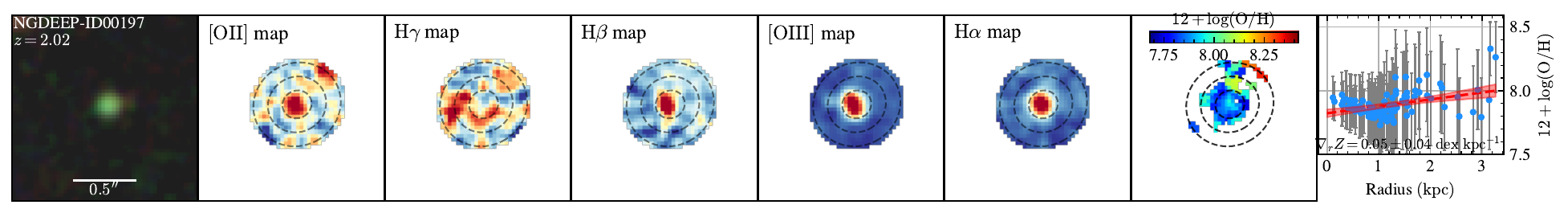}
\includegraphics[width=1\linewidth]{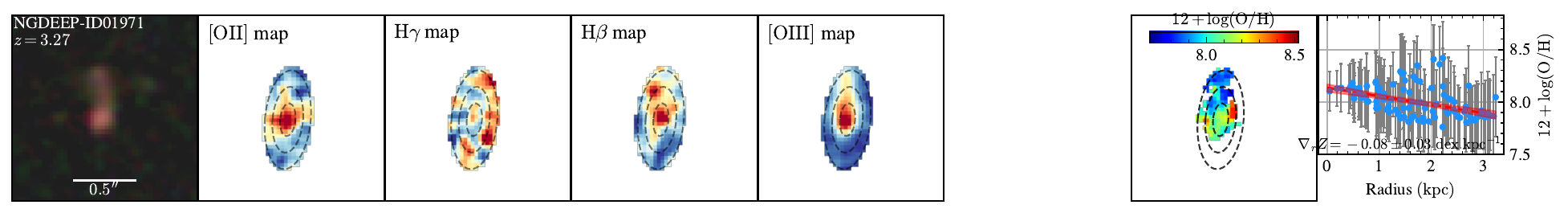}
\includegraphics[width=1\linewidth]{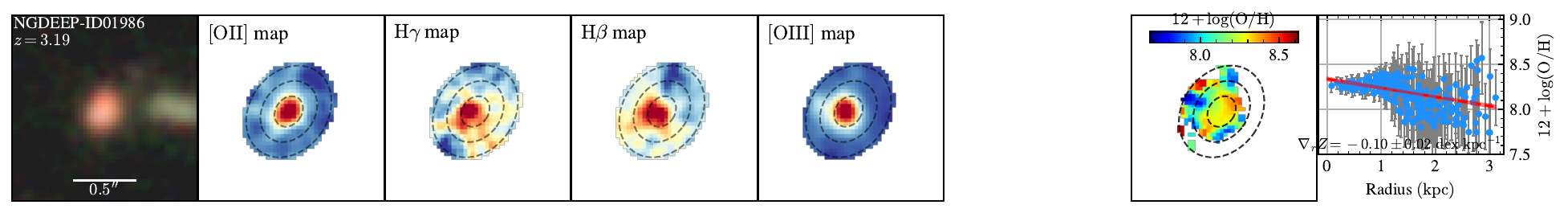}
\includegraphics[width=1\linewidth]{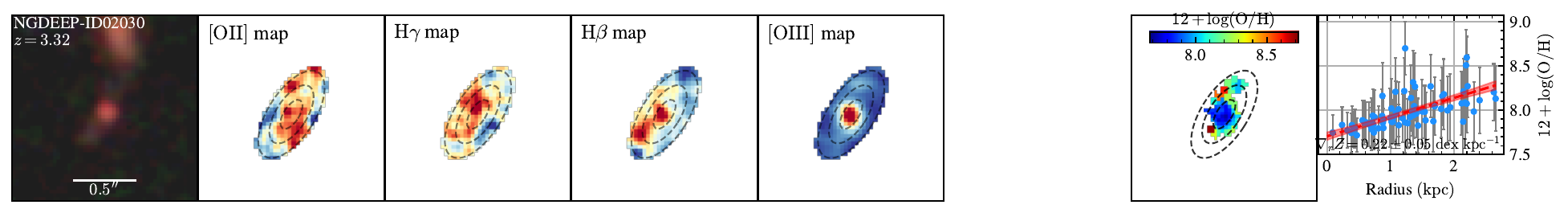}
\includegraphics[width=1\linewidth]{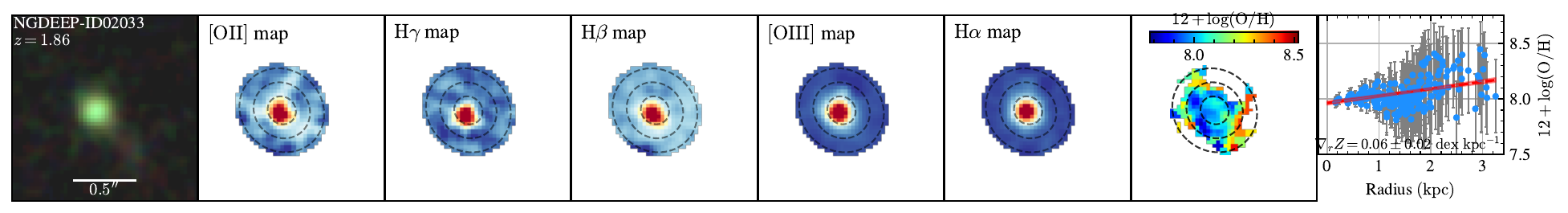}
\caption{Continued.}
\end{figure*}
\addtocounter{figure}{-1}
% 第 5 页
\begin{figure*}[t!]
\centering
\includegraphics[width=1\linewidth]{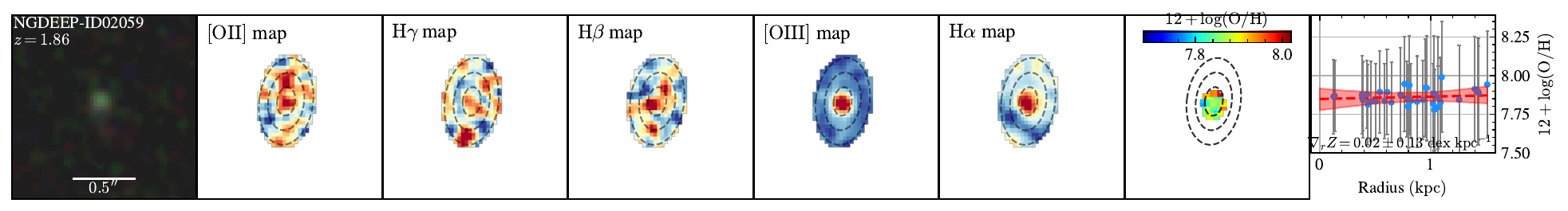}
\includegraphics[width=1\linewidth]{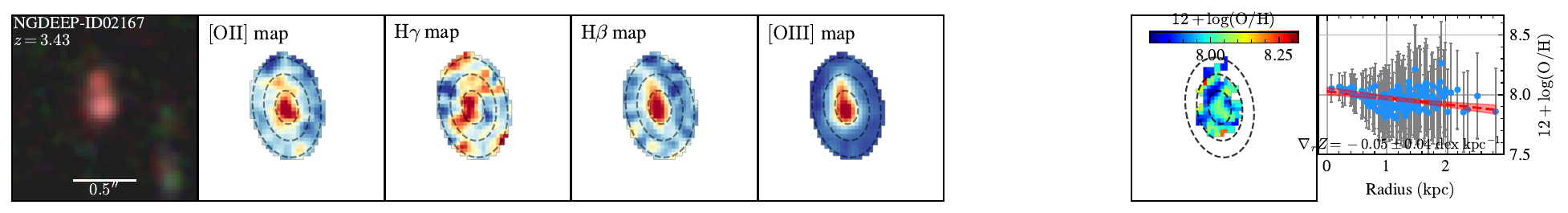}
\includegraphics[width=1\linewidth]{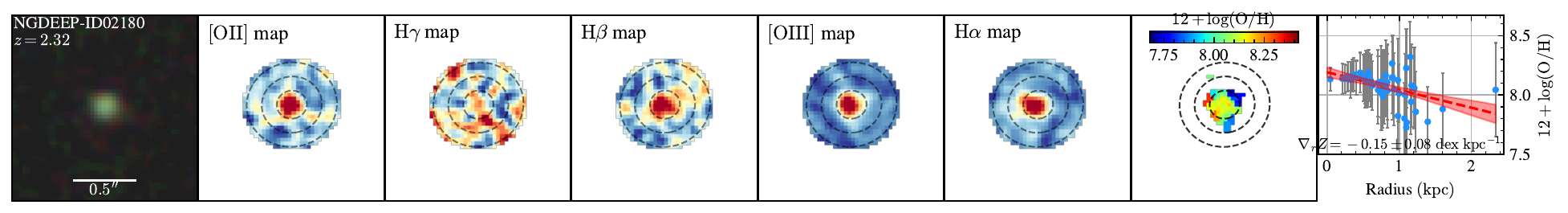}
\includegraphics[width=1\linewidth]{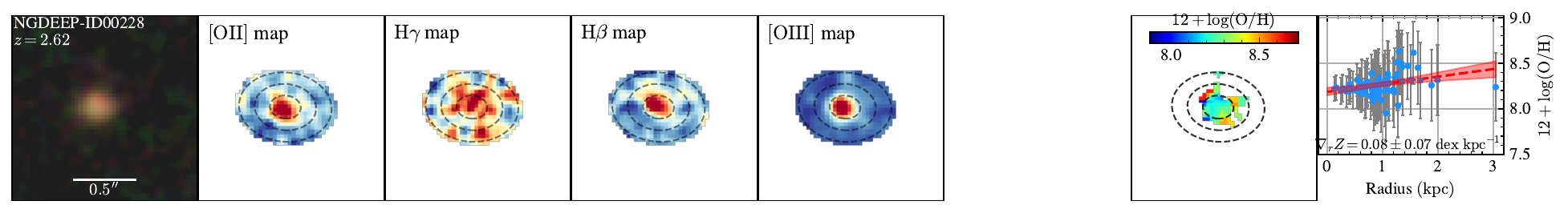}
\includegraphics[width=1\linewidth]{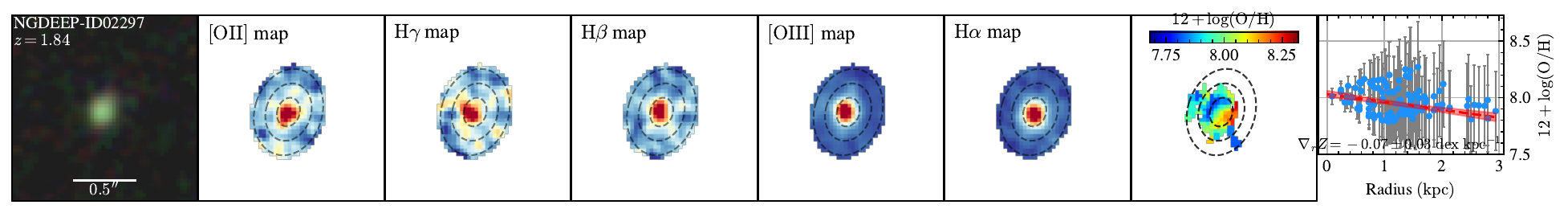}
\includegraphics[width=1\linewidth]{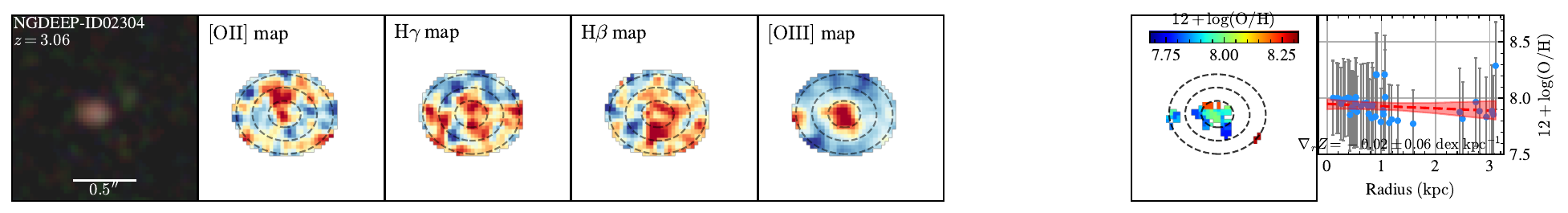}
\includegraphics[width=1\linewidth]{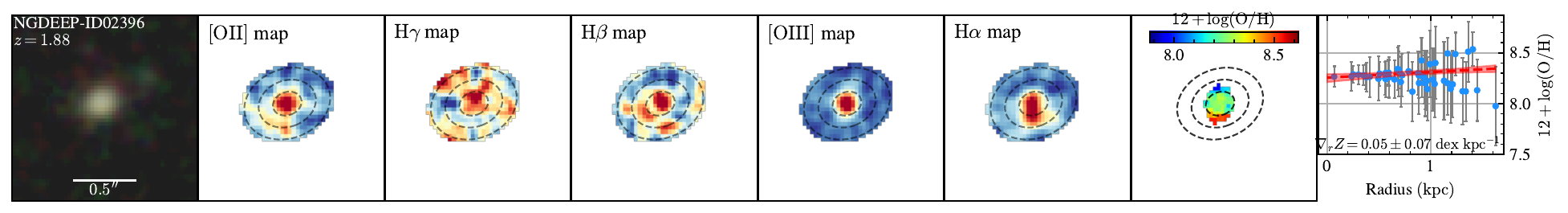}
\includegraphics[width=1\linewidth]{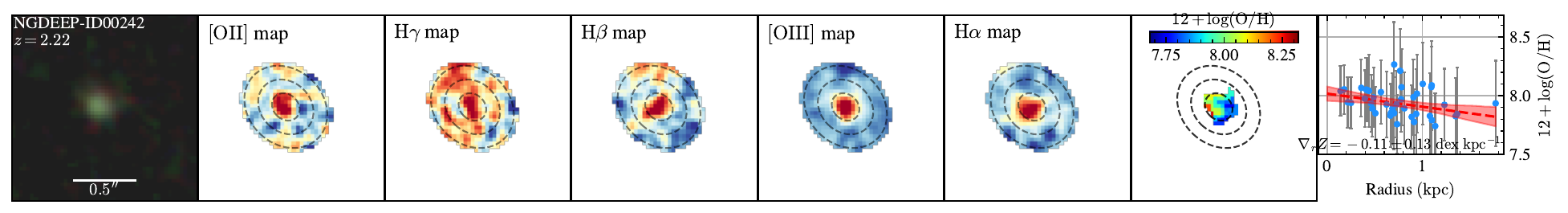}
\includegraphics[width=1\linewidth]{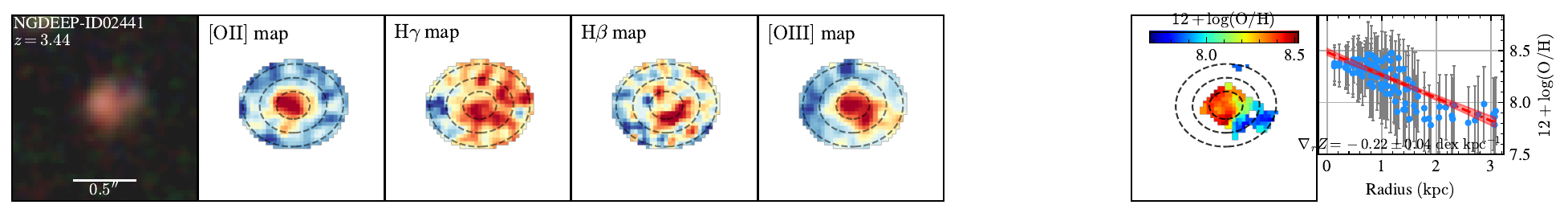}
\caption{Continued.}
\end{figure*}
\addtocounter{figure}{-1}
% 第 6 页
\begin{figure*}[t!]
\centering
\includegraphics[width=1\linewidth]{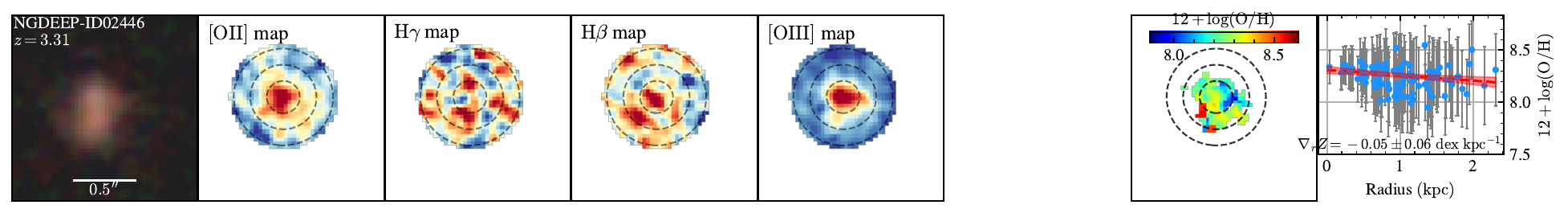}
\includegraphics[width=1\linewidth]{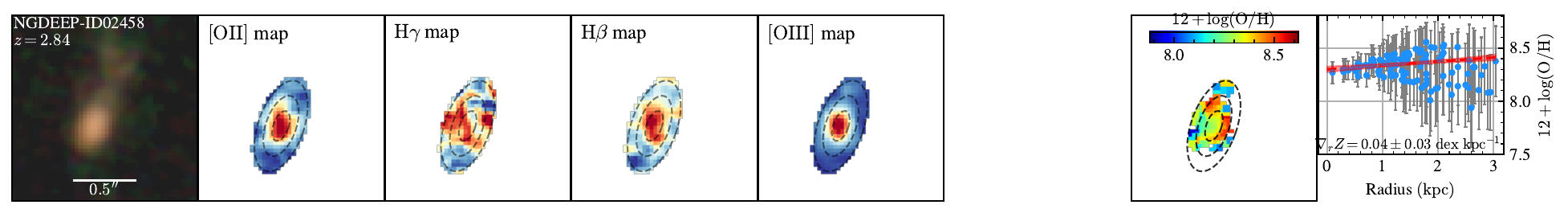}
\includegraphics[width=1\linewidth]{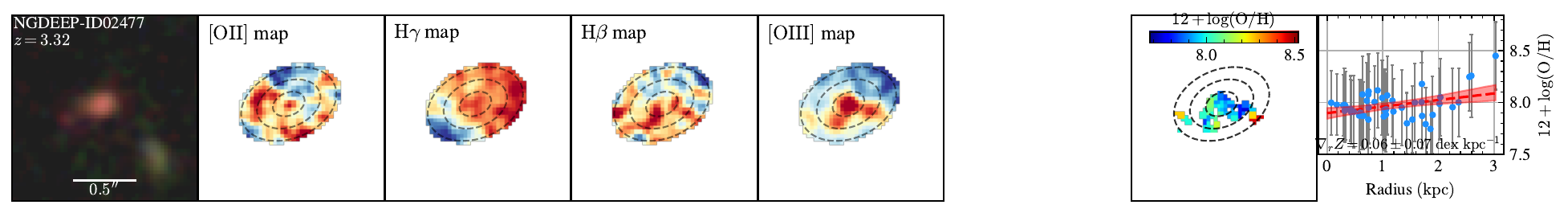}
\includegraphics[width=1\linewidth]{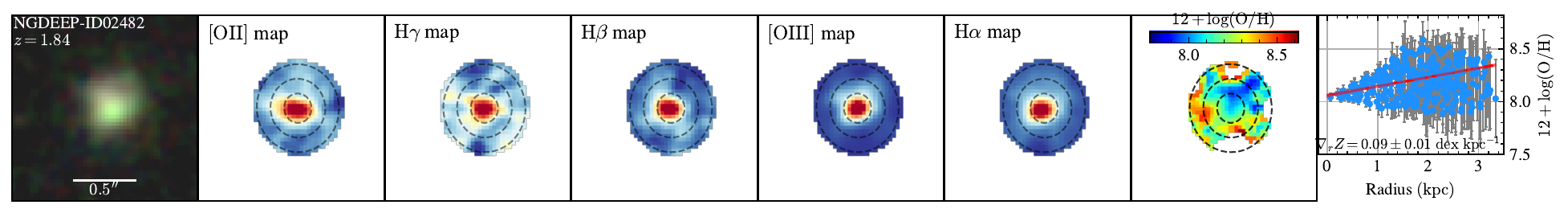}
\includegraphics[width=1\linewidth]{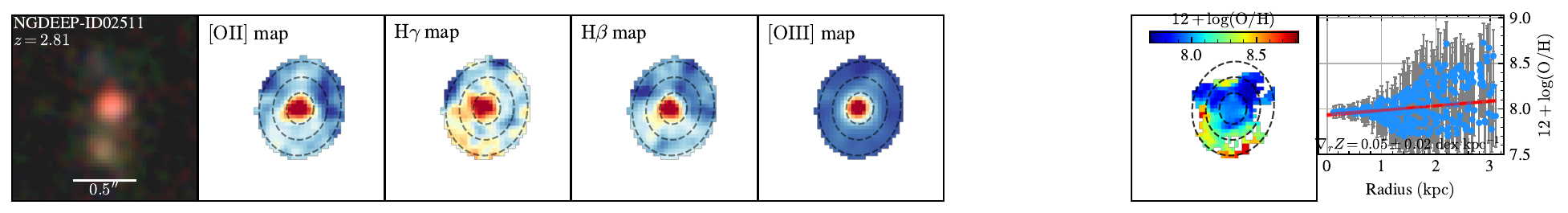}
\includegraphics[width=1\linewidth]{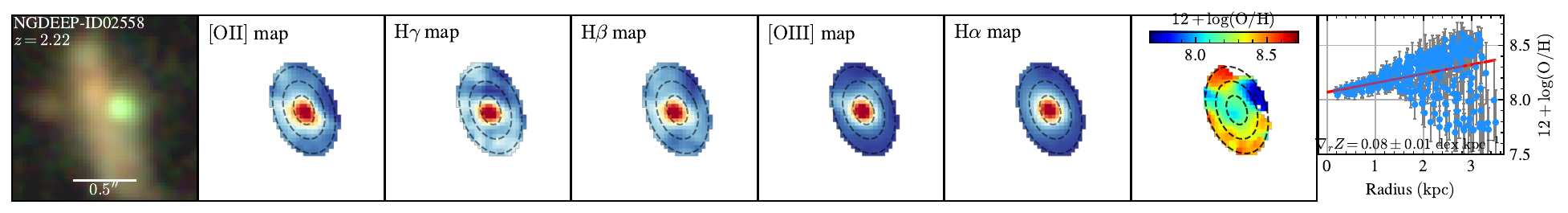}
\includegraphics[width=1\linewidth]{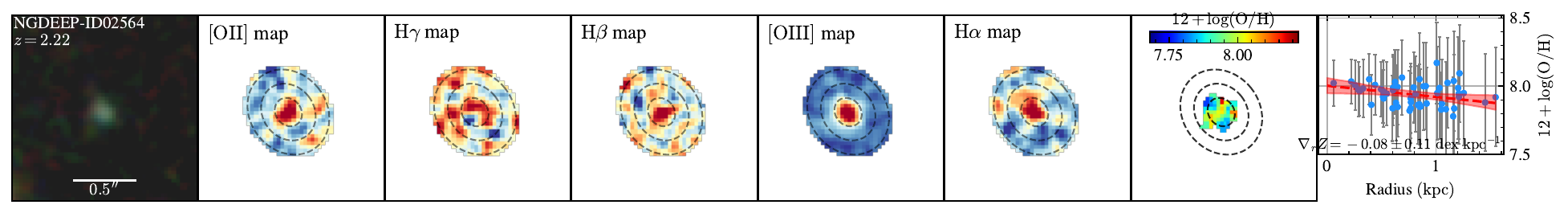}
\includegraphics[width=1\linewidth]{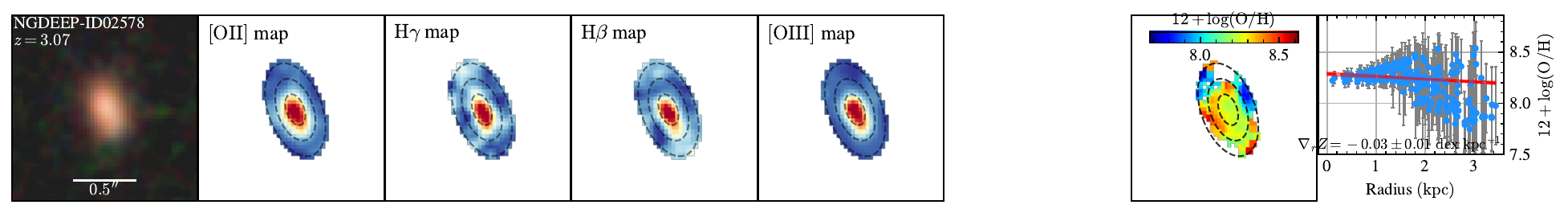}
\includegraphics[width=1\linewidth]{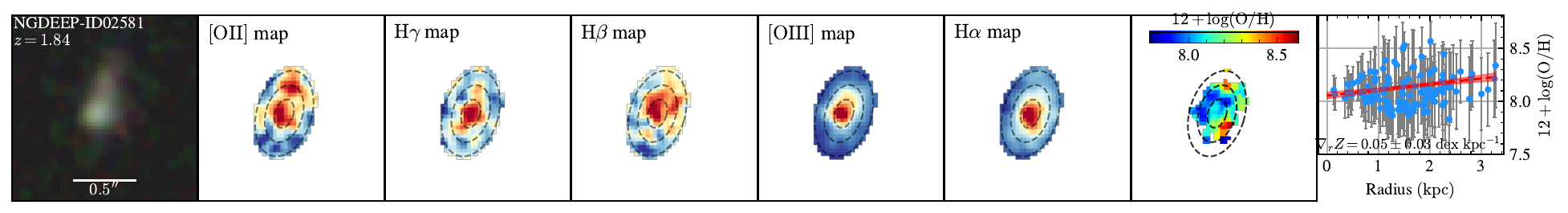}
\caption{Continued.}
\end{figure*}
\addtocounter{figure}{-1}
% 第 7 页
\begin{figure*}[t!]
\centering
\includegraphics[width=1\linewidth]{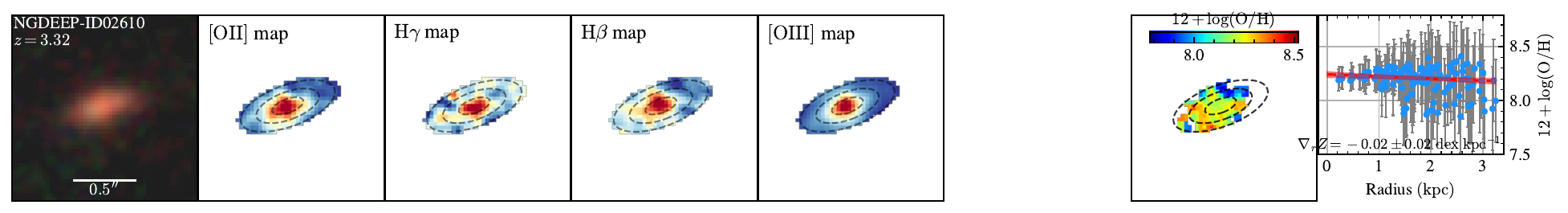}
\includegraphics[width=1\linewidth]{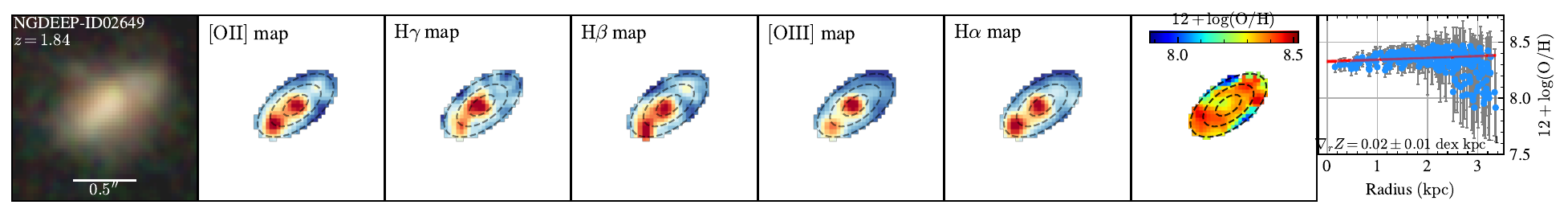}
\includegraphics[width=1\linewidth]{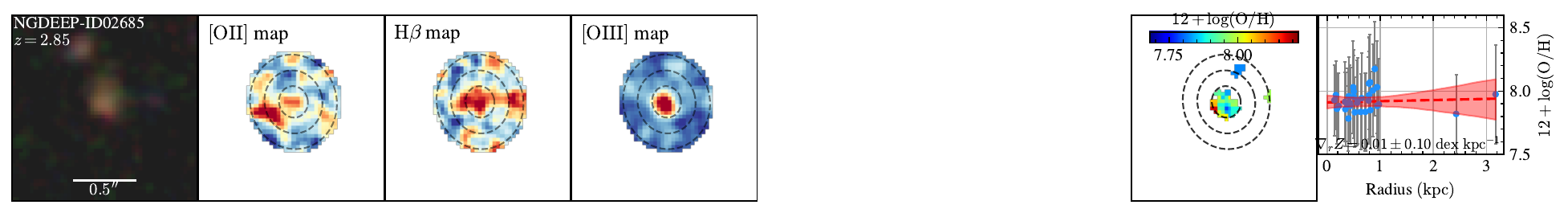}
\includegraphics[width=1\linewidth]{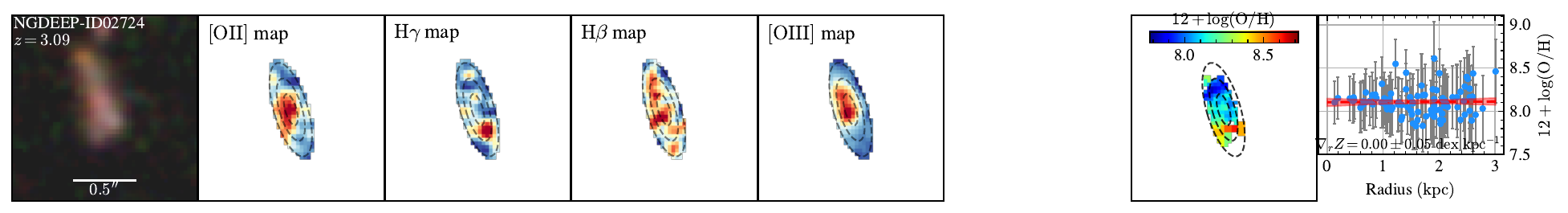}
\includegraphics[width=1\linewidth]{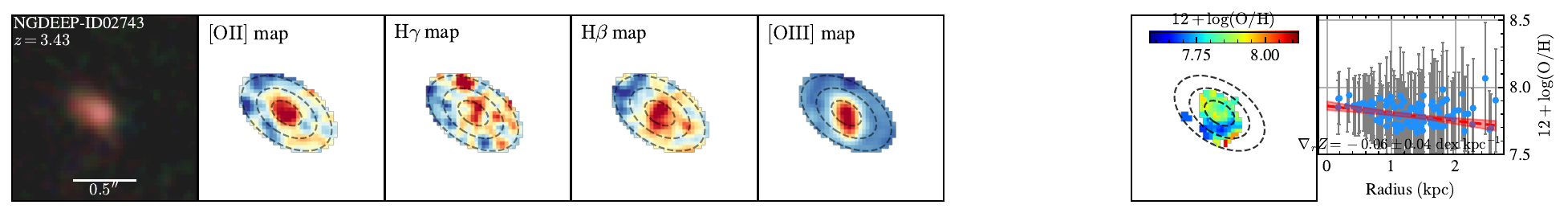}
\includegraphics[width=1\linewidth]{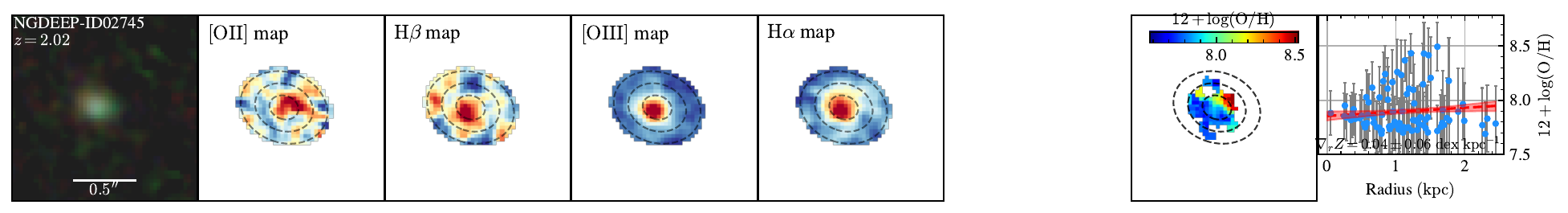}
\includegraphics[width=1\linewidth]{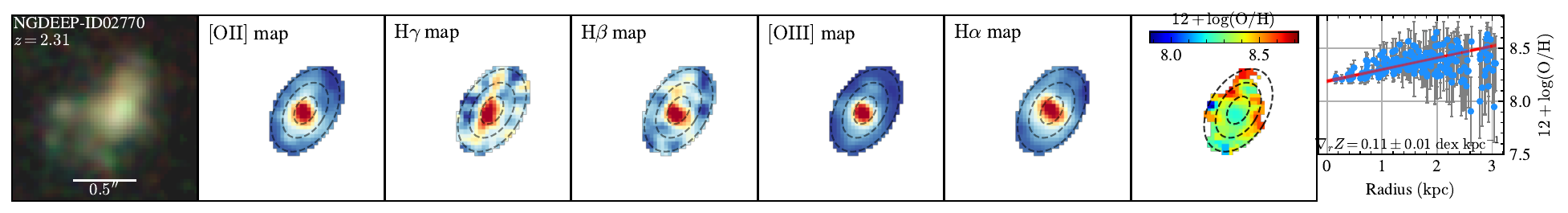}
\includegraphics[width=1\linewidth]{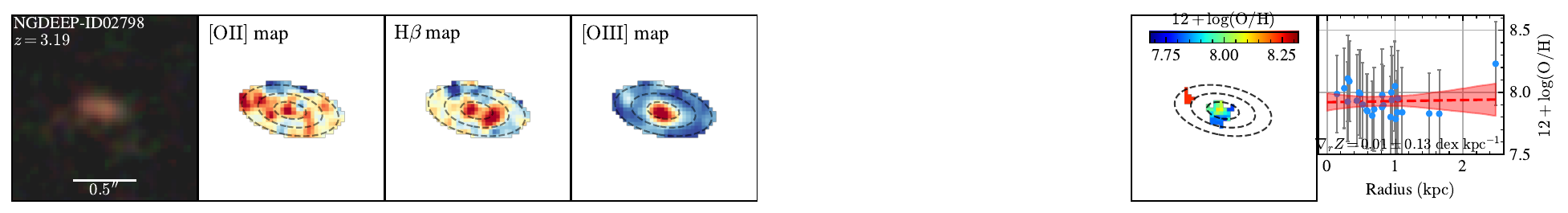}
\includegraphics[width=1\linewidth]{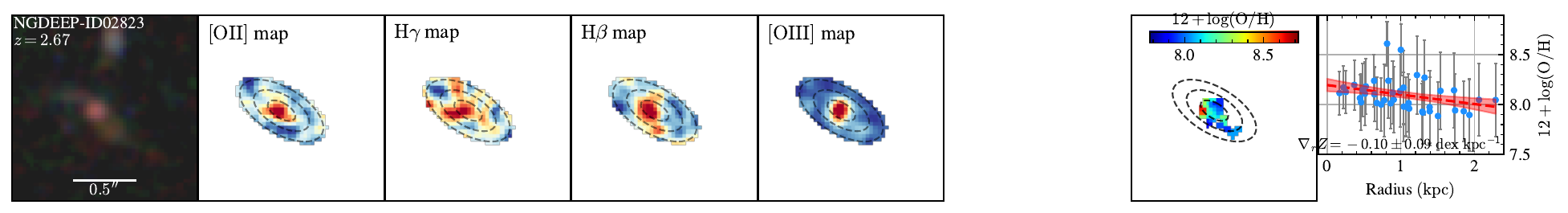}
\caption{Continued.}
\end{figure*}
\addtocounter{figure}{-1}
% 第 8 页
\begin{figure*}[t!]
\centering
\includegraphics[width=1\linewidth]{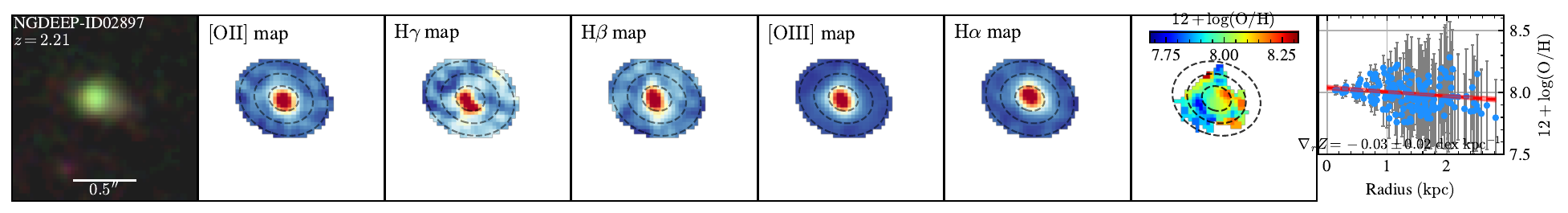}
\includegraphics[width=1\linewidth]{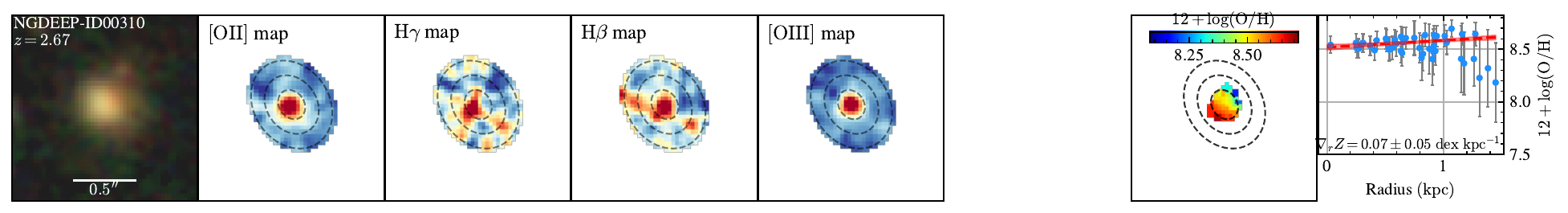}
\includegraphics[width=1\linewidth]{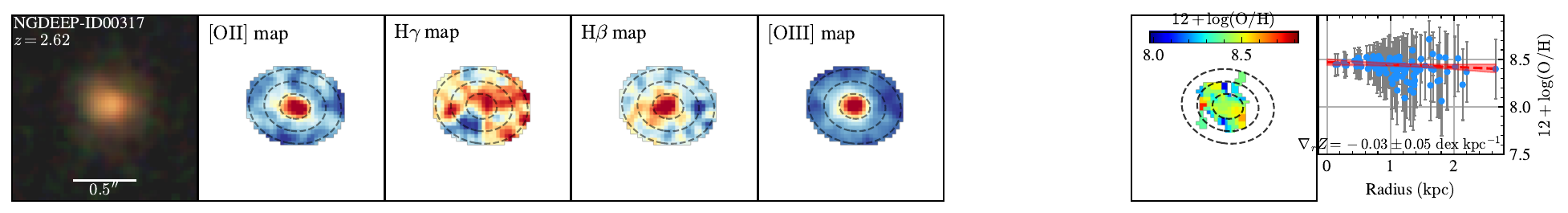}
\includegraphics[width=1\linewidth]{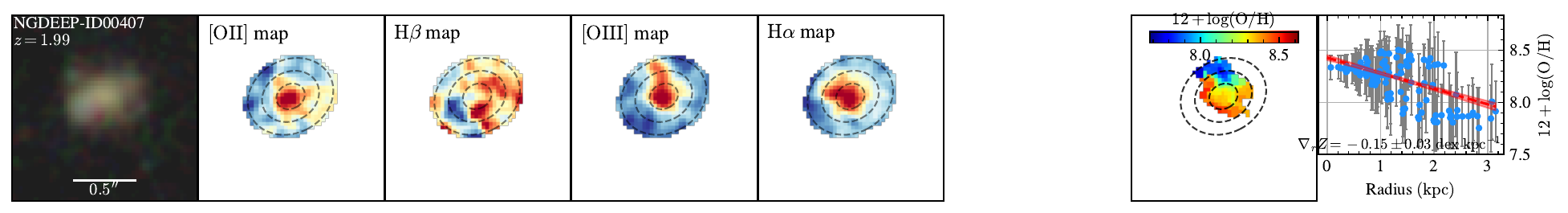}
\includegraphics[width=1\linewidth]{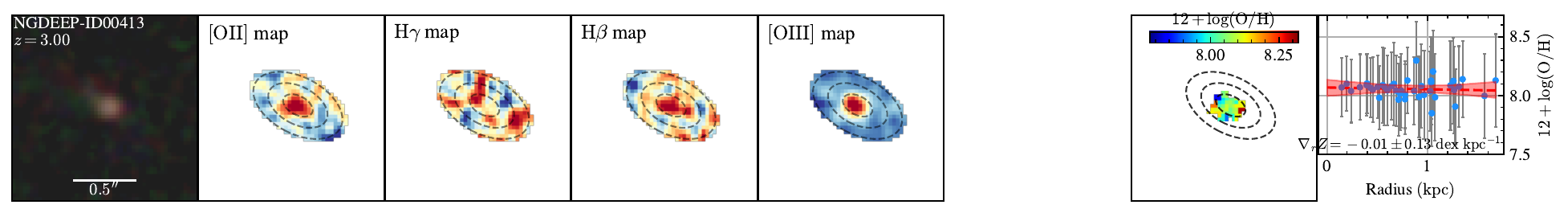}
\includegraphics[width=1\linewidth]{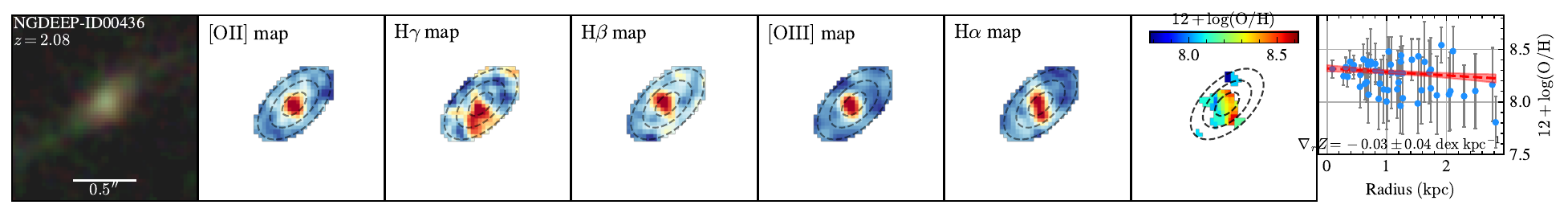}
\includegraphics[width=1\linewidth]{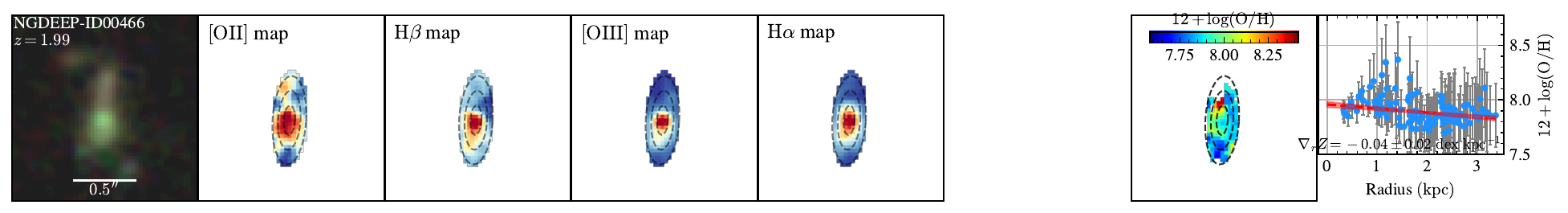}
\includegraphics[width=1\linewidth]{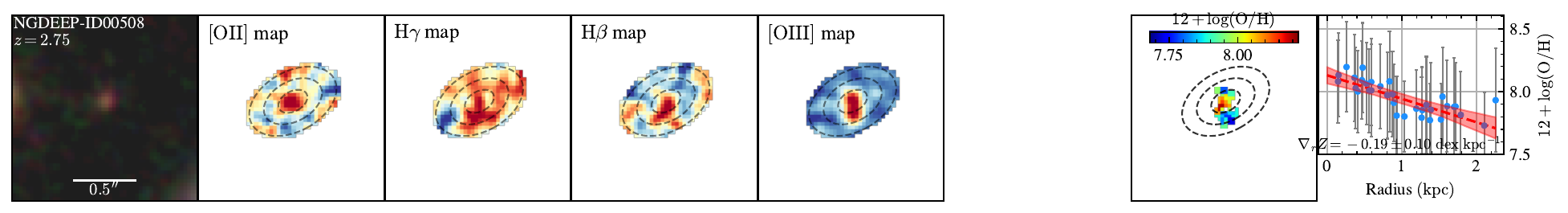}
\includegraphics[width=1\linewidth]{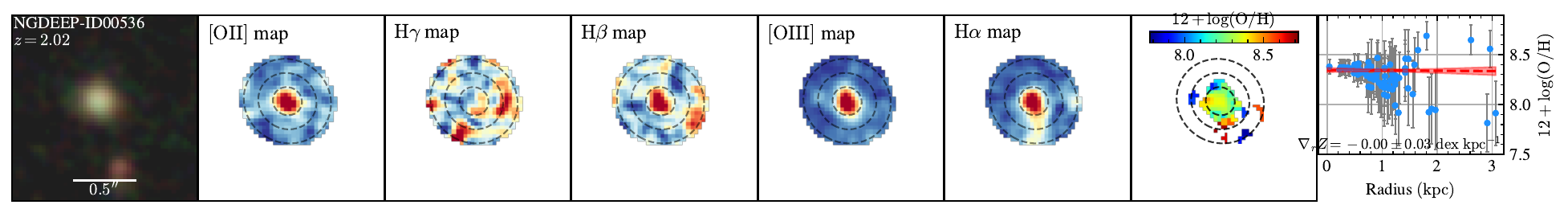}
\caption{Continued.}
\end{figure*}
\addtocounter{figure}{-1}
% 第 9 页
\begin{figure*}[t!]
\centering
\includegraphics[width=1\linewidth]{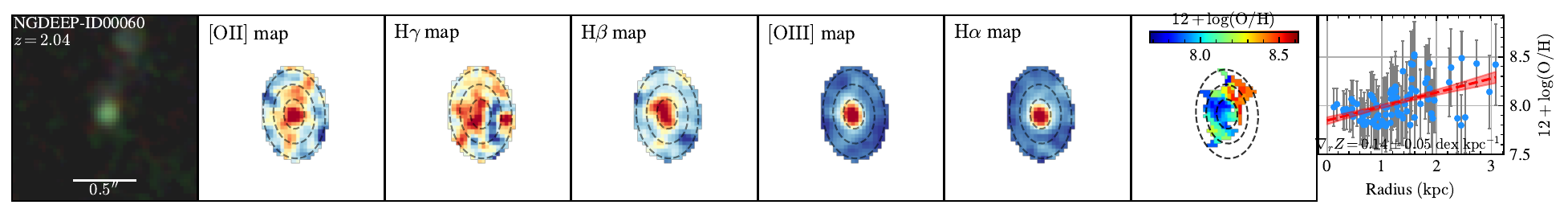}
\includegraphics[width=1\linewidth]{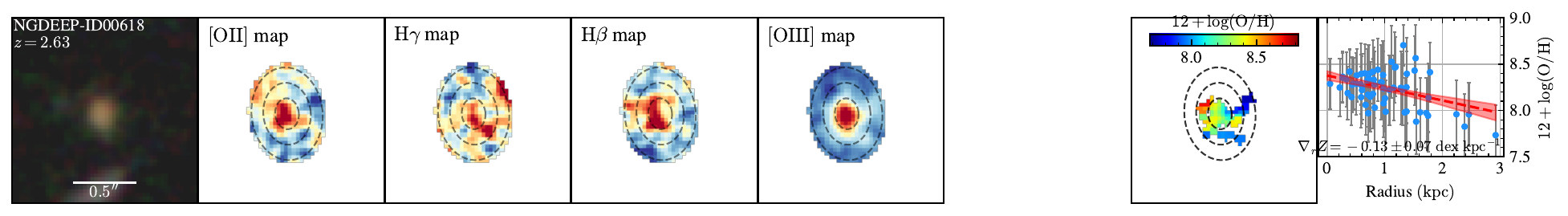}
\includegraphics[width=1\linewidth]{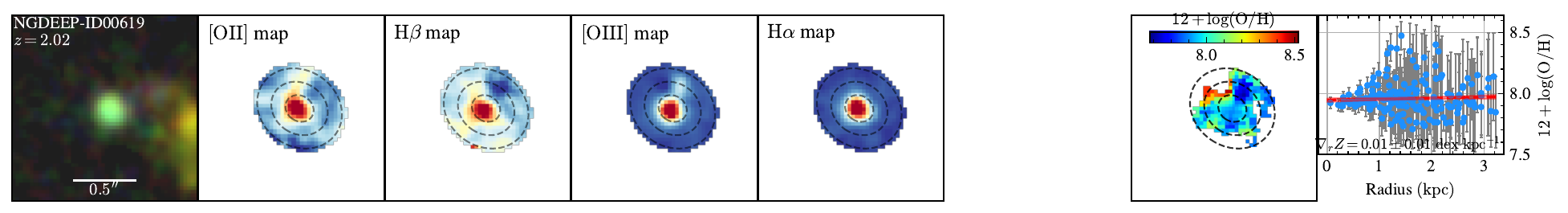}
\includegraphics[width=1\linewidth]{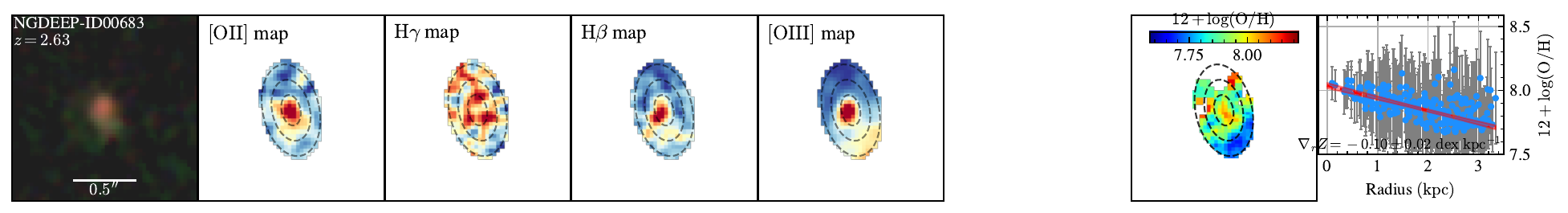}
\includegraphics[width=1\linewidth]{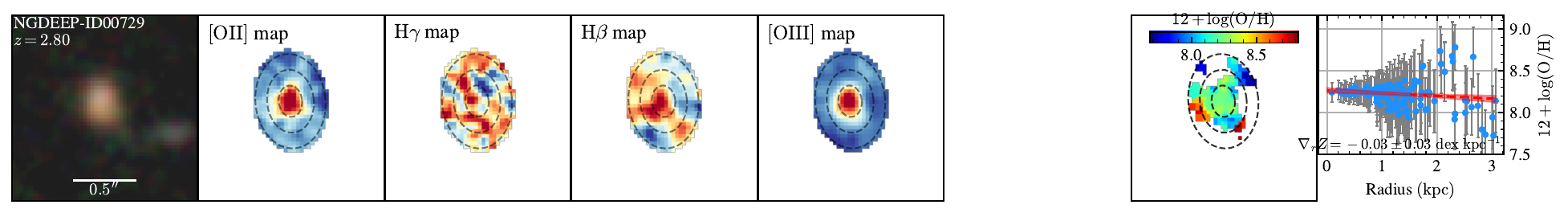}
\includegraphics[width=1\linewidth]{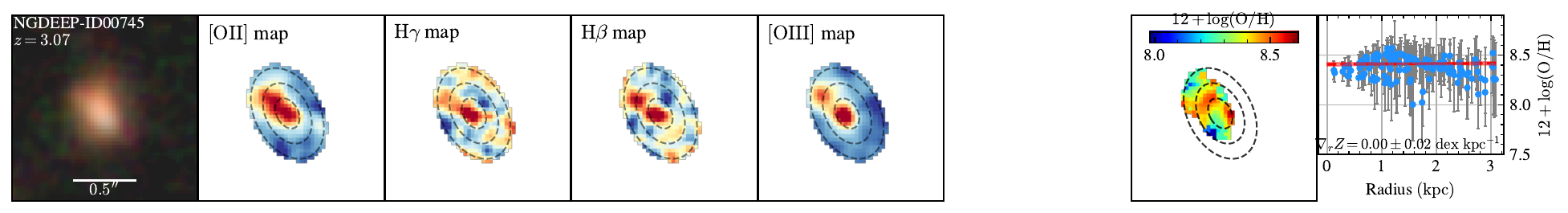}
\includegraphics[width=1\linewidth]{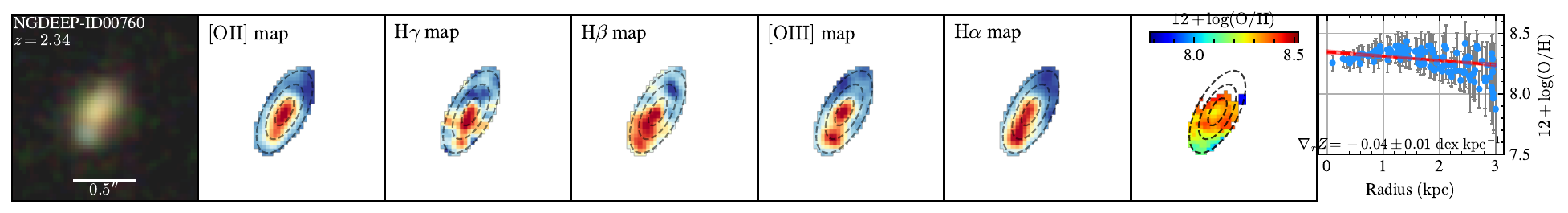}
\includegraphics[width=1\linewidth]{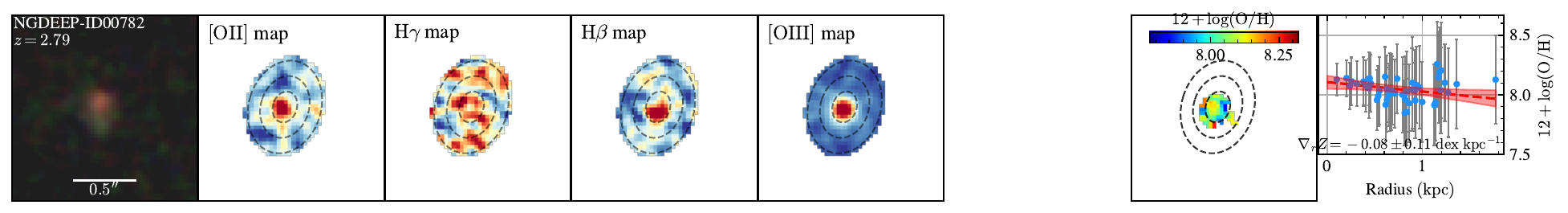}
\includegraphics[width=1\linewidth]{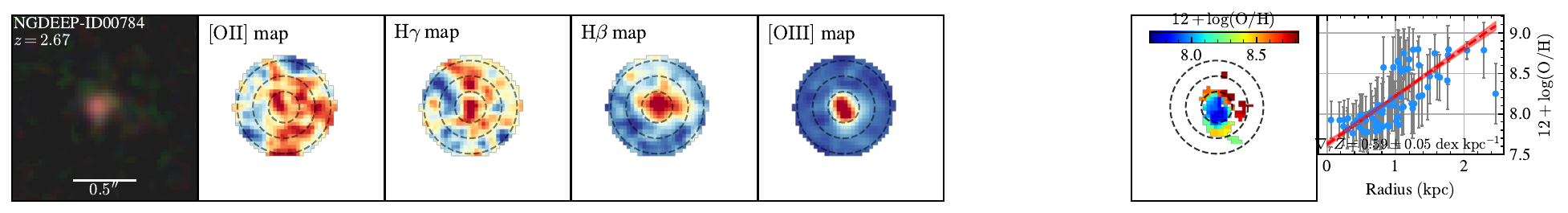}
\caption{Continued.}
\end{figure*}
\addtocounter{figure}{-1}
% 第 10 页（最后一页 8 张图）
\begin{figure*}[t!]
\centering
\includegraphics[width=1\linewidth]{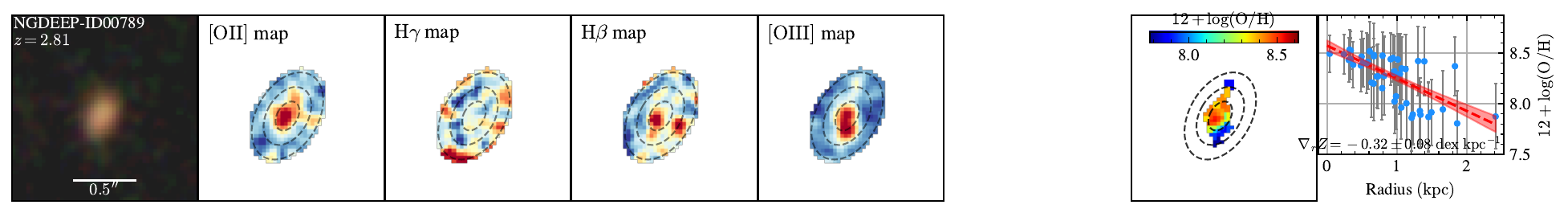}
\includegraphics[width=1\linewidth]{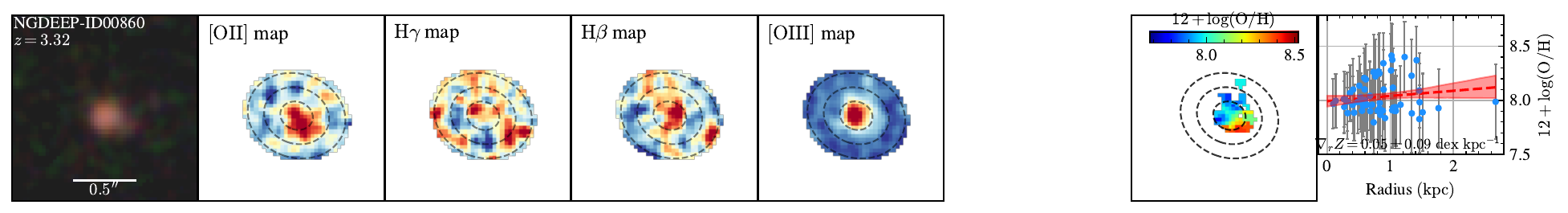}
\includegraphics[width=1\linewidth]{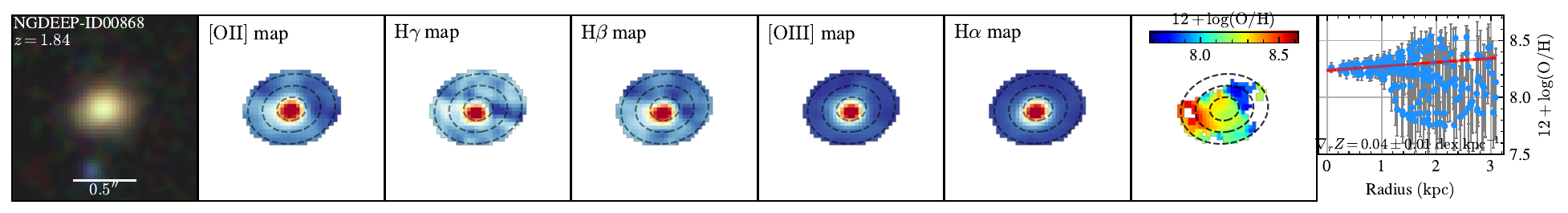}
\includegraphics[width=1\linewidth]{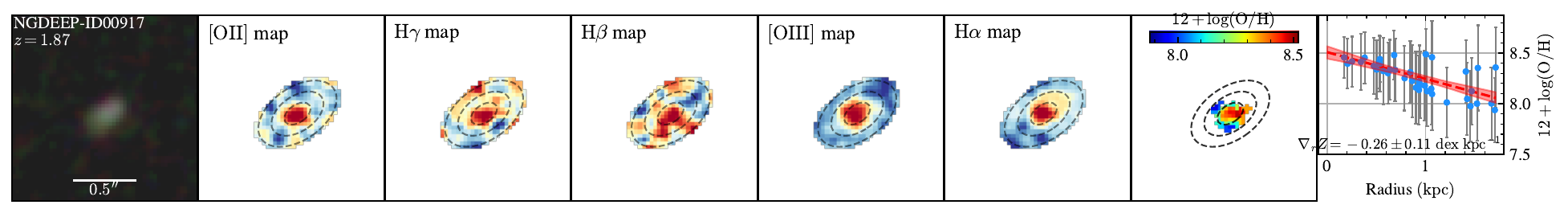}
\includegraphics[width=1\linewidth]{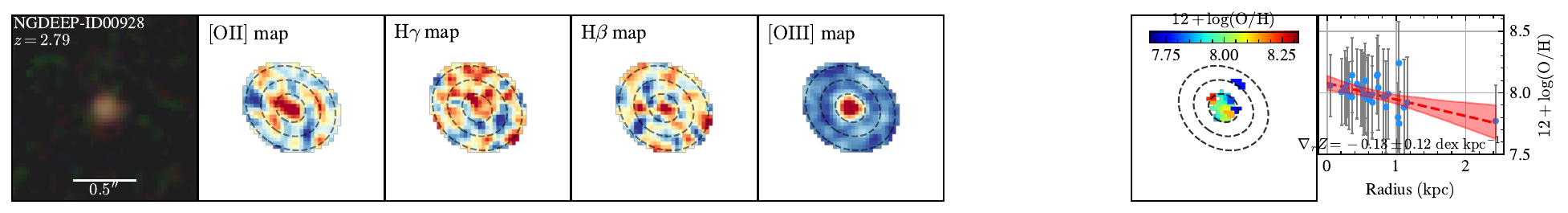}
\includegraphics[width=1\linewidth]{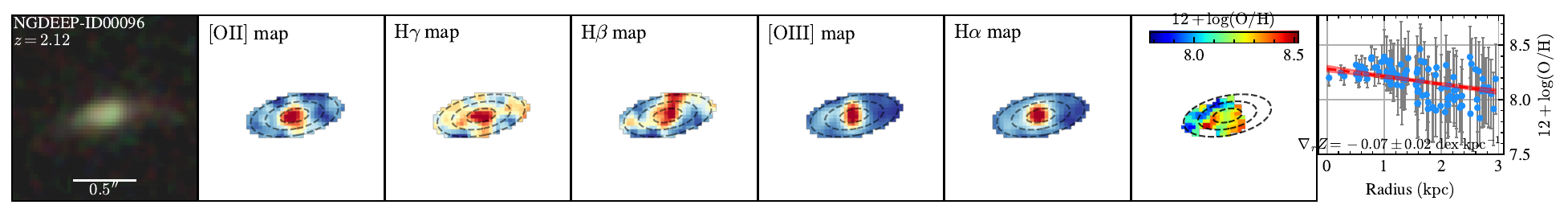}
\includegraphics[width=1\linewidth]{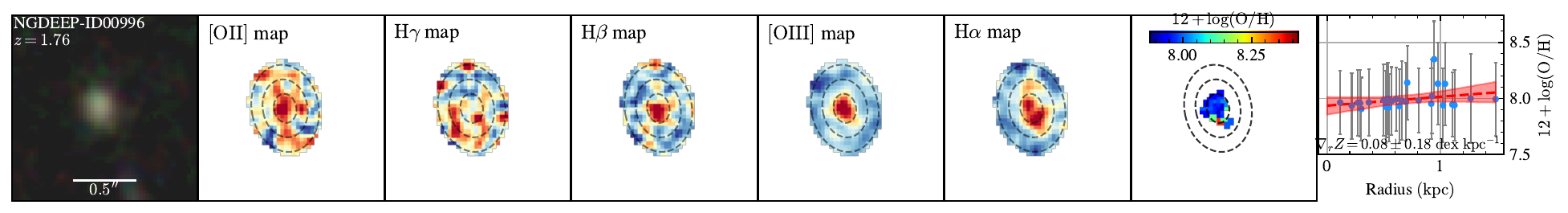}
\caption{Continued.}
\end{figure*}
\addtocounter{figure}{-1}

% \immediate\write18{ls NGDEEP_fig1/*.pdf > images-list.txt}

% \newcounter{allimagecounter}
% \newcounter{imagecounter}
% \newcounter{pagecounter}
% \setcounter{pagecounter}{1}
% \setcounter{imagecounter}{0}
% \setcounter{allimagecounter}{0}

% \immediate\write18{wc -l < images-list.txt > file-length.txt}
% \newread\file
% \openin\file=file-length.txt
% \read\file to \filelength
% \closein\file

% \newread\myfile
% \openin\myfile=images-list.txt

% \loop
%     \read\myfile to \filename
%     \ifeof\myfile\else
%         \ifnum\value{imagecounter}=0
%             \begin{figure*}[t!]
%                 \centering
%         \fi
%         \includegraphics[width=1\linewidth]{\filename}    
%         \stepcounter{imagecounter}
%         \stepcounter{allimagecounter}
%         \ifnum\value{imagecounter}=9
%             \ifnum\value{pagecounter}=1
%                 \caption{\textbf{The false-color image, line maps, metallicity maps, and metallicity gradients for each source in the NGDEEP sample.}}\label{fig:ngdeep-ind}
%             \else
%                 \caption{Continued.}
%             \fi
%             \end{figure*}
%             % \clearpage
%             \addtocounter{figure}{-1}
%             \setcounter{imagecounter}{0} % 重置图片计数器
%             \stepcounter{pagecounter} % 增加页码计数器 
%         \else
%             \ifnum\value{allimagecounter}=\filelength
%             \caption{Continued}
%             \end{figure*}
%             % \addtocounter{figure}{-1}
%             \fi
%         \fi
%     \repeat

% \closein\myfile

\end{appendix}

%% This command is needed to show the entire author+affiliation list when
%% the collaboration and author truncation commands are used.  It has to
%% go at the end of the manuscript.
%\allauthors

%% Include this line if you are using the \added, \replaced, \deleted
%% commands to see a summary list of all changes at the end of the article.
%\listofchanges

\end{document}